\documentclass[a4paper,10pt,fleqn]{article}
\pdfoutput=1
\usepackage{jheppub}
\usepackage[T1]{fontenc}

\usepackage{setspace}

\usepackage{xcolor}
\usepackage{booktabs}
\usepackage{graphicx}
\usepackage{dcolumn}
\usepackage{bm}
\usepackage{hyperref}
\usepackage{subcaption}
\usepackage{slashed}

\renewcommand{\vec}[0]{\boldsymbol}
\renewcommand{\triangle}{\mathrm{tri}}

\newcommand{\tr}{\mathrm{tr}}
\newcommand\Tr{\mathrm{Tr}}

\newcommand{\Eucl}[1]{\mathring{#1}}
\newcommand{\m}[1]{m_\mathrm{#1}}
\newcommand{\QEDL}{\mathrm{QED}_L}
\newcommand{\QEDinf}{\mathrm{QED}_\infty}
\newcommand{\Gam}[1]{\Gamma^{(#1)}}

\def\M{\mathcal M}
\def\real{\mathrm{real}}
\def\virt{\mathrm{virt}}
\def\GF{G_\mathrm{F}}
\def\Vud{V_\mathrm{ud}}

\def\max{\mathrm{max}}
\def\had{\mathrm{had}}
\def\J{\mathcal J}
\def\mW{m_W}
\def\up{\mathrm{u}}
\def\down{\mathrm{d}}

\def\Qu{Q_\up}
\def\Qd{Q_\down}
\def\out{\mathrm{out}}
\def\fm{\mathrm{fm}}
\def\mt{\m{\tau}}
\def\mg{\m{\gamma}}

\def\MS{{\overline{\mathrm{MS}}}}
\def\RI{\mathrm{RI}}
\def\Qbar{\overline{Q}}
\def\scheme{{\mathsf{s}}}
\def\O{\mathcal{O}}
\def\QCD{\mathrm{QCD}}

\def\C{\mathfrak{C}}
\def\X{\mathfrak{X}}
\def\D{\mathfrak{D}}

\def\win{\mathrm{win}}

\def\P{\mathcal P}

\def\rhopos{\rho^+}
\def\rhoneg{\rho^-}

\title{Isospin-breaking effects in inclusive hadronic $\tau$ data for the muon $(g-2)$ from first principles}

\author[a,b]{M.~Bruno,}

\author[e]{T.~Izubuchi,}

\author[c]{C.~Lehner,}

\author[d]{A.~S.~Meyer,}

\author[c]{J.~Parrino,}

\author[e]{X.~Tuo}

\affiliation[a]{Dipartimento di Fisica ``Giuseppe Occhialini'', Universit\`a degli Studi di Milano-Bicocca, Piazza della Scienza 3, 20126 Milan, Italy}
\affiliation[b]{Istituto Nazionale di Fisica Nucleare (INFN), Sezione di Milano-Bicocca, Piazza della Scienza 3, 20126 Milan, Italy}
\affiliation[c]{{Universit\"at Regensburg, Fakult\"at f\"ur Physik, Universit\"atsstra\ss e 31, 93040 Regensburg, Germany}}
\affiliation[d]{{Nuclear and Chemical Sciences Division, Lawrence Livermore National Laboratory, Livermore, California 94550, USA}}
\affiliation[e]{{Physics Department, Brookhaven National Laboratory, Upton, NY 11973, USA}}

\abstract{The knowledge of isospin-breaking effects in hadronic $\tau$ decays is required for a high-precision determination of the Hadronic-Vacuum-Polarization contribution to $(g-2)_\mu$ from experimental $\tau$ data. In this work we present a strategy for their calculation in a fully inclusive setup from first-principles Lattice QCD+QED simulations. 
We separate radiative corrections in three infrared safe classes, which we study individually. We provide analytic expressions for their effects in the initial state and propose a strategy for final-state corrections directly in Euclidean space. We also examine the non-factorizable contributions and highlight the challenges associated with their analytic continuation from Euclidean to Minkowski space. By studying short-distance corrections in the context of momentum schemes, we provide a prescription for the renormalization of the individual terms at first order in the ispospin-breaking parameters.}

\begin{document}

\maketitle

\def\gem{Cirigliano:2001er,Cirigliano:2002pv,Flores-Baez:2006yiq,Miranda:2020wdg}

\def\latib{Westin:2019tgc,Blum:2018mom,Giusti:2019xct,Chakraborty:2017tqp,Djukanovic:2024cmq,Parrino:2025afq,MILC:2024ryz,Ray:2022ycg,Altherr:2025kqw,Bruno:2026kqf,Borsanyi:2020mff,Boccaletti:2024guq}

\def\HVPlat{RBC:2018dos,Giusti:2019xct,Borsanyi:2020mff,Lehner:2020crt,Wang:2022lkq,Aubin:2022hgm,Ce:2022kxy,ExtendedTwistedMass:2022jpw,RBC:2023pvn,Kuberski:2024bcj,Boccaletti:2024guq,Spiegel:2024dec,RBC:2024fic,Djukanovic:2024cmq,ExtendedTwistedMass:2024nyi,MILC:2024ryz,Bazavov:2024eou}

\def\ILT{Hansen_2017,Bulava:2019kbi,HLT,Bailas:2020qmv,Gambino:2020crt,Bulava:2021fre,DelDebbio:2022qgu,Boito:2022njs,Gambino:2022dvu,alexandrou_2023,Barone:2023tbl,Bergamaschi:2023xzx,Bennett:2024cqv,Bruno:2024fqc,Patella:2024cto,Davier:2023cyp,DelDebbio:2024lwm,Tsuji:2026zku,Giusti:2026mcy,Abbott:2026wdw,Jay:2026qoh}

\def\HLbLlat{Blum:2019ugy,Chao:2021tvp,Chao:2022xzg,Blum:2023vlm,Fodor:2024jyn}
\def\HLbLdisp{Colangelo:2015ama,Masjuan:2017tvw,Colangelo:2017fiz,Hoferichter:2018kwz,Eichmann:2019tjk,Bijnens:2019ghy,Leutgeb:2019gbz,Cappiello:2019hwh,Masjuan:2020jsf,Bijnens:2020xnl,Bijnens:2021jqo,Danilkin:2021icn,Stamen:2022uqh,Leutgeb:2022lqw,Hoferichter:2023tgp,Hoferichter:2024fsj,Estrada:2024cfy,Ludtke:2024ase,Deineka:2024mzt,Eichmann:2024glq,Bijnens:2024jgh,Hoferichter:2024bae,Holz:2024diw,Cappiello:2025fyf}

\section{Introduction}

The anomalous magnetic moment of the muon, $a_\mu=(g-2)_\mu/2$, 
being one of the most precise experimental measurements in particle physics, is currently playing a fundamental r\^ole in establishing the limits of validity of the Standard Model.
Theoretical predictions, given the precise knowledge of the QED and weak contributions, are dominated by QCD uncertainties coming from the (leading order) Hadronic Vacuum Polarization (HVP) and the Hadronic Light-by-Light contributions. 
For the latter first-principles calculations from Lattice QCD and dispersive analysis are in very good agreement~\cite{\HLbLlat,\HLbLdisp}, while for the former tensions are present. We refer the interested reader to the two White Papers written by the ``Muon g-2 theory initiative'' for all details and references on the topic~\cite{Aoyama:2020ynm,Aliberti:2025beg}.\\

From unitarity and analyticity, the Leading-Order (LO) HVP may be related to hadronic cross-sections measured in $e^+e^-$ collisions, through a dispersive integral involving a specific kernel function~\cite{Bouchiat:1961lbg,Brodsky:1967sr,Lautrup:1969fr,Gourdin:1969dm}. 
Among all channels, the $\pi^-\pi^+$ cross-section dominates both the signal and the error of the HVP contribution to $a_\mu$, but at present, several significant tensions among different experimental determinations, in particular among BaBar, KLOE and CMD-3 \cite{BaBar:2009wpw,BaBar:2012bdw,CMD-3:2023alj,KLOE:2008fmq,KLOE:2010qei,KLOE:2012anl}, are limiting the attainable accuracy (see for instance Ref.~\cite{Davier:2023fpl}). 
Considering in addition the ongoing worldwide experimental effort in the direct determination of the muon anomaly~\cite{Muong-2:2021ojo,Muong-2:2023cdq,Abe:2019thb}, it is particularly timely to consider the possibility of using hadronic $\tau$ decays as an alternative source of experimental data, as proposed long ago in Ref.~\cite{Alemany:1997tn}. 

Being mediated by weak interactions, one should first isolate the underlying vector spectral density. In the isospin limit, it coincides with the isovector component of the spectral density defined from $e^+e^-$ hadronic cross-sections but an accurate prediction of the HVP contribution to $a_\mu$ requires the appropriate knowledge of isospin-breaking corrections. The latter are predicted from the SM and are particularly delicate given the low-energy non-perturbative nature of the processes involved.
Short-distance electro-weak (EW) effects, addressed in perturbation theory, are sufficiently under control for the current target precision~\cite{Aliberti:2025beg}. The two contributions where non-perturbative physics is predominant are long-distance radiative effects~\cite{\gem} and isospin-breaking corrections to the spectral functions, and more specifically to the two-pion charged and neutral form factors. At present they have been calculated using different phenomenological approaches based on chiral perturbation theory, resonance chiral theory and resonance models~\cite{Alemany:1997tn,Cirigliano:2001er,Cirigliano:2002pv,Davier:2002dy,Flores-Baez:2006yiq,Flores-Tlalpa:2006snz,Flores-Baez:2007vnd,Davier:2010fmf,Davier:2023fpl,Castro:2024prg,Allen:2026iad}. The importance of a model-independent calculation of these effects, given the relevance of $a_\mu$ in the current particle physics panorama, has been highlighted in the latest White Paper~\cite{Aliberti:2025beg}.
Thanks to significant progress in the dispersive treatment of long-distance radiative corrections~\cite{Monnard:2021pvm,Colangelo:2025ivq}, the need for a first-principles determination is now largely confined to the form-factor difference, which remains the leading source of uncertainty in the current White Paper~\cite{Aliberti:2025beg}.
The goal of this work is to develop a strategy based on lattice QCD, the only first-principles framework capable of providing a non-perturbative determination directly from the QCD Lagrangian. By analyzing the complete problem, we introduce a separation between factorizable and non-factorizable radiative contributions. For the former, we formulate a strategy to compute the difference between neutral and charged spectral densities -- and, consequently, their contribution to $a_\mu$ -- which is currently the most pressing theoretical input. For the latter, we find that their determination from an Euclidean formulation is considerably more challenging, suggesting that a  synergy with the dispersive approach may be needed for a first determination~\cite{Colangelo:2025ivq}.\\

Over the last years enormous progress has been achieved in the prediction of the HVP contribution to $a_\mu$ using Lattice QCD, both in our formal understanding of discretization~\cite{Ce:2021xgd,Husung:2019ytz,Husung:2022kvi,Sommer:2022wac} and finite-volume errors~\cite{Hansen:2019rbh,Hansen:2020whp}, but more importantly in its numerical determination of the short, intermediate and long-distance Euclidean windows~\cite{\HVPlat}.
Such a community effort has been primarily focused on the isospin symmetric limit, where up and down quarks are degenerate in mass and the electromagnetic (EM) coupling, $e$, is set to zero. However, reliable and accurate predictions from the Standard (SM) at the sub-percent level require in addition the inclusion of isospin-breaking effects, another topic where substantial progress has been achieved~\cite{\latib} but additional effort is still necessary~\cite{Aliberti:2025beg}\footnote{An extension of the long-distance reconstruction method to the QED corrections has recently appeared \cite{Lehner:2025qrl}.}.
By leveraging this vast progress, in this work we discuss a strategy to predict the (LO) HVP contribution to $a_\mu$ starting from experimental information on hadronic $\tau$ decays, supplemented by the appropriate isospin-breaking corrections, whose calculation from Lattice QCD simulations is outlined.
Contrary to the typical approach to this problem where the two-pion channel (and its isospin-breaking effects) is studied in an exclusive manner, here we describe a strategy which is fully inclusive. While from the experimental point of view this may be less desirable, such an approach has significant benefits from the Lattice QCD perspective. In fact, already in the isospin limit, the exclusive study of the two-pion spectral density above the first open inelastic threshold (four pions) is currently difficult. The finite-volume method~\cite{Luscher:1985dn,Luscher:1986pf,Lellouch:2000pv,Meyer:2011um} successfully used in the past (see e.g. Refs.~\cite{Feng:2014gba,Erben:2019nmx,Andersen:2018mau}) is limited to the elastic kinematic region. A promising alternative receiving significant attention is the direct extraction of spectral densities from lattice Euclidean correlators~\cite{\ILT}. In a few words, it allows to bypass the problem of analytic continuation to estimate convolutions of spectral densities with arbitrary kernels and it can be used for the study of the exclusive two-pion channel in the inelastic regime~\cite{Bulava:2019kbi,Bruno:2020kyl,Patella:2024cto,Morandi:2026nll}. As we will see below, such a technology plays an important r\^ole in a specific contribution to radiative corrections in $\tau$ decays, making evident how the current study is well aligned also with these important recent developments.

This manuscript is organized as follows. In Section~\ref{sec:anatomy} we introduce the notation and define the basic currents and correlators needed in our study, while briefly reviewing hadronic $\tau$ decays at Leading Order (LO) and the HVP calculation from Lattice QCD in the isosymmetric limit. We conclude Section~\ref{sec:anatomy} by presenting the classification of radiative corrections and by summarizing the main strategy to calculate $a_\mu$ from $\tau$-data.
The first ingredient to consider is short-distance corrections covered in Section~\ref{sec:short-distance}. For the matching with the Standard Model we rely on recent perturbative results in momentum schemes, which we then adopt also for the renormalization of the typical composite operator used in the Fermi Lagrangian. 
Radiative corrections of various nature are then discussed sequentially in three separate Sections: more specifically initial-state, non-factorizable and final-state corrections are discussed in Sections~\ref{sec:init-state}, \ref{sec:non-fact} and \ref{sec:fin-state} respectively. After the conclusions, several useful algebraic steps are collected in the Appendices together with our conventions.

\section{Anatomy of radiative corrections}
\label{sec:anatomy}

Being mediated by weak interactions, the natural theoretical framework to study hadronic $\tau$ decays is an Effective Field Theory (EFT) where the $W$ bosons are integrated out. The process that we study is the decay of the $\tau$ lepton into hadronic light vector states (typically referred to as ``non-strange'')
\begin{equation}
    \tau^- (P) \overset{W^-}{\to} \nu_\tau(q) \, \had(p) \,.
\end{equation}
We denote with $p$ the total four momentum of the final hadronic state,
\begin{equation}
    p_\mu = (E, \vec p) \,, \quad
    E = \sum_{i=1}^{n_f} \omega_i \,, \quad \vec p = \sum_{i=1}^{n_f} \vec p_i \,,
\end{equation}
where the energy of the $i$-th particle is $\omega_i = \sqrt{m_i^2 + |\vec p_i|^2}$ and the sums run over the particle content of a given final-state (light vector) channel $f$, e.g. $\pi^-\pi^0$ or $\pi^- \, 3\pi^0$. 
The various on-shell relations read
\begin{equation}
    P^2 = \mt^2 \,, \quad q^2 =0 \,, \quad p^2 = E^2 - |\vec p|^2 = s \,,
    \label{eq:onshell}
\end{equation}
with $\mt$ being the $\tau$ mass and $s$ the invariant mass of the hadronic system. In our notation $\omega_P=\sqrt{\mt^2 + |\vec P|^2}$ and $\omega_q = |\vec q|$, with $P_\mu$ and $q_\mu$ the momenta of the lepton and neutrino respectively.
The electromagnetic (up to a factor $-ie$) and charged hadronic currents relevant for our study are
\begin{align}
    \J^\gamma_\mu(x) = & \, \Qu (\bar u \gamma_\mu u)(x) + \Qd (\bar d \gamma_\mu d)(x) \,, \\
    \J^+_\mu(x) = & \, \frac{1}{\sqrt 2} (\bar d \gamma_\mu u)(x) \,, \quad \J^-_\mu(x) = \frac{1}{\sqrt 2} (\bar u \gamma_\mu d)(x) \,, \\
    \J^{L,-}_\mu(x)= & \, (\bar u \gamma_\mu^L d)(x) \,,
\end{align}
where as usual $\Qu = \frac23 \,, \Qd = - \frac13 $  and $\gamma_\mu^L = \gamma_\mu (1 - \gamma_5)/2$ (similarly we have $\gamma_\mu^R = \gamma_\mu(1 + \gamma_5)/2$). The fine structure constant $\alpha$ is given in terms of the electric charge according to $e^2 = 4 \pi \alpha$. Notice that the charged currents $\J^\pm_\mu$ have been normalized such that their two-point spectral density coincides with the isovector neutral one, see Subsection~\ref{sec:hvp}.
In our notation we use $\Eucl x$ to denote an object with Euclidean signature. For currents instead we use the letter $j$ and for both neutral and charged vector currents the relations $j_0(t,\vec x) = \J_0(-it, \vec x)$ and $j_k(t, \vec x) = i \J_k(-it, \vec x)$ hold for $t \in \mathbb R$ (c.f. Appendix~\ref{app:conventions}). Currents projected to definite momentum are defined as
\begin{equation}
    \widetilde {\mathcal J}_\mu(x_0, \vec p) = \int d^3 \vec x \, e^{- i \vec p \cdot \vec x} \mathcal J_\mu(x_0, \vec x) \,,
\end{equation}
with $\vec p$ leaving the operator. 
For simplicity we formulate our strategy targeting discretizations retaining good chiral properties at finite lattice spacing and defer adaptations to other choices to future studies. In addition we consider only local currents, which at finite lattice spacing require a multiplicative factor which is discussed in Section~\ref{sec:short-distance} and omitted in this Section for a better readability.

\subsection{Leading-order hadronic decays}

To better comprehend isospin-breaking corrections we start from the leading order amplitude describing the hadronic on-shell decay in the isospin limit (repeated indices are summed over)
\begin{equation}
    i \M_f(P, q, p_1 \cdots p_{n_f}) = -\frac{4 i \GF}{\sqrt 2} \Vud^\ast \, \bar u (q) \gamma^\mu_L u (P) \,
    \langle \out, \vec p_1 \cdots \vec p_{n_f} | \J^{L,-}_\mu (0) | 0 \rangle  \,, 
\end{equation}
where the intermediate weak boson has been integrated out in favor of $\GF$. We omit from the notation the explicit dependence on the spin indices of the lepton-neutrino pair. Since we are ultimately interested in the HVP contribution to $a_\mu$ we select only hadronic vector states, denoted by $\langle \out, \vec p_1 \cdots \vec p_{n_f}|$ throughout the manuscript. It follows that at this order the axial component is projected away leading to 
\begin{equation}
    \M_f(P, q, p_1 \cdots p_{n_f}) = -2 \GF \Vud^\ast \, \bar u (q) \gamma^\mu_L u (P) \,
    \langle \out, \vec p_1 \cdots \vec p_{n_f} | \J^-_\mu (0) | 0 \rangle  \,.
    \label{eq:Mf}
\end{equation}
Higher order terms in the EFT are sufficiently suppressed and can be ignored\footnote{This may be understood by examining the large $M_W$ limit of the weak propagator at the typical scale of the problem, $q^2 = \mt^2$,
\begin{equation}
    \frac{1}{q^2 - M_W^2} \to \GF \left[ 1 + O\left(\frac{\mt^2}{M_W^2}\right) \right]  \,.
\end{equation}
}.
By denoting with $\Gamma$ the inclusive hadronic rate at leading order, the differential decay reads
\begin{equation}
    d\Gamma \equiv \frac{1}{2 \mt} \sum_f \frac{d^3 \vec q}{(2\pi)^3 2 \omega_q}  \,
    d \Phi_f(P-q) \, \frac12 \sum_\mathrm{spin} |\M_f(P, q, p_1 \cdots p_{n_f})|^2 \,,
\end{equation}
with $d \Phi_f$ the phase-space factor for a fixed final hadronic state $f$
\begin{equation}
    d\Phi_f(p) \equiv (2\pi)^4 \delta^4(p-p_1 \dots -p_{n_f})
    S_f \prod_{i=1}^{n_f} \frac{d^3 \vec p_i}{(2\pi)^3 2 \omega_i} \,,
\end{equation}
with the symmetry factor $S_f$ taking into account identical hadrons.
As the notation suggests, by treating the hadronic system as a combined object, the inclusive decay may be efficiently described as a $1 \to 2$ process, despite involving several particles in the final state. 
To achieve this we start from the identity
\begin{equation}
    \int \frac{d^3 \vec p}{(2\pi)^3 2 \omega_p} f(p) = \int \frac{d^4 p}{(2\pi)^4} (2\pi) \, \delta(p^2-s) \, \theta(p_0) \, f(p)\,,
    \label{eq:rev_uni}
\end{equation}
which leads us to the following manipulation
\begin{equation}
    d\Phi_f(P-q) = \int ds \, \frac{d^3 \vec p}{(2\pi)^3 2 E} (2\pi)^4 \delta^4(P-q-p) \frac{1}{2\pi} \, d\Phi_f(p) \,.
    \label{eq:dPhif}
\end{equation}
Now with little algebra one readily arrives at
\begin{equation}
    \frac{d\Gamma}{ds} = \frac{\GF^2 |\Vud|^2}{\mt} \int \frac{d^3 \vec q}{(2\pi)^3 2 \omega_q} \frac{d^3 \vec p}{(2\pi)^3 2 E} \, (2\pi)^4 \delta^4(P-q-p) \, \mathcal L^{\mu\nu}(P,q) \, \rho_{\mu\nu}(p) \,,
\end{equation}
with the well-known leptonic tensor
\begin{equation}
    \mathcal L^{\mu\nu}(P,q) = 2 \big[ P^\mu q^\nu + q^\mu P^\nu - g^{\mu\nu} (P\cdot q) - i \epsilon^{\mu\nu\rho\sigma} P_\rho q_\sigma \big] \,,
\end{equation}
and the inclusive (tensorial) spectral density~\cite{Gasser:1983yg}
\begin{align}
    \rho_{\mu\nu}(p) = \frac{1}{2\pi} \int d^4x \, e^{ipx} \, \langle 0 \vert \J^+_\mu(x) \, \J^-_\nu(0) \vert 0 \rangle = \sum_f \int \frac{d \Phi_f(p)}{2\pi} \, \mathcal H_{f,\mu\nu}(p_1, \dots  , p_{n_f}) \notag \,,
    \label{eq:rho_munu}
\end{align}
defined in terms of the hadronic tensor
\begin{equation}    
    \mathcal H_{f,\mu\nu} (p_1, \dots  , p_{n_f}) = 
    \langle 0 \vert \J^+_\mu(0) \vert \text{out}, \vec p_1 \cdots \vec p_{n_f}  \rangle  \langle \text{out} ,
    \vec p_1 \cdots \vec p_{n_f} \vert \J^-_\nu(0) \vert 0 \rangle \,.
    \label{eq:H_munu}
\end{equation}
Alternatively we could have used the notation with Dirac's $\delta$ (e.g. adopted in Refs.~\cite{HLT,Bruno:2020kyl})
\begin{equation}
    \rho_{\mu\nu}(p) = \langle 0 \vert \widetilde \J^+_\mu(0, \vec p) \, \delta(\hat H - E) \, \J^-_\nu(0) \vert 0 \rangle \,,
\end{equation}
in terms of the QCD Hamiltonian $\hat H$.
From Lorentz and vector current conservation, one may express $\rho_{\mu\nu}(p)$ in terms of a single (scalar) spectral density $\rho(s)$
\begin{equation}
    \rho_{\mu\nu}(p) = \, (p_\mu p_\nu - g_{\mu\nu} s) \rho(s) \,.
    \label{eq:rho_munu}
\end{equation}
After performing the integral over the phase space, one easily arrives at the following functional form for the differential decay rate
\begin{equation}
    \frac{d\Gamma}{ds} = \frac{\GF^2 |\Vud|^2 \mt^3}{8\pi} \, \kappa(\hat s) \, \rho(s) \,, \quad \kappa(\hat s) = 
    \left(1 + 2 \hat s\right) \left(1 - \hat s \right)^2 \, \quad \hat s = \frac{s}{\mt^2} \,.
    \label{eq:dGammads}
\end{equation}
In Appendix~\ref{app:lips} we show the connection with the commonly studied two-pion case, explaining also the apparent different normalization w.r.t. the typical literature on the topic, where the symbol $v_1(s)$ is often used to indicate the hadronic density.

Following common practice from the experimental community we normalize the differential decay by the total electronic rate (at LO)
\begin{equation}
    \Gamma(\tau^- \to e^- \nu_e \nu_\tau) = \Gamma_e = \frac{\mathcal B_e\, \Gamma}{\mathcal B}= \frac{\GF^2 \mt^5}{192 \pi^3} \,.
\end{equation}
By expressing it in terms of the inclusive hadronic and electronic branching ratios $\mathcal B$ and $\mathcal B_e$, one finds the well-known representation of the spectral density in terms of experimental data
\begin{equation}
    \rho^{\exp}(s) = \frac{\mt^2}{24 \pi^2 |\Vud|^2\, \kappa(\hat s)} \left[ \frac{\mathcal B}{\mathcal B_e} \frac{1}{\Gamma} \frac{d\Gamma}{ds} \right]^{\exp} \,.
    \label{eq:rho_exp}
\end{equation}
Note that with our notation we want to distinguish two objects which are different from the theoretical point of view: $\rho(s)$ in Eq.~\eqref{eq:rho_munu} denotes the spectral density as an expectation value of well-defined operators defined for all values of $s$, while $\rho^{\exp}(s)$ refers to the extraction of the spectral density from the differential decay rate, which is taken from experimental measurements and therefore has support up to the $\tau$ mass. $\rho(s)$ can be calculated from the SM, and in the kinematic regime $s < \mt^2$ it coincides with $\rho^{\exp}(s)$ (in the absence of strong effects from new physics).

\subsection{The Hadronic-Vacuum-Polarization and Euclidean correlators}
\label{sec:hvp}

In this Section we discuss the HVP contribution to $a_\mu$ and establish proper connections between the results seen above and the corresponding Euclidean calculations. We adopt the Time-Momentum representation and perform the Wick rotation to real Euclidean time $t$
\begin{equation}
    G^W(t) \equiv \frac13 \sum_k \int d^3 \vec x \langle \J_k^{+}(-i t,\vec x) \J_k^{-}(0) \rangle = \int ds \, \frac{\sqrt s}{2} \, e^{-\sqrt s t} \, \rho(s)  \,.
    \label{eq:GWt}
\end{equation}
The K\"allen-Lehmann spectral decomposition, via the Laplace transform, provides the connection between the spectral density $\rho(s)$ introduced above and the lattice correlator (with Euclidean currents, see Appendix~\ref{app:conventions})
\begin{equation}
    G^W(t) = - \frac13 \sum_k \int d^3 \vec x \, \langle j^{+}_k(t,\vec x) j^{-}_k(0) \rangle \,.
\end{equation}
The HVP contribution to $a_\mu$ may be calculated entirely with Euclidean signature~\cite{Blum:2002ii,Bernecker:2011gh},
\begin{equation}
    a_\mu = \int_0^\infty dt \,  w(t,m_\mu) \, G^\gamma (t),
    \label{eq:amu_tmr}
\end{equation}
from the two-point Euclidean correlator
\begin{equation}
    G^\gamma (t) \equiv -\frac{1}{3} \sum_k \int d^3 \vec x \ \langle
    j_k^\gamma (t, \vec x) j_k^\gamma (0) \rangle  \,.
    \label{eq:Ggamma}
\end{equation}
In Eq.~\eqref{eq:amu_tmr} the weights $w(t,m_\mu)$~\cite{Blum:2002ii,Bernecker:2011gh} play the r\^ole of the muon kernel typically given as a function of $\sqrt s$ in the dispersive framework~\cite{Bouchiat:1961lbg,Brodsky:1967sr,Lautrup:1969fr,Gourdin:1969dm}.
Recently, the splitting of the integral in Eq.~\eqref{eq:amu_tmr}  in different Euclidean-time windows has been proposed~\cite{Blum:2018mom}. The community has agreed to focus onto three specific intervals, defining a short-distance, an intermediate and a long-distance window. Here we concentrate mostly on the intermediate window
\begin{equation}
    a_\mu^\win = \int_0^\infty dt \, w(m_\mu,t) \, G^\gamma(t) \, \Theta^\win(t) \,, \quad \Theta^\win(t) = \frac12\left(\tanh\left(\frac{t-t_0}{\Delta}\right) - \tanh\left(\frac{t-t_1}{\Delta}\right)\right),
\end{equation}
for $t_0=0.4~\fm$, $t_1=1.0~\fm$, $\Delta=0.15~\fm$ but our derivation can be easily extended to any other case.
Such a division turned out to be instrumental for a better scrutiny of $a_\mu$, and led to the identification of tensions between dispersive and lattice determinations of the same windows~\cite{Aliberti:2025beg}. Such tensions with the data-driven approach represent at the moment one of the most intriguing questions, and with this (and future) work based on $\tau$ data we aim at shedding some light onto this puzzle.

In the isospin limit, from Wick's theorem, one obtains quark-connected and quark-disconnected correlation functions. Since such calculations are performed with Lattice QCD, we write their expression in terms of Euclidean Dirac's matrices $\Eucl \gamma_\mu$, Euclidean coordinates $\Eucl x_\mu$ and light quark propagators\footnote{
In this work we consider only light degenerate quarks and we omit color and spin indices.} 
$D^{-1}_r(\Eucl x,\Eucl y)$ (with $r=\up,\down$) defined with Euclidean signature
\begin{align}
    \label{eq:C}
    (c)(\Eucl x_4) = & \, \frac{1}{3} \sum_{k} \int d^3\vec x \
    \langle \tr \big[
    \Eucl \gamma_k D^{-1}_r (\Eucl x,0) \Eucl \gamma_k D^{-1}_r (0,\Eucl x)
    \big] \rangle \,, \\
    (d)(\Eucl x_4) = & \, \frac{1}{3} \sum_{k} \int d^3\vec x \ 
    \langle \tr \big[
    \Eucl \gamma_k D^{-1}_r (\Eucl x,\Eucl x) \big] \tr \big[
    \Eucl \gamma_k D^{-1}_r (0,0) \big] \rangle \,.
    \label{eq:D}
\end{align}
In our notation only the factor $1/3$ is part of the definition of a ``diagram'', while minus signs due to fermionic traces or charge factors are kept explicit (see Eq.~\eqref{eq:Ggamma-iso2} below).
For later convenience, we decompose the electromagnetic current into its two isospin components, (with $I=0$ and 1 and $I_3=0$)
\begin{equation}
    \J_\mu^{\gamma,0} = \frac{\Qu+\Qd}{2} \big(\bar u \gamma_\mu u + \bar d \gamma_\mu d \big) \,, \quad
    \J_\mu^{\gamma,1} = \frac{\Qu-\Qd}{2} \big(\bar u \gamma_\mu u - \bar d \gamma_\mu d \big) \,.
    \label{eq:jmu-isospin}
\end{equation}
By introducing the auxiliary correlation function with the corresponding Euclidean currents
\begin{equation}
    G_{II'}^\gamma(t) \equiv -\frac{1}{3} \sum_k \int d^3 \vec x \ \langle
    j_k^{\gamma,I} (t, \vec x) j_k^{\gamma,I'} (0)
    \rangle \,,
\end{equation}
we decompose the correlator introduced in Eq.~\eqref{eq:Ggamma} accordingly
\begin{equation}
    G^\gamma(t) \equiv G_{00}^\gamma(t) + 2 G_{01}^\gamma(t) + G_{11}^\gamma(t) \,.
    \label{eq:Ggamma-iso}
\end{equation}
In the isospin limit $G_{01}^\gamma(t)$ vanishes and $G^\gamma(t)$ simplifies to (omitting the trivial time dependence)
\begin{equation}
    G^\gamma(t) = \frac{(\Qu + \Qd)^2}{2} ((c) - 2(d))(t)
    + \frac{(\Qu - \Qd)^2}{2} (c)(t)
 \,.
    \label{eq:Ggamma-iso2}
\end{equation}
Moreover, since $\Qu-\Qd=1$, $G^W(t) = G^\gamma_{11}(t) = \frac12 (c)(t)$ and $\rho(s)$ coincides, at LO, with the isovector component of the neutral spectral density. The normalization of the currents $\J^\pm_\mu$ has been chosen to achieve this simple relation.
Starting from the representation of the spectral density given in Eq.~\eqref{eq:rho_exp} we can formulate the problem entirely in Euclidean space-time by adopting the correlator
\begin{equation}
    G^{\exp}(t) = \mathfrak{L}(t) \cdot \left[\frac{1}{\Gamma_e} \frac{d \Gamma}{ds} \right]^{\exp},
\end{equation}
and the linear operator
\begin{equation}
    \mathfrak{L} (t) \cdot f = \frac{\mt^2}{24 \pi^2 |\Vud|^2} \int_0^\infty ds \, \frac{\sqrt s}{2} e^{-\sqrt s t} \, \frac{\theta(\hat s-1)}{\kappa(\hat s)}  f(s) \,.
\end{equation}

\subsection{Euclidean correlators and inverse problems} 
\label{sec:inverse}
Before continuing the analysis on radiative corrections, we examine the relation between smeared spectral densities and Euclidean correlators, to establish a framework which will be useful later.
The HVP contribution to the muon anomalous moment is directly calculable in Euclidean space-time~\cite{Blum:2002ii,Meyer:2011um}, but another similar and interesting quantity, the smeared R-ratio, is not. 
To understand the difference let us consider the situation of a generic spectral density $\dot \rho(\omega^2)$, with support over $[\omega_0, \infty)$, with $\omega_0>0$, defined such that the integral 
\begin{equation}
    \dot \rho_K = \int d\omega \, K(\omega) \, \dot\rho(\omega^2) \,, \quad \text{with } K(\omega) = \frac{1}{\omega - \bar\omega - i \sigma} \,,
\end{equation}
exists. The corresponding Euclidean correlator is $\dot G(t) = \int d\omega \, e^{- \omega t} \dot \rho(\omega^2)$.
By introducing an integral representation of the smearing kernel $K$, we re-write the physical smeared density as
\begin{equation}
    \dot \rho_K = \lim_{T \to \infty} \int d\omega  \int_0^T dt \, e^{-(\omega - \bar\omega - i \sigma) t} \, \dot \rho(\omega) = \lim_{T \to \infty} \int_0^T dt \, e^{(\bar\omega + i\sigma) t} \, \dot G(t) \,. 
\end{equation}
Evidently the limit $T\to\infty$ exists only if $\bar\omega<\omega_0$, namely if $\bar\omega$ is below the support of $\dot \rho(s)$, which starts typically at a $n$-particle production threshold, once poles of bound states are subtracted. This simple picture offers a very useful guideline on inverse problems in the context of Lattice field theories: there is an interplay between the analytic structures of the kernel and the spectral density, and in our example if poles are located on the left of the particle-production branch cut then a direct analytic continuation is possible, in the sense that $\dot\rho_K$ is calculable from an integral over Euclidean time (like the HVP contribution to $a_\mu$) with coefficients $g(t) = e^{(\bar\omega + i\sigma) t}$.
In the opposite situation a direct continuation is not possible, and finding the coefficients $g(t)$ (formally) given by
\begin{equation}
    K(\omega) = \int_0^\infty dt \, e^{-\omega t} \, g(t) \,,
    \label{eq:Kg}
\end{equation}
defines an inverse problem, namely the inverse Laplace transform problem. Their explicit calculation requires a regularization procedure and the regulator may be removed after integration~\cite{HLT,Bruno:2024fqc}.

Recently, within the Lattice QCD community, there has been significant activity on the development of numerical and formal methods to attack the inverse Laplace transform~\cite{\ILT}. While in practice several strategies are viable, in this work we take a continuum and infinite-volume point of view and discuss problems related to the analytic continuation by studying equations of the form \eqref{eq:Kg}.

\subsection{Summary of the strategy}
\label{sec:summary}

Having established the connection between Minkowski and Euclidean quantities, we now outline our strategy built around the calculation of $G^W$ from $G^{\exp}$ including isospin-breaking effects, with the ultimate goal of using it for the HVP contribution to $a_\mu$.
We adopt a perturbative approach in the isospin-breaking parameters, as in Refs.~\cite{deDivitiis:2013xla,Giusti:2017dmp}. Expectation values $\langle \cdot \rangle$ are always non-perturbative in the strong coupling and taken in the isosymmetric background. Observables with the superscript ${}^\QCD$ denote the absence of QED and strong-isospin breaking (SIB) effects.\\

\begin{figure}[ht]
    \centering
    \includegraphics[width=.45\textwidth]{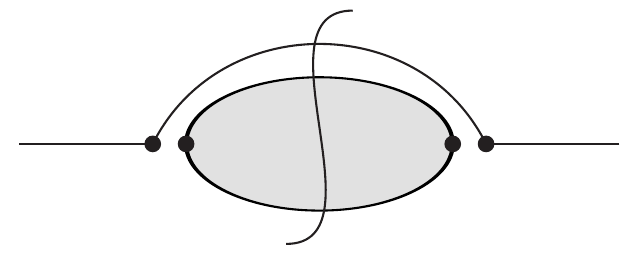}
    \caption{Diagrammatic representation of the LO contribution to the decay rate. The vertical squiggle line stands for a final-state cut, i.e. particles crossing it are on-shell.}
    \label{fig:tau-rate}
\end{figure}

\begin{figure}[ht]
    \centering
    \begin{subfigure}[t]{.23\textwidth}
        \centering
        \includegraphics[width=\textwidth]{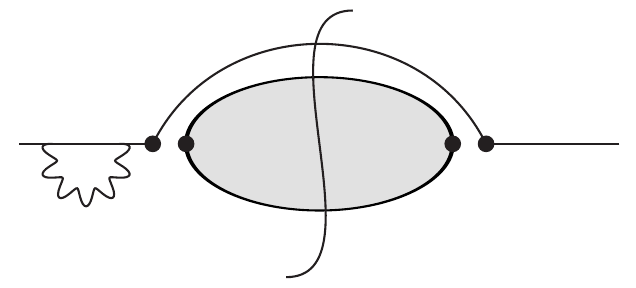}
        \caption{$\tau$ self-energy.}
    \end{subfigure}
    \hspace{2em}
    \begin{subfigure}[t]{.23\textwidth}
        \centering
        \includegraphics[width=\textwidth]{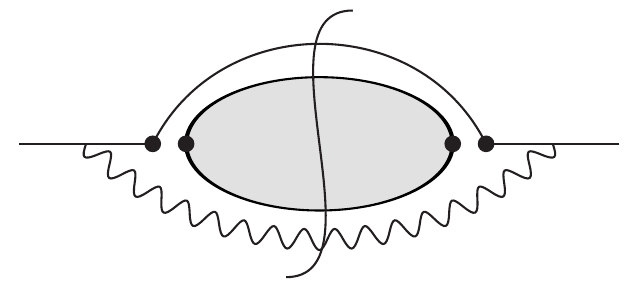}
        \caption{Real photon emission.}
    \end{subfigure}
    \caption{Initial-state radiative corrections, discussed in Section~\ref{sec:init-state}. The sum of these contributions defines $d\Gamma^\mathrm{init}/ds$. We do not draw the hermitian conjugate diagrams.}
    \label{fig:init-state}
\end{figure}

\begin{figure}[ht]
    \centering
    \begin{subfigure}[t]{.23\textwidth}
        \centering
        \includegraphics[width=\textwidth]{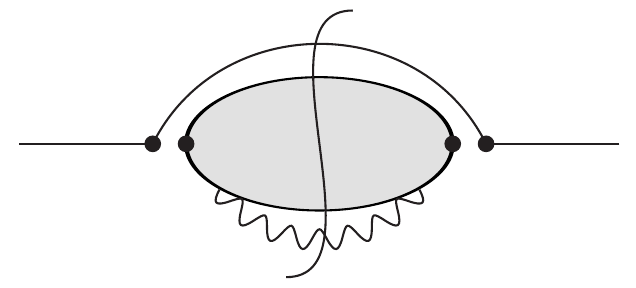}
        \caption{Real photon emission.}
    \end{subfigure}
    \begin{subfigure}[t]{.23\textwidth}
        \centering
        \includegraphics[width=\textwidth]{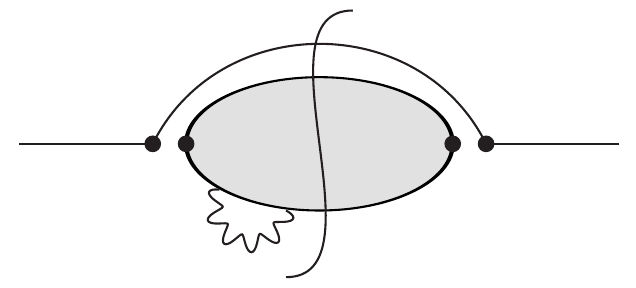}
        \caption{Quark self energies.}
    \end{subfigure}
    \begin{subfigure}[t]{.23\textwidth}
        \centering
        \includegraphics[width=\textwidth]{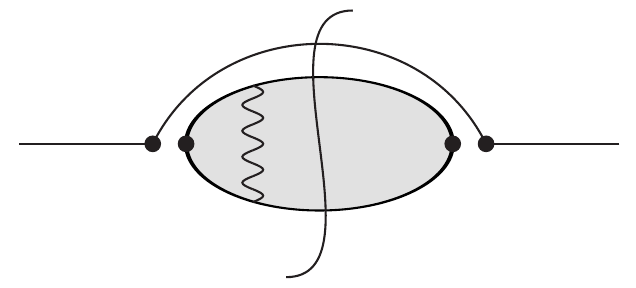}
        \caption{Vertex corrections.}
    \end{subfigure}
    \caption{Final-state radiative corrections. Their sum defines $d\Gamma^\mathrm{fin}/ds$ and these are pictorial representations. Precise diagrams and definitions are given in Section~\ref{sec:fin-state} for the corresponding Euclidean correlators.}
    \label{fig:fin-state}
\end{figure}

\begin{figure}[ht]
    \centering
    \begin{subfigure}[t]{.23\textwidth}
        \centering
        \includegraphics[width=\textwidth]{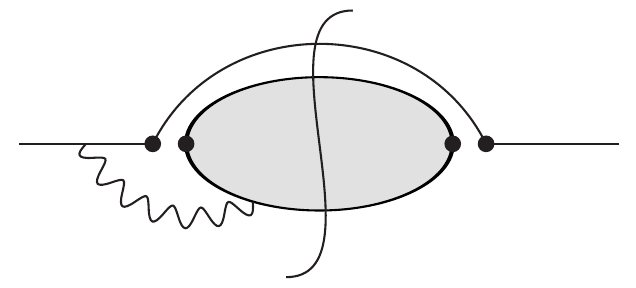}
        \caption{Loop diagram with a virtual photon.}
        \label{fig:virt_loop}        
    \end{subfigure}
    \hspace{2em}
    \begin{subfigure}[t]{.23\textwidth}
        \centering
        \includegraphics[width=\textwidth]{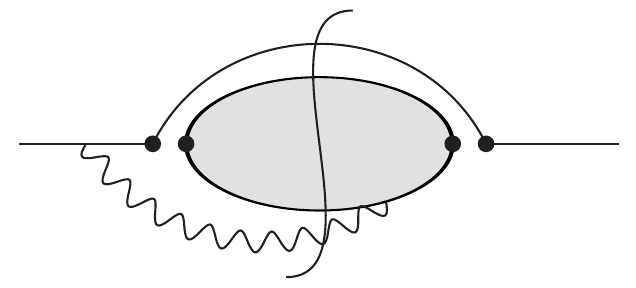}
        \caption{Interference of the real photon emission.}
    \end{subfigure}
    \caption{Non-factorizable radiative corrections, discussed in Section~\ref{sec:non-fact}. The sum of these effects, properly renormalized, defines $d\Gamma^\mathrm{nonf}/ds$.}
    \label{fig:non-fact}
\end{figure}

We begin, by classifying the radiative corrections based on the IR behavior. By using Eq.~\eqref{eq:rev_uni} the phase-space factor (and decay rate) in the isospin limit may be represented by a loop diagram, with a cutting rule setting the intermediate particles on shell~\cite{Sterman_1993}, as in Fig.~\ref{fig:tau-rate}.
With this representation, when we look at radiative corrections we can identify, by visual inspection, sets of diagrams where IR divergences cancel out. The main idea is to combine diagrams where only one of the two photon vertices is reflected across the final-state cut (i.e. the vertical squiggle line), like in the contributions reported in Fig.~\ref{fig:non-fact}.
With this method we address different sources of isospin breaking by grouping the radiative contributions into three infrared-safe classes of diagrams which we label as initial-state, final-state and non-factorizable contributions, and collect schematically in Fig.~\ref{fig:init-state}, \ref{fig:fin-state} and \ref{fig:non-fact} respectively.

Despite being IR safe, they individually generate UV divergences in the EFT which have to be taken into account. In particular leaving aside corrections on the external legs, i.e. wave-function renormalization effects, the most important non-trivial UV contribution arises from the non-factorizable loop diagram in Fig.~\ref{fig:virt_loop}, later also referred to as the ``triangle'' diagram, discussed in Section~\ref{sec:short-distance}.
With this classification, a key observation in our strategy is that final-state radiative corrections as sketched in Fig.~\ref{fig:fin-state} and discussed in Section~\ref{sec:fin-state} are calculated directly in Euclidean space-time and they completely define $G^W(t,\mu)$ at first order in $\alpha$
\begin{equation}
\begin{split}
    G^W(t,\mu) = &\, - \frac13 \sum_k \int d^3 \vec x \bigg\{ \langle j^{+}_k(t,\vec x) j^{-}_k(0) \rangle \\ & - \frac{e^2}{2}  \int d^4 \Eucl y \, d^4 \Eucl z \langle j^{+}_k(t,\vec x) j^\gamma_\alpha(\Eucl y) j^\gamma_\beta(\Eucl z) j^{-}_k(0) \rangle \Eucl \Delta_{\alpha\beta}(\Eucl y,\Eucl z) \bigg|_\mu + \mathrm{c.t.}(\mu) + O(\alpha^2) \bigg\} \,.
    \label{eq:GWtmu}
\end{split}
\end{equation}
The symbols $|_\mu$ and $\mathrm{c.t.}(\mu)$ indicate a generic subtraction point and the corresponding counter terms, required to regulate the UV divergences of the four-point function.
This may be achieved by regulating the photon propagator with a Pauli-Villars (PV) mass equal to $\mu$, for example~\cite{Biloshytskyi:2022ets}, or via a momentum subtraction point as discussed in Section~\ref{sec:short-distance}.
For these reasons, starting from Eq.~\eqref{eq:GWtmu} we assign to $G^W(t,\mu)$ the generic functional dependence on the renormalization scale $\mu$, with the understanding that there is an implicit dependence also on the scheme (generically denoted by $\scheme$) and, being defined from charged operators, also on the QED gauge.
Throughout the manuscript we adopt Feynman's gauge with photon propagator
\begin{equation}
    i\Delta_{\alpha\beta}(k) = \frac{-i g_{\alpha\beta}}{k^2 + i \varepsilon}\,,
\end{equation}
and, where needed, a small photon mass $\mg$ will be used to regulate intermediate steps of the calculation. 
The Euclidean photon propagator is denoted by $\Eucl \Delta_{\alpha\beta}$. 

To understand why $G^W(t,\mu)$ is useful, we decompose the neutral correlation function (needed for $a_\mu$) according to
\begin{equation}
    G^\gamma(t) = G_{00}^\gamma(t) + 2 G_{01}^\gamma(t) + \delta G_{11}(t,\mu) + G^W(t,\mu) \,, \quad \text{with} \quad \delta G_{11}(t,\mu) \equiv G_{11}^\gamma(t) - G^W(t,\mu) \,,
    \label{eq:dG11}
\end{equation}
and the equation above is exact.
At first order in the isospin-breaking parameters, $\delta G_{11}(t,\mu)$ acquires a non-trivial expectation value. Being defined from charged currents, $G^W(t,\mu)$ is scale, scheme and gauge dependent and for consistency $\delta G_{11}(t,\mu)$ has to be calculated with the same prescriptions such that their sum is free from these ambiguities, contrary to $G_{00}^\gamma$ and $G_{01}^\gamma$.

Our strategy may be summarized in the following steps.
\begin{enumerate}
    \item Multiply the LO inclusive decay with the Wilson coefficient (squared) $[C^\scheme(\mu)]^2$, the generic short-distance correction to the electro-weak effective Hamiltonian operator $\O$, defined and discussed in Section~\ref{sec:short-distance}. While in general one may adopt a generic scheme $\scheme$ we will specialize the calculation to the RI-SMOM scheme and use simply $C(\mu)$ in our notation.
    \item Starting from the normalized experimental data for the inclusive (vector) decay rate, such as the ALEPH or Belle compilations~\cite{aleph13,Hayashii:2009zz}, separate the three IR-safe classes at first order in isospin-breaking, introduced in the previous Subsection,
    \begin{equation}
        \left[\frac{1}{\Gamma_e} \frac{d\Gamma}{ds} \right]^{\exp} = \frac{1}{\Gamma_e(1 + \delta \Gamma_e)} \bigg([C(\mu)]^2 \frac{d\Gamma}{ds} + \frac{d\Gamma^\mathrm{init}}{ds}(\mu) + \frac{d\Gamma^\mathrm{fin}}{ds}(\mu) + \frac{d\Gamma^\mathrm{nonf}}{ds}(\mu) \bigg) \,.
    \end{equation}
    The l.h.s. is assumed to include all radiative events. $\delta \Gamma_e$ denotes the first order correction to the electronic rate. 
    The last three terms, $d\Gamma^X/ds$ ($X=\mathrm{init},\mathrm{fin},\mathrm{nonf}$) are of order $\alpha$, and are scheme and scale dependent (we indicate only the latter for simplicity). Such a dependence cancels out in the total sum that defines the physical amplitude (up to remaining perturbative truncation errors). The Wilson coefficient $C(\mu) = 1 + O(\alpha)$ and higher-order effects in the EM coupling are neglected in the current setup.

    \item Calculate the initial-state correction, schematically given by Fig.~\ref{fig:init-state}, following the analytical derivation in Section~\ref{sec:init-state}.
    \item Subtract the effects of non-factorizable contributions, following the derivation in Section~\ref{sec:non-fact}, thereby arriving at the scale and gauge dependent correlator
    \begin{equation}
        G^{W,\exp}(t,\mu) \equiv \mathfrak{L}(t) \cdot \left( \frac{1 + \delta \Gamma_e}{[C(\mu)]^2} \left[\frac{1}{\Gamma_e} \frac{d\Gamma}{ds} \right]^{\exp} - \frac{1}{\Gamma_e} \frac{d\Gamma^\mathrm{init}}{ds}(\mu) - \frac{1}{\Gamma_e} \frac{d\Gamma^\mathrm{nonf}}{ds}(\mu) \right) \,.
    \end{equation} 
    The equation above defines how we intend to estimate $G^W(t,\mu)$ starting from experimental data and (theoretical) isospin-breaking corrections. Notice that by subtracting $d\Gamma^\mathrm{init}/ds$ and $d\Gamma^\mathrm{nonf}/ds$ we are ``leaving behind'' the LO contribution $d\Gamma/ds$ and all radiative corrections involving only the hadronic sector, which almost correspond (see next point) to the first and second line of Eq.~\eqref{eq:GWtmu} respectively.

    \item We stress once again that $G^W(t,\mu)$ is defined as a purely theoretical correlator, hence without a physical cut at the $\tau$ mass. From the previous point it is evident that we must supplement the missing contributions from the mass spectrum to estimate
    \begin{equation}
        G^{W,>}(t,\mu) \equiv G^W(t,\mu) - G^{W,\exp}(t,\mu) \,.
    \end{equation}
    For this, one could use $e^+e^-$ experimental data (only from isovector channels)
    \begin{equation}
        G^{W,>}(t,\mu) = \int ds \, \frac{\sqrt s}{2} \, e^{-\sqrt s t} \, \rho^{\exp,ee}(s) \, \theta(\sqrt s - \mt) + O(\alpha) \,,
    \end{equation}
    or alternatively, by solving the corresponding inverse problem, one could estimate it from the correlator $(c)(t)$ in the isosymmetric limit
    \begin{equation}
        G^{W,>}(t,\mu) = \int ds \, \frac{\sqrt s}{2} \, e^{-\sqrt s t} \, \rho(s) \, \theta(\sqrt s - \mt) + O(\alpha) \,.
    \end{equation}
    In both cases this introduces an ambiguity at $O(\alpha)$ (and makes the $\mu$ dependence obsolete), an effect which should be estimated and added as a systematic uncertainty. For the $e^+e^-$ case there would be a double counting of isospin-breaking effects, while for $\rho(s)$ they would be missing. The relevance of $G^{W,>}(t,\mu)$ is expected to be larger for the short-distance window and more suppressed for the long-distance one, implying that the former is less suitable for a determination of the HVP from $\tau$ data~\cite{Allen:2026iad}. For this reason a study of the intermediate and long-distance windows seems more appropriate. Evidently, solving the inverse problem with the $\theta$-function on the full QCD+QED correlator would remove this ambiguity and to this end the QED quenched limit should be sufficient for the current precision.
    
    \item With all these ingredients the separation of the neutral HVP correlator introduced in Eq.~\eqref{eq:dG11} is replaced with
    \begin{equation}
    \begin{split}
        G^\gamma(t) = & G_{00}^\gamma(t) + 2 G_{01}^\gamma(t) & \quad \text{Lattice QCD+QED}\\
        & + \delta G_{11}(t,\mu) & \quad \text{Lattice QCD+QED}\\
        & + G^{W,\exp}(t,\mu) & \quad \text{Exp. data and theory} \\
        & + G^{W,>}(t,\mu) & \quad \text{Exp. data or Lattice QCD}
    \end{split}
    \end{equation}
    and is used to calculate $a_\mu$ or $a_\mu^\mathrm{win}$ by integrating it over $t$ with the appropriate kernel, c.f. Eq~\eqref{eq:amu_tmr}.
    The idea is to estimate $G_{00}^\gamma(t), G_{01}^\gamma(t)$ and $\delta G_{11}(t,\mu)$ (and possibly also $G^{W,>}(t,\mu)$) directly in Euclidean space-time from fully inclusive correlators. 
    
    \item The isovector component $G_{11}^\gamma(t)$ is effectively replaced by $\delta G_{11}(t,\mu) + G^{W,\exp}(t,\mu) + G^{W,>}(t,\mu)$, which is $\mu$ independent and it provides useful intermediate information. In fact it is mostly dominated by the two-pion channel and with dedicated studies it could be compared with previous phenomenological estimates~\cite{Aliberti:2025beg}. Alternatively calculating $G^W(t,\mu)$ directly from Lattice QCD+QED simulations would provide a direct test of the $\tau$ data itself and, with minor extensions of our definitions to the strange sector, it could be used as a probe for $V_\mathrm{us}$ from inclusive decays including isospin-breaking effects~\cite{ETMCtau23,ETMCtau24,DiCarlo:2026kpv}.
\end{enumerate}

\section{Short-distance matching}
\label{sec:short-distance}

The effective Hamiltonian and the corresponding four-fermion operator are
\begin{equation}
    \frac{4 \GF}{\sqrt 2} \Vud^\ast \O(x) \,, \quad \text{with} \quad \O(x) = \big(\overline \nu_\tau(x) \gamma^\mu_L \tau(x) \big) \, \J_\mu^{L,-}(x) \,.
\end{equation}
When addressing short-distance corrections, the first step consists in matching the EFT to the Standard Model, i.e. calculate the effects induced by the degrees of freedom that we are integrating out, many of which are universal and typically absorbed in the definition of $\GF$~\cite{Marciano:1988vm} taken from the muon life time (see Refs.~\cite{Steinhauser:1999bx,vanRitbergen:1999fi} for a recent calculation at two loops in QED). Since here we use the electronic decay rate of the $\tau$ lepton as a normalization, they are automatically canceled, leaving us with the process-specific corrections of the hadronic and electronic decays. For the latter, we briefly review them towards the end of this Section.
For the former, such effects are encoded in the Wilson coefficient $C^\scheme(\mu_W)$ of the operator $\O$, evaluated at the scale $\mu_W$ where the matching with the SM is performed. By renormalizing the operator in the same scheme, the  effective Hamiltonian reads $C^\scheme(\mu_W) \O^\scheme(\mu_W)$ and the scale dependence cancels out between the two terms.

From the one-loop correction generated by the exchange of a photon, the operator develops a non-trivial anomalous dimension which induces a logarithmic dependence on the scale. This effect has been traditionally addressed in the $W$-regularization scheme~\cite{Marciano:1988vm,Sirlin:1977sv,Sirlin:1981ie}, which consists of splitting the photon propagator in the EFT into two pieces
\begin{equation}
    \frac{1}{k^2} \to \frac{1}{k^2 - \mW^2} - \frac{\mW^2}{k^2(k^2 - \mW^2)},
\end{equation}
and using only the second term. Finite scheme-dependent terms have been addressed in Ref.~\cite{Braaten:1990ef} in dimensional regularization for the inclusive rate, and only recently the full two-loop anomalous dimension up to $O(\alpha_s\alpha)$ has been completed in the $\MS$ scheme~\cite{Brod:2008ss,Gorbahn:2022rgl,Erler:2002mv,Cirigliano:2023fnz}.
The initial conditions for the Wilson coefficient are provided for example in Ref.~\cite{Gorbahn:2022rgl} together with the one-loop conversion from the $W$-regularization to the $\MS$ scheme\footnote{By extending the methodology developed in Ref.~\cite{Bruno:2017iwk} one could improve also the matching by means of Lattice QCD simulations.}.

When running the Wilson coefficient to lower scales, large logarithms can be resummed via the (renormalization-group improved) scale-evolution operator $U(\mu_1,\mu_2)$. The latter may be studied non-perturbatively using Lattice QCD simulations and step-scaling techniques (see e.g.~\cite{Arthur:2010ht,Bruno:2017iwk}). The present available knowledge is purely perturbative and we leverage the important results of Ref.~\cite{Gorbahn:2022rgl} (and \cite{Boyle:2026xls}) which provide the finite conversion factor from the $\MS$ scheme to (several) Regularization-Invariant Momentum (RI-MOM) schemes, suitable for a Lattice QCD calculation, thereby arriving to a full $O(\alpha_s \alpha)$ Wilson coefficient $C^\RI(\mu)$.

At this point, our strategy consists in renormalizing the operator with the same $\RI$ conditions, thereby defining $Z_\O^\RI(\mu)$, such that the combination $C^\RI(\mu) Z_\O^\RI(\mu) \O(x)$ is scale independent up to truncation errors and can be inserted in physical matrix elements. 
As an alternative, one could consider the Schr\"odinger functional~\cite{Luscher:1991wu,Luscher:1993gh,deDivitiis:1994yz,Jansen:1995ck,Sint:2010eh} or position-space schemes~\cite{Gimenez:2004me,Costa:2021iyv}, extensively used in the Lattice QCD community, for which, however, the additional matching to the $\MS$ scheme of the Wilson Coefficient would be required\footnote{The interested reader may consult Refs.~\cite{Bruno:2017gxd,DallaBrida:2022eua} and Refs.~\cite{Dawson:1997ic,Tomii:2016xiv,Korcyl:2015xmd,Lin:2024mws} for a few examples of recent calculations based on the SF formalism and position-space schemes respectively. For an example of conversion factors between position-space and $\MS$ scheme see Ref.~\cite{Chetyrkin:2010dx}.}. 

In our definition of the scheme, we take the massless limit and the interested reader may check Ref.~\cite{Boyle:2016wis} for extensions to the massive case. In the next Subsections we will explicitly use the Euclidean signature for the calculation of $Z_\O^\RI(\mu)$ including isospin-breaking effects (for the first discussion on this topic see Refs.~\cite{DiCarlo:2019thl,DiCarlo:2019knp}) and since we focus on a single scheme, the RI-SMOM one, for better readability we drop the superscript ${}^\RI$ from the rest of the manuscript.

\subsection{The quark propagator}

We use $S_r^\QCD(p)$ to identify the expectation value of a gauge-fixed quark propagator with (Euclidean) momentum $\Eucl p^2>0$ in the isosymmetric limit 
\begin{equation}
    S_r^\QCD(\Eucl p) \equiv \int d^4 \Eucl x \, e^{-i \Eucl p \Eucl x} \, \langle \psi_r(\Eucl x) \overline \psi_r(0) \rangle \,, 
\end{equation}
with $r=\up,\down$ and $S_\up^\QCD(p) = S_\down^\QCD(p)$. Fields and correlation functions are intended as bare.

\def\sep{0.16\textwidth}
\def\hf{\textwidth}
\begin{figure}[ht]
    \centering
    \begin{subfigure}[t]{\sep}
        \centering
        \includegraphics[width=\hf]{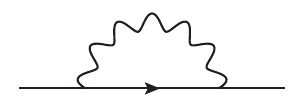}
        \caption{$S^{(S)}$}
        \label{fig:mom-prop-S}
    \end{subfigure}
    \begin{subfigure}[t]{\sep}
        \centering
        \includegraphics[width=\hf]{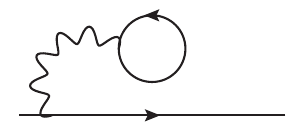}
        \caption{$S^{(T)}$}
        \label{fig:mom-prop-T}
    \end{subfigure}
    \begin{subfigure}[t]{\sep}
        \centering
        \includegraphics[width=\hf]{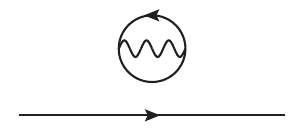}
        \caption{$S^{(D1)}$}
        \label{fig:mom-prop-D1}
    \end{subfigure}
    \begin{subfigure}[t]{\sep}
        \centering
        \includegraphics[width=\hf]{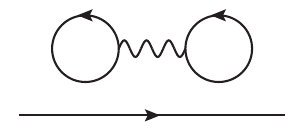}
        \caption{$S^{(D2)}$}
        \label{fig:mom-prop-D2}
    \end{subfigure}
    \begin{subfigure}[t]{\sep}
        \centering
        \includegraphics[width=\hf]{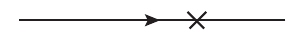}
        \caption{$S^{(M)}$}
        \label{fig:mom-prop-M}
    \end{subfigure}
    \begin{subfigure}[t]{\sep}
        \centering
        \includegraphics[width=\hf]{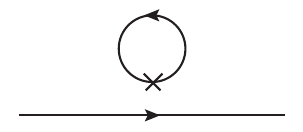}
        \caption{$S^{(R)}$}
        \label{fig:mom-prop-R}
    \end{subfigure}
    \caption{QED and strong isospin breaking contributions to the quark propagator. If a conserved lattice current is coupled to the photon additional topologies appear, see e.g. Ref.~\cite{deDivitiis:2013xla}.}
    \label{fig:mom-prop}
\end{figure}

From the operator product expansion of two EM currents, we know that a shift in the bare quark mass and strong coupling is generated (see e.g. Ref.~\cite{deDivitiis:2013xla}). 
By inserting one photon propagator and the scalar operator, we find the diagrams illustrated in Fig.~\ref{fig:mom-prop} and collected in the factor $\delta S_r(\Eucl p^2)$, leading to (we omit the trivial  dependence on the scale)
\begin{equation}
\begin{split}
    \delta S_r = \, - e^2 (Z_V^\QCD)^2 \Big[ Q_r^2 S_r^{(S)} - Q_r \Qbar S_r^{(T)} - \frac12 \Qbar_2 S_r^{(D1)} + \frac12 \Qbar^2 S^{(D2)} \Big] - \Delta m_r S_r^{(M)} + \overline{\Delta m} S_r^{(R)} \,,
    \label{eq:mom-prop-corr}
\end{split}
\end{equation}
with the following intermediate quantities originating from sea-quark loops
\begin{equation}
    \Qbar = \sum_{r=\up,\down} Q_r \,, \quad \Qbar_2 = \sum_{r=\up,\down} Q_r^2 \,, \quad \overline{\Delta m} = \sum_{r=\up,\down} \Delta m_r \,,
    \label{eq:overlineQ}
\end{equation}
and the prefactor $(Z_V^\QCD)^2$ for local vector currents (at non-zero lattice spacing) defined below.

Here we consider a setup where the strange quark (and charm) is quenched from the QED point of view, but our formulae can be easily extended to include these effects provided that such a precision is needed.
For disconnected quark loops the exchange of gluons is always understood and diagrams should be intended without minus signs or pre-factors proportional to the quark charges, which we keep explicit in Eq.~\eqref{eq:mom-prop-corr}. 
For example the self-energy diagram $S_r^{(S)}$ in Fig.~\ref{fig:mom-prop-S} is defined by the (QCD gauge-fixed) expectation value
\begin{equation}
    S_r^{(S)}(\Eucl p) = \int d^4 \Eucl x \, d^4 \Eucl y \, d^4 \Eucl z \, e^{-i \Eucl p \Eucl x} \,  \langle D_r^{-1}(\Eucl x,\Eucl y) \Eucl \gamma_\alpha D_r^{-1}(\Eucl y, \Eucl z) \Eucl \gamma_\beta D_r^{-1}(\Eucl z,0)  \rangle \, \Eucl \Delta_{\alpha\beta}(\Eucl y,\Eucl z)  \,.
\end{equation}
We remind the reader that $D_r(\Eucl x,\Eucl y)$ denotes the (Euclidean) Dirac operator for flavor $r$ and color and spin indices are suppressed from the notation.

The  full propagator $S_r(\Eucl p) = S_r^\QCD(\Eucl p) + \delta S_r(\Eucl p)$  is renormalized by the mass counter terms together with the wave function factor $Z_r(\Eucl p^2) = Z_r^\QCD(\Eucl p^2) + \delta Z_r(\Eucl p^2)$, which may be estimated using
\begin{equation}
    \frac{[Z_r(\Eucl p^2)]^{-1}}{12 \Eucl p^2} \tr\big[i [S_r(\Eucl p) ]^{-1}\, \Eucl{\slashed p} \big] = 
    \frac{[Z_r^\QCD(\Eucl p^2)]^{-1}}{12 \Eucl p^2} \tr\big[ i [S_r^\QCD(\Eucl p)]^{-1} \, \Eucl{ \slashed p} \big]= -1 \,,
    \label{eq:Zr}
\end{equation}
where by construction $Z_\up^\QCD(\Eucl p^2) = Z_\down^\QCD(\Eucl p^2)$.
The discussion of a specific procedure to implement the re-tuning of the bare masses and strong coupling (or lattice spacing) goes beyond the scope of the paper. Here we simply assume that the values of $\Delta m_\up$ and $\Delta m_\down$ have been properly fixed for example using hadronic renormalization schemes; those are typically given in terms of mesonic or baryonic mass differences, which are in general preferable for a Lattice QCD calculation as suggested in Refs.~\cite{Aoyama:2020ynm,Aliberti:2025beg} or in the FLAG report~\cite{FLAG:2024oxs}.\\

\subsection{Quark bilinears}

Before discussing the renormalization of $\O$ we focus on the following charged and neutral vector bilinear operators, for reasons that will become evident later,
\begin{align}
    \Gamma_{\mu,sr}^{V,\QCD}(\Eucl p,\Eucl p') = \int d^4 \Eucl x \, d^4 \Eucl y \, e^{-i \Eucl p \Eucl x} \, e^{i \Eucl p' \Eucl y} \, \langle \psi_s(\Eucl x) \, [\overline \psi_s \Eucl \gamma_\mu \psi_r](0) \, \overline \psi_r(\Eucl y) \rangle \,,  \\
    \Gamma_{\mu,sr}^{\gamma,1,\QCD}(\Eucl p,\Eucl p') = \int d^4 \Eucl x \, d^4 \Eucl y \, e^{-i \Eucl p \Eucl x} \, e^{i \Eucl p' \Eucl y} \, \langle \psi_s(\Eucl x) \, j^{\gamma,1}_\mu(0) \, \overline \psi_r(\Eucl y) \rangle \,.
\end{align}
Amputated Green's functions are labeled by $\Lambda^{\dots,\QCD}_{\mu,sr}(\Eucl p,\Eucl p')$ for the various operators and are given by
\begin{equation}
    \Lambda^{\dots,\QCD}_{\mu,sr}(\Eucl p,\Eucl p') = [S_s^\QCD(\Eucl p)]^{-1} \, \Gamma^{\dots,\QCD}_{\mu,sr}(\Eucl p,\Eucl p') \, [S_r^\QCD(\Eucl p')]^{-1} \,.
    \label{eq:GammaQCD}
\end{equation}
Thanks to the isospin-1 structure, in isosymmetric QCD the disconnected diagrams that appear in $\Gamma^{\gamma,1,\QCD}_{\mu,ss}$ cancel out, leading to the identity $ \Gamma^{V,\QCD}_{\mu,\up\down} = \Gamma^{V,\QCD}_{\mu,\down\up} =  2 \Gamma_{\mu,\up\up}^{\gamma,1,\QCD} = - 2 \Gamma_{\mu,\down\down}^{\gamma,1,\QCD}$.
In MOM schemes~\cite{Martinelli:1994ty}, renormalization conditions on amputated off-shell Green's amplitudes are typically imposed using suitably chosen projectors in spin and color space, which together with the momenta (and the QCD gauge) fully specify the scheme. 
In our work we are mostly interested in SMOM schemes defined by the symmetric kinematic configuration~\cite{Sturm:2009kb}
\begin{equation}
    \Eucl p^2 = \Eucl p'^2 = (\Eucl p - \Eucl p')^2 = \mu^2 > 0\,.
\end{equation}

By introducing a projector $[P_\mu]^{\alpha\beta,ab}$, normalized such that $\tr[P^\mu \Eucl \gamma_\mu]=1$, we impose the following renormalization condition\footnote{Above we used an open-index notation both in spin ($\alpha,\beta$) and color space ($a,b$), in fact, for example, by $\Lambda^{\dots,\QCD}_{\mu,sr}(\Eucl p,\Eucl p')$ we effectively mean $[\Lambda^{\dots,\QCD}_{\mu,sr}(\Eucl p,\Eucl p')]^{\alpha\beta,ab}$. Contrary to the typical literature on the topic, notice that we do not specify the scheme with an additional superscript ${}^\RI$.}
\begin{equation}
    Z_V^\QCD \frac12 \frac{\tr[P^\mu (\Lambda^{V,\QCD}_{\mu,\up\down}(\Eucl p,\Eucl p') + \Lambda^{V,\QCD}_{\mu,\down\up}(\Eucl p,\Eucl p'))]}{\sqrt{Z_\up^\QCD(\mu^2) Z_\down^\QCD(\mu^2)}} = 1 \,,
\end{equation}
where we use the fact that the positive and negative bilinears renormalize in the same way.
Similarly, the renormalization condition of the isospin-1 current requires a projector in flavor space that respects the same structure, leading to ($Q_\up - Q_\down = 1$)
\begin{equation}
    Z_{\gamma,1}^\QCD \left[\frac{\tr[P^\mu \Lambda^{\gamma,1,\QCD}_{\mu,\up\up}(\Eucl p,\Eucl p')]}{Z_\up^\QCD(\mu^2)} - \frac{\tr[P^\mu \Lambda^{\gamma,1,\QCD}_{\mu,\down\down}(\Eucl p,\Eucl p')]}{Z_\down^\QCD(\mu^2)} \right] = 1 \,.
    \label{eq:Zgamma1_QCD}
\end{equation}
Using $Z_\up^\QCD(\mu^2) = Z_\down^\QCD(\mu^2)$ and the relation among the amplitudes, we readily find the expected result, $Z_V^\QCD = Z_{\gamma,1}^\QCD$. 
For conserved currents, the Ward identity protects their renormalization implying that $Z_V^\QCD=1$.

At this point we switch on QED and consider the first-order contributions to $\Gamma_{\mu,sr}^{V,\QCD}(\Eucl p,\Eucl p')$ in $\alpha$ and $\Delta m$, which are reported in Figs.~\ref{fig:mom-qed} and \ref{fig:mom-sib} respectively. For example, the Green function in Fig.~\ref{fig:mom-qed-V}, in our notation is given by
\begin{equation}
\begin{split}
    \Gamma^{V,(V)}_{\mu,sr} (\Eucl p,\Eucl p') = \int d^4 \Eucl x \, d^4 \Eucl y \, & \,  d^4 \Eucl x' \, d^4 \Eucl y' \, e^{-i \Eucl p \Eucl x} \, e^{i \Eucl p' \Eucl x'} \Eucl \Delta_{\alpha\beta}(\Eucl y,\Eucl y') \\ & \times \langle  D_s^{-1}(\Eucl x,\Eucl y) \Eucl \gamma_\alpha D_s^{-1}(\Eucl y,0) \Eucl \gamma_\mu D_r^{-1}(0,\Eucl y') \Eucl \gamma_\beta D_r^{-1}(\Eucl y', \Eucl x') \rangle \,.
\end{split}
\end{equation}

\def\sep{0.19\textwidth}
\def\hf{4em}
\begin{figure}[ht]
    \centering
    \begin{subfigure}[t]{\sep}
        \centering
        \includegraphics[height=\hf]{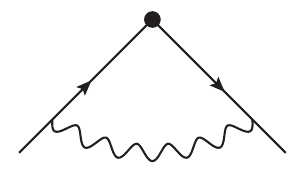}
        \caption{$\Gam{V}$}
        \label{fig:mom-qed-V}
    \end{subfigure}
    \begin{subfigure}[t]{\sep}
        \centering
        \includegraphics[height=\hf]{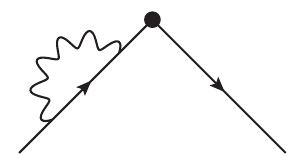}
        \caption{$\Gam{S}$}
        \label{fig:mom-qed-S}
    \end{subfigure}
    \begin{subfigure}[t]{\sep}
        \centering
        \includegraphics[height=\hf]{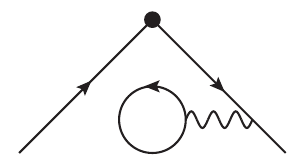}
        \caption{$\Gam{T}$}
    \end{subfigure}
    \begin{subfigure}[t]{\sep}
        \centering
        \includegraphics[height=\hf]{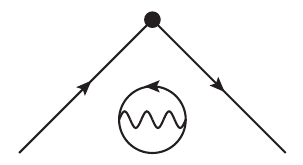}
        \caption{$\Gam{D1}$}
    \end{subfigure}
    \begin{subfigure}[t]{\sep}
        \centering
        \includegraphics[height=\hf]{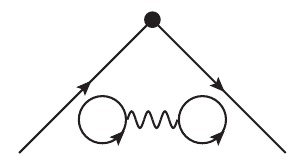}
        \caption{$\Gam{D2}$}
    \end{subfigure}
    \caption{QED contributions to the quark bilinear Green's function between off-shell external quark states.}
    \label{fig:mom-qed}
\end{figure}

\begin{figure}
    \centering
    \begin{subfigure}[t]{\sep}
        \centering
        \includegraphics[height=\hf]{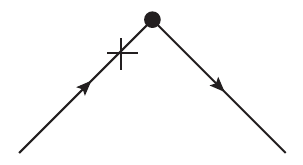}
        \caption{$\Gam{M}$}
    \end{subfigure}
    \hspace{2em}
    \begin{subfigure}[t]{\sep}
        \centering
        \includegraphics[height=\hf]{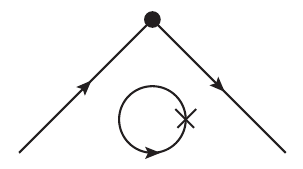}
        \caption{$\Gam{R}$}
    \end{subfigure}
    \caption{Strong-isospin breaking corrections to the quark bilinear.}
    \label{fig:mom-sib}
\end{figure}

\begin{figure}[ht]
    \centering
    \includegraphics[width=.7\textwidth]{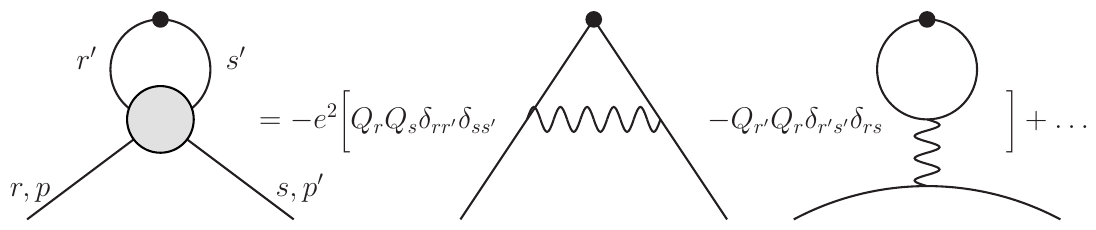}
    \caption{Diagrams defining $\Gamma^{\X}$, namely corrections to the quark bilinear from the four-leg vertex $\X$. Higher orders in the isospin breaking are neglected. The two terms on the r.h.s. are labeled by $\Gam{V}$ and $\Gam{F}$ respectively.}
    \label{fig:gamma-x}
\end{figure}

From the depicted topologies it is clear that we would benefit from a separation between radiative corrections that involve both external legs against those involving only a single leg.
Taking the self-energy diagram or its mass counter term as examples, we note that for generic kinematics the amplitudes with the photonic loop (or scalar operator) inserted on the left or on the right external legs are different. We identify the two cases by adding the corresponding flavor superscript $s$, $\Gamma_{\mu}^{V,(S)s}$, and instead we drop the $sr$ subscripts since their dependence is captured entirely by the quark charges and mass shifts which we keep separate. Then we collect all the factorized (single-leg) corrections
in the quantity $\Gamma_{\mu,sr}^{V,\C}$, which for the charged vector bilinear operator at first order in isospin breaking is given by
\begin{equation}
\begin{split}
    \Gamma^{V,\C}_{\mu,sr} = & \, -e^2 (Z_V^\QCD)^2 \Big[Q_r^2 \, \Gamma^{V,(S)r}_\mu + Q_s^2 \, \Gamma^{V,(S)s}_\mu - \frac12 \overline Q_2 \Gamma^{V,(D1)}_\mu + \frac12 \overline Q^2 \Gamma^{V,(D2)}_\mu \\ & - Q_r \overline Q \Gamma^{V,(T)r}_\mu - Q_s \overline Q \Gamma^{V,(T)s}_\mu \Big]- \Delta m_r \Gamma^{V,(M)r}_\mu - \Delta m_s \Gamma^{V,(M)s}_\mu + \overline{\Delta m} \Gamma^{V,(R)}_\mu   \,.
    \label{eq:Gamma_C}
\end{split}
\end{equation}   
Non-factorizable corrections require a generic four-point vertex (necessarily of $O(\alpha)$ since mediated by a virtual photon exchange), which we denote with $\X$. Once inserted in correlation functions it generates the appropriate Feynman diagrams shown in Fig.~\ref{fig:gamma-x}, leading to the amplitude
\begin{equation}
    \Gamma^{V,\X}_{\mu,sr} = - e^2 (Z_V^\QCD)^2 [ Q_s Q_r \Gamma_{\mu}^{V,(V)} - \delta_{rs} (Q_\up - Q_\down)Q_s \Gamma_\mu^{V,(F)}]
    \label{eq:gammaX}
\end{equation}
which for the flavor-changing quark bilinear amounts to $\Gamma^{V,(V)}_\mu$ only.

Amputated Green functions in QCD+QED for generic bilinears are denoted by $\Lambda_{\mu,sr}^{\dots}(\Eucl p, \Eucl p')$.
At this point we have all the ingredients to study the vector flavor-changing bilinear 
\begin{equation}
    \Lambda_{\mu,\up\down}^{V}(\Eucl p, \Eucl p') = \Lambda_{\mu,\up\down}^{V,\QCD} + \Lambda_{\mu,\up\down}^{V,\C}(\Eucl p, \Eucl p') + \Lambda_{\mu,\up\down}^{V,\X}(\Eucl p, \Eucl p') \,,
\end{equation}
with\footnote{We expanded the inverse quark propagator according to (and neglecting  higher orders)
\begin{equation*}
    [S_r(\Eucl p)]^{-1} = \left[S_r^\QCD(\Eucl p) + \delta S_r(\Eucl p) \right]^{-1} = [S_r^\QCD(\Eucl p)]^{-1} - [S_r^\QCD(\Eucl p)]^{-1} \delta S_r(\Eucl p) [S_r^\QCD(\Eucl p)]^{-1} + \cdots
    \label{eq:mom-prop-corr-inv}
\end{equation*}}
\begin{equation}
\begin{split}
    \Lambda_{\mu,\up\down}^{V,\C}(\Eucl p,\Eucl p') & = [S_\up^\QCD(\Eucl p)]^{-1} \Gamma^{V,\C}_{\mu,\up\down}(\Eucl p,\Eucl p') [S_\down^\QCD(\Eucl p')]^{-1} - [S_\up^\QCD(\Eucl p)]^{-1} \delta S_\up(\Eucl p) \Lambda_{\mu,\up\down}^{V,\QCD}(\Eucl p,\Eucl p') \\ & - \Lambda_{\mu,\up\down}^{V,\QCD}(\Eucl p,\Eucl p') \delta S_\down(\Eucl p') [S_\down^\QCD(\Eucl p')]^{-1} + O(\alpha^2, \alpha \Delta m, \Delta m^2)\,,
\end{split}
\end{equation}
and
\begin{equation}
    \Lambda_{\mu,\up\down}^{V,\X}(\Eucl p,\Eucl p') = [S_\up^\QCD(\Eucl p)]^{-1} \Gamma^{V,\X}_{\mu,\up\down}(\Eucl p,\Eucl p') [S_\down^\QCD(\Eucl p')]^{-1} + O(\alpha^2, \alpha \Delta m, \Delta m^2)\,.
\end{equation}
By expanding all terms to first order one finds that the corrections to the vertex function are amputated by the isosymmetric quark propagators, while the shifts $\delta S_r$ (related to the wave-function renormalization) multiply the isosymmetric bilinear $\Gamma^{V,\QCD}_{\mu,sr}$. The renormalization condition that we impose on the full amplitude is
\begin{equation}
    Z_{V,\up\down}(\mu^2) \frac12 \frac{\tr[P^\mu (\Lambda^V_{\mu,\up\down}(\Eucl p,\Eucl p') + \Lambda^V_{\mu,\down\up}(\Eucl p,\Eucl p'))]}{\sqrt{Z_\up(\mu^2) Z_\down(\mu^2)}} = 1 \,,
\end{equation}
where a scale and flavor dependence in $Z_{V,\up\down}$ is now required due to the absence of a Ward identity for the charged current. Moreover the reader should also keep in mind that an intrinsic dependence on the QED gauge is also present, and in this work we use consistently Feynman's gauge everywhere. With little algebra, by expanding both the numerator and denominator (to first order), we find
\begin{equation}
    \frac{Z_{V,\up\down}(\mu^2)}{Z_V^\QCD} =1 + \frac12 \frac{\delta Z_\up(\mu^2) + \delta Z_\down(\mu^2)}{Z_\up^\QCD(\mu^2)} - \frac{\tr [P^\mu (\Lambda_{\mu,\up\down}^{V,\C} + \Lambda_{\mu,\down\up}^{V,\C} +  \Lambda_{\mu,\up\down}^{V,\X} + \Lambda_{\mu,\down\up}^{V,\X})(\Eucl p,\Eucl p')]}{2 \tr [P^\mu \Lambda_{\mu,\up\down}^{V,\QCD}(\Eucl p,\Eucl p')]}  \,.
    \label{eq:ZV_expanded_ZVQCD}
\end{equation}
As already noted, in the continuum massless theory $Z_V^\mathrm{QCD}=1$ but at finite lattice spacing $Z_V^\mathrm{QCD} \neq 1$ for local currents.
For this reason we keep it explicitly here but we also observe that when it is unity, the relation above further simplifies to
\begin{equation}
    Z_{V,\up\down}(\mu^2) = 1 + \frac{ \delta Z_\up(\mu^2) + \delta Z_\down(\mu^2) - \tr [P^\mu 
    (\Lambda_{\mu,\up\down}^{V,\C} + \Lambda_{\mu,\down\up}^{V,\C} +  \Lambda_{\mu,\up\down}^{V,\X} + \Lambda_{\mu,\down\up}^{V,\X})(\Eucl p,\Eucl p')]}{2 Z_\up^\QCD(\mu^2)}
     \,.
    \label{eq:ZV_expanded}
\end{equation}

\begin{figure}[ht]
    \centering
    \includegraphics[height=6em]{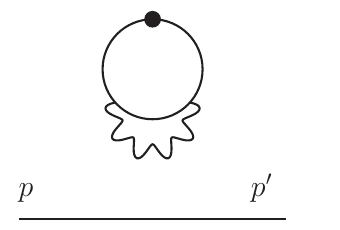}
    \hskip 2ex
    \includegraphics[height=6em]{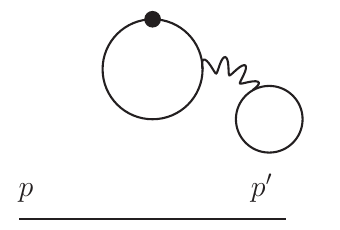}
    \caption{Diagrams contributing to $\Gamma_{\mu,ss}^{\gamma,1,\D}$. The other topologies simplify thanks to the isovector structure.}
    \label{fig:D}
\end{figure}

Isospin-breaking effects in the neutral amplitude, beyond the insertion of the 1PI terms on the individual legs of the connected amplitude, captured by  $\Gamma_{\mu,ss}^{\gamma,1,\C} = \frac12 \Gamma_{\mu,ss}^{V,\C}$, include also disconnected topologies since QED distinguishes the up and down flavors. Specifically $\Gamma^{V,(F)}_{\mu,ss}$ now contributes and in addition, since the cancellation that guarantees $ \Lambda_{\mu,\up\up}^{\gamma,1,\QCD} = \frac12 \Lambda_{\mu,\up\down}^{V,\QCD}$ is not valid in the full theory, we also encounter a new set of disconnected diagrams reported in Fig.~\ref{fig:D} and collectively labeled by $\Gamma_{\mu,ss}^{\gamma,1,\D}$. They are a pure isospin-breaking effect and we observe that $\Gamma_{\mu,\up\up}^{\gamma,1,\D} = \Gamma_{\mu,\down\down}^{\gamma,1,\D}$, since the flavor of the spectator quark line is not resolved by the $j^{\gamma,1}$ current at this order. 
Their amputation is performed as usual and for the up-quark bilinear the following expansion holds
\begin{equation}
    \Lambda_{\mu,\up\up}^{\gamma,1}(\Eucl p, \Eucl p') = \frac12 \Lambda_{\mu,\up\down}^{V,\QCD}(\Eucl p, \Eucl p') + \frac12 \Lambda_{\mu,\up\up}^{V,\C}(\Eucl p, \Eucl p') + \frac12 \Lambda_{\mu,\up\up}^{V,\X}(\Eucl p, \Eucl p') + \Lambda_{\mu,\up\up}^{\gamma,1,\D}(\Eucl p, \Eucl p') \,,
\end{equation}
while for the down quark, taking into account the minus signs due to the isospin-1 projection, one has
\begin{equation}
    \Lambda_{\mu,\down\down}^{\gamma,1}(\Eucl p, \Eucl p') = -\frac12 \Lambda_{\mu,\up\down}^{V,\QCD}(\Eucl p, \Eucl p') - \frac12 \Lambda_{\mu,\down\down}^{V,\C}(\Eucl p, \Eucl p') - \frac12 \Lambda_{\mu,\down\down}^{V,\X}(\Eucl p, \Eucl p') + \Lambda_{\mu,\up\up}^{\gamma,1,\D}(\Eucl p, \Eucl p') \,.
\end{equation}
After imposing the renormalization condition in the full theory
\begin{equation}
    Z_{\gamma,1} \left[\frac{\tr[P^\mu \Lambda^{\gamma,1}_{\mu,\up\up}(\Eucl p,\Eucl p')]}{Z_\up(\mu^2)} - \frac{\tr[P^\mu \Lambda^{\gamma,1}_{\mu,\down\down}(\Eucl p,\Eucl p')]}{Z_\down(\mu^2)} \right] = 1 \,,
    \label{eq:Zgamma1}
\end{equation}
with little algebra we obtain the counter-part of Eq.~\eqref{eq:ZV_expanded_ZVQCD}
\begin{equation}
    \frac{Z_{\gamma,1}}{Z_V^\QCD} = 1 + \frac12  \frac{\delta Z_\up(\mu^2) + \delta Z_\down(\mu^2)}{Z_\up^\QCD(\mu^2)} - \frac{\tr [P^\mu (\Lambda_{\mu,\up\up}^{V,\C} + \Lambda_{\mu,\down\down}^{V,\C} +  \Lambda_{\mu,\up\up}^{V,\X} + \Lambda_{\mu,\down\down}^{V,\X})(\Eucl p,\Eucl p')]}{2 \tr [P^\mu \Lambda_{\mu,\up\down}^{V,\QCD}(\Eucl p,\Eucl p')]} \,.
    \label{eq:Zgamma1_expanded_ZVQCD}
\end{equation}

So far, we have considered only local currents. Let us now briefly discuss the case of conserved currents. For the neutral vector bilinear the Ward identity ($Z_{\gamma,1}=1$) implies that the wave-function renormalization factors may be expressed in terms of the neutral vertex functions (we omit the dependence on momenta),
\begin{equation}
    \delta Z_\up + \delta Z_\down = \tr[P_\mu (\Lambda_{\mu,\up\up}^{V,\C} + \Lambda_{\mu,\down\down}^{V,\C} +  \Lambda_{\mu,\up\up}^{V,\X} + \Lambda_{\mu,\down\down}^{V,\X})] \,.
    \label{eq:ZuZd_neutral}
\end{equation}
Eq.~\eqref{eq:ZuZd_neutral} is the key for the cancellation we are seeking. Once inserted in Eq.~\eqref{eq:ZV_expanded}, thanks to the identities
\begin{align}
    & \tr[P_\mu (\Lambda_{\mu,\up\up}^{V,\C} + \Lambda_{\mu,\down\down}^{V,\C})] = \tr[P_\mu (\Lambda_{\mu,\up\down}^{V,\C} + \Lambda_{\mu,\down\up}^{V,\C})] \,,
    \label{eq:lambdaC_equivalence} \\
    & \tr[P_\mu (\Lambda_{\mu,\up\up}^{V,\X} + \Lambda_{\mu,\down\down}^{V,\X} - \Lambda_{\mu,\up\down}^{V,\X} - \Lambda_{\mu,\down\up}^{V,\X})] = - e^2 (Q_\up - Q_\down)^2 \tr[P_\mu (\Lambda_\mu^{V,(V)} - \Lambda_\mu^{V,(F)})]\,,
\end{align}
it leads to
\begin{equation}
    Z_{V,\up\down}(\mu^2) = 1 - \frac{e^2}{2} (Q_\up - Q_\down)^2 \frac{\tr[P^\mu (\Lambda_\mu^{V,(V)} - \Lambda_\mu^{V,(F)})(\Eucl p,\Eucl p')]}{\tr[P_\mu \Lambda^{V,\QCD}_{\mu,\up\down}(\Eucl p,\Eucl p')] } \,, \quad \text{(cons. cur.).}
    \label{eq:ZVud}
\end{equation}
Remarkably, at first order in isospin breaking the renormalization of the charged quark bilinear is dictated by two diagrams and it is purely electromagnetic.\\

At this point we turn again to local currents, our preferred setup~\cite{Blum:2018mom} for our calculation (to appear in a future publication).
Concerning the internal vertices, conserved EM currents generate photonic tadpole topologies which nevertheless simplify in the difference (they essentially follow topologies $S$ and $D1$) leading again to the survival of $\Gam{V}$ and $\Gam{F}$ alone.
For the external bilinear operators instead we must keep the full expression of $Z_{V,\up\down}(\mu^2)$ and $Z_{\gamma,1}$. However the cancellation exploited above is physically related to the cancellation occurring in the difference between the renormalized $j_\mu^{\gamma,1}$ current and the renormalized vector charged bilinear in $\delta G_{11}(t,\mu)$. In fact, neglecting radiative corrections to the correlators, discussed later in Section~\ref{sec:fin-state}, the short distance effects of $\delta G_{11}(t,\mu)$ are
\begin{equation}
    \frac12 (Z_V^\QCD)^2 \frac{Z_{\gamma,1}^2 - Z_{V,\up\down}^2(\mu^2)}{(Z_V^\QCD)^2} (c)(t) = (Z_V^\QCD)^2 \frac{Z_{\gamma,1} - Z_{V,\up\down}(\mu^2)}{Z_V^\QCD} (c)(t) \,.
\end{equation}
By calculating the difference between Eqs.~\eqref{eq:ZV_expanded_ZVQCD} and \eqref{eq:Zgamma1_expanded_ZVQCD}, we reach the same conclusion
\begin{equation}
    \frac{Z_{\gamma,1} - Z_{V,\up\down}(\mu^2)}{Z_V^\QCD} = \frac{e^2}{2} (Q_\up - Q_\down)^2 (Z_V^\QCD)^2  \frac{\tr[P^\mu (\Lambda_\mu^{V,(V)} - \Lambda_\mu^{V,(F)})(\Eucl p,\Eucl p')]}{\tr[P_\mu \Lambda^{V,\QCD}_{\mu,\up\down}(\Eucl p,\Eucl p')] } \,, \quad \text{(loc. cur.).}
    \label{eq:Zg1_minus_ZVud}
\end{equation}

In massless chirally-symmetric QCD it is possible to prove, for example, that the contribution of $\Gamma_\mu^{V,(M)}$ vanishes after projection, as expected. Since in practice the massless limit is achieved by extrapolating calculations from Lattice QCD at non-zero quark masses, we kept all the diagrams at first order and checked\footnote{For this purpose we have used the automatic Wick contraction tool from Ref.~\cite{giancarlo}.} that the identities above are satisfied.\\

\subsection{Semileptonic operators}

\begin{figure}[ht]
    \centering
    \includegraphics[height=7em]{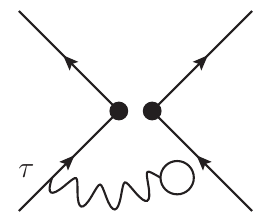}
    \hspace{2em}
    \includegraphics[height=7em]{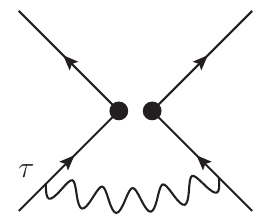}
    \caption{Triangle diagrams where the photon is exchanged between the lepton, indicated by $\tau$, and one of the quarks (legs on the right). The left and right plots define the two topologies $\Lambda_{\up\down}^{\triangle,(d)}(\Eucl p,\Eucl p')$ and $\Lambda_{\up\down}^{\triangle,(c)}(\Eucl p,\Eucl p')$ respectively.}
    \label{fig:triangle-mom}
\end{figure}

With the previous results we have set the stage to discuss the renormalization of the charged weak bilinear operators
\begin{align}
    \Gamma_{\mu,sr}^{A,\QCD}(\Eucl p,\Eucl p') = \int d^4 \Eucl x \, d^4 \Eucl y \, e^{-i \Eucl p \Eucl x} \, e^{i \Eucl p' \Eucl y} \, \langle \psi_s(\Eucl x) \, [\overline \psi_s \Eucl \gamma_\mu \gamma_5 \psi_r](0) \, \overline \psi_r(\Eucl y) \rangle \,,  \\
    \Gamma_{\mu,sr}^{L,\QCD}(\Eucl p,\Eucl p') = \int d^4 \Eucl x \, d^4 \Eucl y \, e^{-i \Eucl p \Eucl x} \, e^{i \Eucl p' \Eucl y} \, \langle \psi_s(\Eucl x) \, j^{L,-}_\mu(0) \, \overline \psi_r(\Eucl y) \rangle \,.
\end{align}
By adopting the same (yet unspecified) projector $P_\mu$, for which we demand the additional constraint $\{P_\mu, \gamma_5\}=0$, we impose as a sensible renormalization condition for the weak operator
\begin{equation}
    Z_L(\mu^2) \frac{\tr[P^\mu P_R \Lambda_{\mu,\up\down}^L(\Eucl p,\Eucl p')]}{\sqrt{Z_\up(\mu^2) Z_\down(\mu^2)}} = \frac12 \,,
    \label{eq:ZL}
\end{equation}
where the r.h.s. reflects the tree-level condition $\tr[P^\mu P_R \gamma_\mu P_L]=\frac12 \tr[P^\mu \gamma_\mu]$.
The final step in our discussion on the quark bilinears is the connection among the charged vector and weak Green functions. By considering momenta sufficiently high to neglect chiral symmetry-breaking effects, one may easily prove that $\Lambda^{A} = \Lambda^{V} \gamma_5$ and hence $\Lambda^L = \Lambda^V P_L$~\cite{Boyle:2026xls}. At finite lattice spacing and in four dimensions, this is valid, for example, for the Domain-Wall formulation of QCD (up to exponentially small errors~\cite{Arthur:2010ht}).
This leads to
\begin{equation}
    \tr[P^\mu P_R \Lambda^{L}_{\mu,\up\down}] = \tr[P^\mu \Lambda_{\mu,\up\down}^V P_L] = \frac12 \tr[P^\mu \Lambda_{\mu,\up\down}^V] \,,
\end{equation}
and $Z_L(\mu^2) = Z_{V,\up\down}(\mu^2)$.
Since in our work we are interested in the renormalization of the operator $\O(x)$, and the corrections that we discussed so far factorize completely from the leptonic sector, we can add it back by considering 
\begin{equation}
    \Lambda_{\mu,sr}^L(\Eucl p,\Eucl p') \otimes \Eucl \gamma_\mu^L = [\Lambda_{\mu,sr}(\Eucl p,\Eucl p')]^{\alpha\beta,ab} [\Eucl \gamma_\mu^L]^{\gamma\delta} \,.
\end{equation}
Since this is a charged process, a crucial contribution to the calculation of the anomalous dimension and matching originates from the triangle exchange diagrams, drawn in Fig.~\ref{fig:triangle-mom}. The quark connected topology is described by the amputated amplitude (omitting only color indices)
\begin{equation}
\begin{split}
    [\Lambda^{\triangle,(c)}_{sr}& \, (\Eucl p,\Eucl p')]^{\gamma\lambda\rho\sigma} = \int d^4 \Eucl x' d^4 \Eucl y\, d^4 \Eucl y' \, d^4 \Eucl z \, \big[\Eucl \gamma_\alpha D^{-1}_\tau(\Eucl x',0) \Eucl \gamma_\mu^L \big]^{\rho\sigma} \\ & \times \Eucl \Delta_{\alpha\beta}(\Eucl x',\Eucl y') \, 
    \big[ S_s(\Eucl p)^{-1} e^{-i\Eucl p \Eucl y} \langle D_s^{-1}(\Eucl y,\Eucl y') \Eucl \gamma_\beta D_s^{-1}(\Eucl y',0) \Eucl \gamma_\mu^L D_r^{-1}(0,\Eucl z) \rangle e^{i\Eucl p' \Eucl z}  S_r(\Eucl p')^{-1} \big]^{\gamma\lambda} \,.
\end{split}
\end{equation}
$D_\tau(\Eucl x,\Eucl y)$ denotes the free leptonic propagator: since we plan to perform this calculation using the lattice regulator, $D_\tau$ is understood as discretized like the quark fields, i.e. with a regularization preserving chiral symmetry to avoid the appearance of additional operators, as in Ref.~\cite{Boyle:2022lsi}. As before, since we work with the light quarks only, the subscripts ${}_{sr}$ can be omitted. The total contribution of the triangle diagrams is given by ($Q_\tau=-1$)
\begin{equation}
    \delta \Lambda_{\up\down}^\triangle(\Eucl p,\Eucl p') = - \frac{e^2}{2} Q_\tau (\Qu + \Qd)  Z_V^\QCD  \big[ \Lambda^{\triangle,(c)}(\Eucl p,\Eucl p') - \Lambda^{\triangle,(d)}(\Eucl p,\Eucl p') \big] \,.
\end{equation}

For the operator $\O$ we need a projector $P$ which we take from Ref.~\cite{Boyle:2026xls} and whose action on a given two-fermion and two-quark amplitude is defined by
\begin{equation}
    \P [ \Lambda ] \equiv  P^{\alpha\beta\gamma\delta,ab} \Lambda_{\alpha\beta\gamma\delta,ab}  \,.
\end{equation}
It obeys $\P \Big[\Eucl \gamma_\nu^L \otimes \Eucl \gamma_\nu^L \Big]=1$ and it has the functional form\footnote{We remind the reader that $\Eucl \gamma_\mu^L = \Eucl \gamma_\mu (1 - \gamma_5)/2$ and therefore $\Tr[\Eucl \gamma_\mu^R \Eucl \gamma_\nu^L] = 2 \delta_{\mu\nu}$. We use the symbol $\Tr$ to indicate the trace over spin indices only.}
\begin{equation}
    P^{\alpha\beta\gamma\delta,ab} = [P^\mu P_R]^{\alpha\beta,ab}  \otimes [\Eucl \gamma_\mu^R]^{\gamma\delta} \,,
    \label{eq:proj}
\end{equation}
particularly convenient since
\begin{equation}
    \mathcal P[\Lambda_{\mu,sr}^L(\Eucl p,\Eucl p') \otimes \Eucl \gamma_\mu^L] = \tr[P^\mu \Lambda_{\mu,sr}^V(\Eucl p,\Eucl p')] \,.
\end{equation}

For the renormalization of $\O$ we need to consider also the wave-function renormalization of the $\tau$ lepton, $Z_\tau(\mu^2)$, known analytically, thereby arriving at the final renormalization condition
\begin{equation}
    Z_\O(\mu^2) \,  \frac{\P \Big[ \Lambda^L_{\mu,\up\down}(\Eucl p,\Eucl p') \otimes \Eucl \gamma_\mu^L + \delta \Lambda^\triangle_{\up\down}(\Eucl p,\Eucl p') + O(\alpha^2,\alpha\Delta m,\Delta m^2) \Big] }{\sqrt{ Z_\up(\mu^2) \, Z_\down(\mu^2) \, Z_\tau(\mu^2)} } 
    =  1 \,,
    \label{eq:ZO}
\end{equation}
where we have again dropped ${}^\RI$ from the notation.
After some algebra, Eq.~\eqref{eq:ZO} can be re-written (up to corrections of order $O(\alpha^2,\alpha \Delta m, \Delta m^2)$) as the product of three factors
\begin{equation}
    Z_\O(\mu^2) = \sqrt{Z_\tau(\mu^2)} Z_{V,\up\down}(\mu^2) Z_\triangle(\mu^2)
    \label{eq:ZO_product}
\end{equation}
with a term arising from the triangle diagram
\begin{equation}
    Z_\triangle(\mu^2) = 1 - \frac{\P [\delta \Lambda^\triangle_{\up\down}(\Eucl p,\Eucl p')]}{\tr [P^\mu \Lambda_{\mu,\up\down}^{V,\QCD}(\Eucl p,\Eucl p')]} \,.
\end{equation}
The splitting in Eq.~\eqref{eq:ZO_product} has been introduced with the goal of identifying the individual renormalization factors of the three classes of radiative corrections discussed in Subsection~\ref{sec:summary}. In practice we will not use $Z_\O(\mu)$ directly but instead we will separate its contributions according to
\begin{equation}
    [Z_\O(\mu^2)]^2 = 1 + (Z_\tau(\mu^2) - 1) + 2 (Z_{V,\up\down}(\mu^2) - 1) + 2 (Z_\triangle(\mu^2) - 1) + O(\alpha^2,\alpha \Delta m,\Delta m^2) \,,
    \label{eq:ZO_separation}
\end{equation}
where the square implies that we are looking at the rate. Every term in parenthesis is $O(\alpha)$.
As we will see in Section~\ref{sec:init-state} initial-state corrections require the inclusion of the wave function renormalization of the $\tau$ lepton, which is precisely the first parenthesis on the r.h.s. of the equation above. Therefore rather than using $Z_\O(\mu^2)$ we will absorb $(Z_\tau(\mu^2) - 1)$ in the definition of $d\Gamma^\mathrm{init}/ds(\mu)$, and proceed similarly for the other parts.

A second important remark concerning the splitting in Eq.~\eqref{eq:ZO_product} is that it makes evident that if $Z_{V,\up\down}^\QCD=1$ (non-perturbatively) then we have
\begin{equation}
    \sqrt{Z_\tau(\mu^2)} = 1 + O(\alpha) \,, \quad Z_{V,\up\down}(\mu^2) = 1 + O(\alpha) + O(\alpha \alpha_s^n) \,, \quad Z_\triangle(\mu^2) = O(\alpha) + O(\alpha \alpha_s^n)
\end{equation}
namely no pure-QCD corrections of $O(\alpha_s^n)$ appear. In continuum QCD, given the underlying Ward identity, such a statement is driven by the interplay between the wave-function renormalization and the choice of projector $P_\mu$, as studied in Ref.~\cite{Boyle:2026xls}. Here we adopted the family of double-trace SMOM projectors proposed in Ref.~\cite{Boyle:2026xls}, which respects this property and can be used with the Wilson coefficients of Ref.~\cite{Gorbahn:2022rgl}.

A third consideration concerns instead the triangle diagrams. Replacing the (massive Euclidean) photon propagator with
\begin{equation}
    \frac{\mu^2 - \mg^2}{(\Eucl k^2+\mg^2)(\Eucl k^2 + \mu^2)}
    \label{eq:PV}
\end{equation}
is effectively regulating its UV behavior and amounts to a generalization of the $W$-regularization scheme, i.e. a typical Pauli-Villars approach \cite{Biloshytskyi:2022ets}. 
Performing such a change only for the triangle diagrams amounts to replacing $Z_\triangle(\mu)$ with $Z_\triangle^\scheme(\mu)$, with $\scheme$ denoting the new scheme; a similar reasoning holds for the other factors as well and in essence these choices result in considering the renormalized operator $Z_\triangle^\scheme(\mu) Z_{\up\down}^{\scheme'}(\mu) \O(x)$, where $\scheme$ and $\scheme'$ may be different (valid only at the order considered here). Since we plan to use the Wilson coefficient in the $\RI$-SMOM scheme a proper conversion factor should be calculated and we defer the study of renormalization schemes more suitable for a comparison with phenomenological calculations \cite{Cirigliano:2026ios} to another publication.
For the triangle diagram such a conversion is given by
\begin{equation}
    \frac{Z_\triangle(\mu^2)}{Z_\triangle^\scheme(\mu^2)} = 1 - \frac{\P [\delta \Lambda^\triangle_{\up\down}(\Eucl p,\Eucl p')]}{\tr [P^\mu \Lambda_{\mu,\up\down}^{V,\QCD}(\Eucl p,\Eucl p')]} \,,
\end{equation}
with $\Eucl p^2=\mu^2$ and with the diagrams in Fig.~\ref{fig:triangle-mom} evaluated using $\Eucl \Delta_{\alpha\beta}(\Eucl x)$ defined from Eq.~\eqref{eq:PV}. Another common prescription to regulate UV divergences, that has different properties compared to the PV case, is to use a hard cutoff in the loop integral. Here we propose a slight modification, with a smooth exponential cutoff function
\begin{equation}
    \frac{1}{\Eucl k^2 + \mg^2} \to \frac{\Theta_\Delta(\mu - |\vec k|)}{\Eucl k^2 + \mg^2} \,, \quad \Theta_\Delta(x) = \frac12 (1 + \tanh(x/\Delta)) \,.
    \label{eq:expscheme}
\end{equation}
Clearly several similar families of smeared step functions may be defined. The relevance of the smoothness and in general of this prescription will be more evident in Section~\ref{sec:non-fact}.
From the perspective of Lattice QCD adopting one of the regularizations proposed above seems rather convenient because it allows to split the calculation in two terms: the scheme conversion factor which can be evaluated in Euclidean space by means of external off-shell quark states, and the physical contribution to the rate from the triangle diagram (regulated by either the PV or cutoff prescriptions). The remaining problem to be solved is the analytic continuation to Minkowski signature of the latter, the topic of Section~\ref{sec:non-fact}.

Finally we note that for the concrete implementation and calculation of the diagrams shown above in Lattice QCD+QED simulations, additional considerations on the precise definition of the photon are needed. In general for a local theory the UV renormalization factors are chosen to be independent from the IR regulator, typically given by a small photon mass $\mg$ or by the box size $L$.

\subsection{Radiative corrections to the leptonic rate}

We conclude this discussion by considering the fact that the experimental data are normalized by the total electronic rate, as mentioned earlier. The latter is represented by the same operator $\mathcal{O}(x)$, with $\J^{L,-}_\mu(x)$ replaced by the electronic current and the net effect of this normalization is the simplification of $\GF$, including its universal radiative corrections. To first order in the EM coupling, the remaining process-specific corrections to $\Gamma_e$ have been calculated in Refs.~\cite{Kinoshita:1958ru,Marciano:1988vm} and are
\begin{equation}
    \Gamma_e = \frac{\GF^2 \mt^5}{192 \pi^3} \left[1 + O\left(\frac{\mt^2}{m_W^2}\right) + O\left(\frac{m_e^2}{\mt^2}\right) \right] \left[ 1 + \delta \Gamma_e \right] \,,
\end{equation}
with
\begin{equation}
    \delta \Gamma_e = \frac{\alpha}{2\pi} \left(\frac{25}{4} - \pi^2 \right) = - 0.0042 \,.
\end{equation}
Hence a 0.4\% correction is typically added~\cite{Marciano:1988vm,Erler:2002mv}. The two-loop correction is also available \cite{vanRitbergen:1999fi} but leads to a marginal effect of less than 0.04\%, that we discard together with the mass dependent terms.

\section{Initial-state corrections}
\label{sec:init-state}

Here we cover the contribution from all diagrams where a photon is exchanged only within the leptonic sector, see Fig.~\ref{fig:init-state}. 
The first piece is the familiar lepton self-energy which is UV and IR divergent, and which we take into account by multiplying the tree-level decay with $\frac12 [Z_\tau^\scheme(\mu^2) - 1]$, where $Z_\tau^\scheme(\mu)$ is the scale and scheme dependent wave-function renormalization factor 
\begin{equation}
    Z_\tau^\scheme(\mu^2) =1 + \frac{\alpha}{4\pi} \Big[-\log \frac{\mu^2}{\mt^2} + 4 \log \frac{\mt}{\mg} + z_\tau^\scheme + x_\tau^\scheme(\xi) \Big] \,.
\end{equation}
Note the presence of the photon mass $\mg$ to regulate the IR logarithm and the QED gauge parameter $\xi$. 
In the $\MS$ scheme we have (see for instance Ref.~\cite{Jegerlehner:2008zza})
\begin{equation}
    z_\tau^\MS = -4 \,, \quad x_\tau^\MS(\xi) = (1-\xi) \left(1- \log \frac{\mg^2}{\mu^2} \right) + \xi \log \xi \,,
\end{equation}
and the conversion factor to the MOM scheme is $1 - \frac{\alpha}{4\pi} \xi$~\cite{Chetyrkin:1999pq}.

The real emission of a photon from the $\tau$ lepton produces a UV-safe but IR-divergent term, which cancels the IR logarithm in $Z_\tau^\scheme(\mu)$. Therefore 
by separating the IR piece from the rest of the wave function renormalization, we define the short-distance initial-state correction as
\begin{equation}
    \frac{d\Gamma}{ds} \frac{\alpha}{4\pi} \delta Z_\tau^\mathsf{s}(\mu^2) \,, \quad \text{with} \quad 
    \delta Z_\tau^\mathsf{s}(\mu^2) \equiv -\log \frac{\mu^2}{\mt^2} + z_\tau^\mathsf{s} + x_\tau^\scheme(\xi=1) \,.
\end{equation}
This term can be recognized inside Eqs.~\eqref{eq:ZO} and \eqref{eq:ZO_separation}. 

The second contribution where the photon runs across the final-state cut, is related to the squared real radiative emission and it is UV safe. Clearly we must adopt the same regulator to properly cancel the IR divergent part of $Z_\tau$, so we give a small mass $\mg$ to the photon and keep only singular and finite terms, i.e. we neglect all parts vanishing for $\mg \to 0$. To calculate the photon Bremsstrahlung effect, we start from the amplitude
\begin{equation}
\begin{split}
    \M^{\text{real},\tau}_f( P, q, p_1 \cdots  \, p_{n_f}) = -2 \GF \Vud^\ast \, \bar u (q) \gamma^\mu_L \, [\varepsilon^\alpha(k)]^\ast & \,  \frac{i(\slashed P - \slashed k + \mt)}{(P-k)^2-\mt^2} (-ie  \gamma_\alpha) u (P) \\ & \times  
    \langle \out, \vec p_1 \cdots \vec p_{n_f} | \J^-_\mu (0) | 0 \rangle  \,, 
    \label{eq:M_real_tau}
\end{split}
\end{equation}
where we introduced the four-momentum of the photon $k_\mu$, obeying $k^2=\mg^2$, and its polarization vector $\varepsilon_\alpha(k)$. To simplify the notation and the calculation we introduce three Mandelstam invariants
\begin{equation}
    t = (P-k)^2 \,, \quad u=(q+k)^2 \,, \quad v=(P-q)^2
\end{equation}
while we continue to treat the hadronic system as a combined object, with invariant squared mass $s$. Four-momentum conservation relates $t,u,v$ and $s$, and it is used to eliminate $v$. The denominator from the lepton propagator
\begin{equation}
    \frac{1}{(P-k)^2-\mt^2} = \frac{1}{-2P \cdot k + k^2} = \frac{1}{t-\mt^2} \,,
    \label{eq:den}
\end{equation}
generates the known infrared divergence in the limit $t\to \mt^2$.
The contribution of the matrix element to the rate 
\begin{equation}
\begin{split}
    \frac{1}{4 \mt} \sum_{\mathrm{spin,pol}} & \left\vert \M_f^{\real,\tau} \right\vert^2 = \frac{e^2}{4 \mt} 4 \GF^2 |\Vud|^2 \frac{\widetilde {\mathcal L}^{\mu\nu}(P,q,k)}{(k^2 -2P \cdot k)^2} \mathcal H_{f,\mu\nu}(p_1, \dots p_{n_f}) 
\end{split}
\end{equation}
is given in terms of the modified leptonic tensor 
\begin{equation}
    \widetilde {\mathcal L}^{\mu\nu}(P,q,k) = \mathrm{Tr}\left[\gamma_{L}^\mu(\slashed{P}-\slashed{k}+m_\tau)\gamma^\alpha(\slashed{P}+m_\tau)\gamma_\alpha(\slashed{P}-\slashed{k}+m_\tau)\gamma_L^\nu\slashed{q}\right] \,.
\end{equation}
Above we used the fact that the sum over the photon polarizations is proportional to $g_{\alpha\beta}$,
\begin{equation}
    \sum_\text{pol} \varepsilon^\ast_\alpha(k) \varepsilon_\beta(k) = - g_{\alpha\beta} + k_\alpha f_\beta + k_\beta f_\alpha \,,
\end{equation}
up to the gauge-dependent terms $f_\alpha$, which we neglect since they do not contribute to the physical rate.  
The phase-space factor includes also the photon ($\omega_k=\sqrt{\mg^2 + |\vec k|^2}$)
\begin{equation}
    \frac{d^3 \vec q}{(2\pi)^3 2 \omega_q} \frac{d^3 \vec k}{(2\pi)^3 2 \omega_k}  \, 
    d \Phi_f(P-q-k) \,,
\end{equation}
and after factorizing the hadronic contributions in the spectral density, as we did for the leading order (see Eq.~\eqref{eq:dPhif}), we calculate the Jacobian for the change of integration variables to the three invariants introduced above, leading to $(256 \pi^4 \mt^2)^{-1} \, dt \, du \, ds $.
The integral over the phase space reduces to an integral over $u$ and $t$, at fixed $s$. By choosing a frame where $q_\mu=(\omega_q,0,0,\omega_q)$ and $\cos\theta = \frac{\vec q \cdot \vec k}{\omega_q |\vec k|}$, the integration range for $u$ at fixed $s,t$ is determined from $\cos\theta \in [-1,1]$, while the integration range for $t$ is $[(\mt-\mg)^2, s]$. The limit $\mg \to 0$ is taken at the very end of the calculation. 

\begin{figure}
    \centering
    \includegraphics[width=0.7\linewidth]{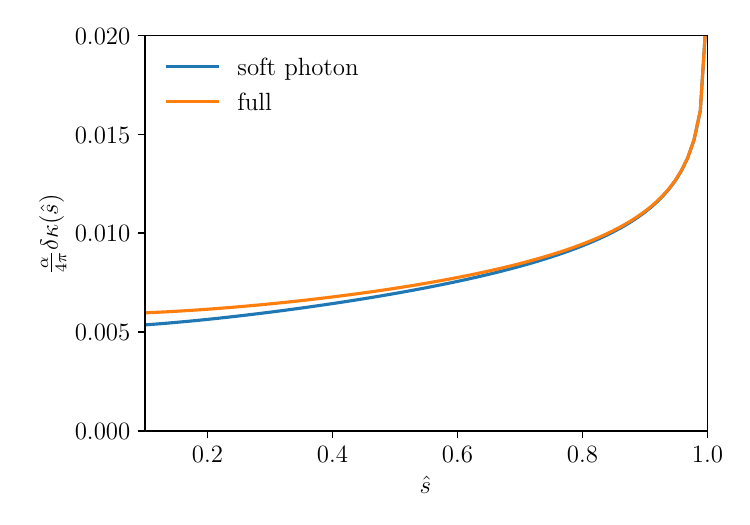}
    \caption{Initial-state corrections in the soft-photon limit and for the full case.}
    \label{fig:dkappa}
\end{figure}

We organize our result for the real emission off the initial-state $\tau$ lepton as a relative correction to the leading rate
\begin{equation}
    \frac{\alpha}{4\pi} \, \frac{d\Gamma}{ds} \big[ B_{\log}(\hat s,\mg) + B_1(\hat s) \log(\hat s)+ B_2(\hat s)  \big]
\end{equation}
where $B_{\log}(\hat s, \mg)$, given by
\begin{equation}
    B_{\log}(\hat s, \mg) = 4 \log \frac{\mg}{\mt} - 4 \log (1- \hat s) \,,
\end{equation}
contains the IR logarithm, while $B_1(\hat s)$ and $B_2(\hat s)$ are $\mg$ independent. Hence, at this order in the EM coupling and by noting that the self-energy contributes twice, we can explicitly check the cancellation of IR divergences in the combination
\begin{equation}
    [Z_\tau-1] + \frac{\alpha}{4 \pi} B_{\log}(\hat s,\mg)
\end{equation}
and we can formulate the fully-factorized (IR-safe) initial-state correction as
\begin{equation}
    \frac{1}{\Gamma_e} \frac{d\Gamma^\mathrm{init}}{ds}(\mu) \equiv \frac{1}{\Gamma_e} \frac{d\Gamma}{ds} \frac{\alpha}{4\pi} \big[\delta Z_\tau(\mu^2) + \delta \kappa(s) \big] \,, \quad
    \delta \kappa(s) \equiv B_{\log}(\hat s,\mt) + B_1(\hat s) \log(\hat s)+ B_2(\hat s) \,.
\end{equation}
For the calculation of $B_1$ and $B_2$, by taking the soft-photon approximation (Low's theorem) we restrict ourselves to the most singular part of the amplitude
\begin{equation}
    \M^{\text{real},\tau}_f \simeq 
    e \, \M_f \frac{2 P \cdot [\varepsilon(k)]^\ast}{-2P \cdot k + k^2} \,,
\end{equation}
 where we also used the equation of motion in the numerator of Eq.~\eqref{eq:M_real_tau}, $(\slashed P + \mt ) \gamma_\alpha u_\tau(P) = 2 P_\alpha u_\tau (P)$. After some algebra, we find
\begin{align}
    B_1(s) =  \frac{2 \hat s^2 (4 \hat s-3)}{\kappa(\hat s)} \,, \quad
    B_2(s) = \frac{(1-\hat s)(9 +5 \hat s - 12 \hat s^2)}{\kappa(\hat s)}
    \,.
\end{align}
Instead, by using the full amplitude without the soft-photon approximation we obtain 
\begin{align}
    B_1(\hat s) = \frac{(3 + 10 \hat s) \hat s^2}{\kappa(\hat s)} \,, \quad 
    B_2(\hat s) = \frac{(1 - \hat s)(61 + 37 \hat s - 20 \hat s^2)}{6 \kappa(\hat s)} \,.
\end{align}
These effects have been previously calculated in Refs.~\cite{Cirigliano:2001er,Cirigliano:2002pv} for two final-state pions. Here we have generalized this result to arbitrary hadronic final states by factorizing their effect in the spectral density $\rho(s)$. We have checked numerically that our formulae agree with the Ref.~\cite{Cirigliano:2001er} and we plot them in Fig.~\ref{fig:dkappa} for both the soft-photon and full amplitudes\footnote{Results for the first derivation of these equations have been presented in Ref.~\cite{Bruno:lat2024,Bruno:2024} and recently an independent derivation, in agreement with ours, has been given in Ref.~\cite{DiCarlo:2026kpv}.}.

\section{Non-factorizable corrections}
\label{sec:non-fact}

The second class of infrared safe diagrams that we consider, are obtained by the exchange of a photon between the $\tau$ lepton and the hadronic sector, in all possible ways as depicted in Fig.~\ref{fig:non-fact}. Our goal is to define a calculation suitable for Euclidean field theories but for these contributions we find a problem with the analytic continuation, as outlined in Subsection~\ref{sec:inverse}. 
We begin with the calculation of the contributions to the rate with Minkowski signature and by isolating the hadronic parts we highlight the nature of the problem.

\subsection{Bremsstrahlung interference}

Using a notation similar to other Sections, we introduce the relevant matrix element corresponding to the emission of a real photon ($k^2 = \mg^2$) from the hadronic system (open spin indices)
\begin{equation}
\begin{split}
    \M^{\real,\had}_f(P, q, k, p_1 \cdots  \, p_{n_f}) =  -(-ie) &\, \frac{4 \GF \Vud^\ast}{\sqrt 2} \, \bar u (q) \, \gamma^{\mu,L} \,  u (P) \, [\varepsilon^\alpha(k)]^\ast \\ \times &
    \int d^4 x \, e^{i k x} \, \langle \out, \vec p_1 \cdots \vec p_{n_f} | T\big\{ \J^\gamma_\alpha(x) \J^{L,-}_\mu (0) \big\} | 0 \rangle  \,.
    \label{eq:M_real_had}
\end{split}
\end{equation}
Notice that despite the selection over final vector states, now the full weak current is required.
The interference between Eq.~\eqref{eq:M_real_had} and the real-photon emission amplitude $\M_f^{\real,\tau}$ given in Eq.~\eqref{eq:M_real_tau}, gives
\begin{equation}
\begin{split}
    \frac{1}{4 \mt} \sum_{\mathrm{spin,pol}} & \M_f^{\real,\had} \left[\M_f^{\real,\tau} \right]^\dagger = i \frac{e^2}{4 \mt} \frac{8 \GF^2 |\Vud|^2}{\sqrt 2} \frac{\mathcal L^{\mu\nu\alpha}(P,q,k)}{-2P \cdot k + k^2} \\ & \times \int d^4 x \, e^{i k x} \langle 0 \vert \J_\mu^+(0) \vert \vec p_1 \cdots \vec p_{n_f} \,, \out \rangle
    \langle \out, \vec p_1 \cdots \vec p_{n_f} | T\big\{ \J_\alpha^\gamma(x) \J^{L,-}_\nu (0) \big\} | 0 \rangle
    \label{eq:interf_M_real}
\end{split}
\end{equation}
with the extended leptonic tensor being
\begin{equation}
    \mathcal L^{\mu\nu\alpha}(P,q,k) = 2 \mathcal L^{\mu\nu}(P,q) P^\alpha - \Tr[\gamma^{\mu,L} \slashed q \gamma^{\nu,L} \slashed P \gamma^\alpha \slashed k] \,.
    \label{eq:L_mu_nu_alpha} 
\end{equation}
The soft-photon limit is obtained from
\begin{equation}
    \frac{e}{4\mt} \sum_{\mathrm{spin,pol}} \left[ \M_f^{\real,\had} \M_f^\dagger \frac{2 P \cdot \varepsilon(k)}{-2P \cdot k + k^2} \right]
\end{equation}
and in practice it is recovered by neglecting the second term of $\mathcal L^{\mu\nu\alpha}$.
Similarly to the real photon emission from the initial state, via Eq.~\eqref{eq:dPhif} the Lorentz-invariant phase-space factor is written as
\begin{equation}
    \frac{d^3 \vec k}{(2\pi)^3 2 \omega_k} \frac{d^3 \vec q}{(2\pi)^3 2 \omega_q} \frac{d^3 \vec p}{(2\pi)^3 2 E} \, (2\pi)^4 \delta^4(P-q-p-k) \, \frac{ds}{2\pi} \, d\Phi_f(p) \,. 
    \label{eq:lips_3}
\end{equation}
After integrating over $d\Phi_f(p)$ and summing over all final states $f$, we find a new inclusive hadronic spectral density
\begin{equation}
\begin{split}
    R_{\mu|\nu\alpha}(p,k_0,\vec k) = & \sum_f \int \frac{d\Phi_f(p)}{2\pi} \mathcal H_{f,\mu|\nu\alpha}(p_1 \dots p_{n_f},k)
    \\ = & \int d x_0 \, e^{i k_0 x_0} \, \langle 0 \vert \widetilde \J_\mu^+(0, \vec p) \, \delta(\hat H - p_0) \,  T\big\{ \widetilde \J_\alpha^\gamma(x_0, \vec k) \J^{L,-}_\nu (0) \big\} | 0 \rangle \,,
\end{split}
\end{equation}
with
\begin{equation}
\begin{split}
    \mathcal H_{f,\mu|\nu\alpha}(p_1 \dots p_{n_f},k) = \int d x_0 \, e^{i k_0 x_0} & \, \langle 0 \vert \J_\mu^+(0) \vert \vec p_1 \cdots \vec p_{n_f} \,, \out \rangle
    \\ & \times \langle \out, \vec p_1 \cdots \vec p_{n_f} \vert  T\big\{ \widetilde \J_\alpha^\gamma(x_0, \vec k) \J^{L,-}_\nu (0) \big\} | 0 \rangle \,.
    \label{eq:H_mu_nu_alpha}
\end{split}
\end{equation}
To make further progress in our derivation we restrict ourselves to the hadronic rest frame where $p_\mu=(\sqrt s, \vec 0)$. A minor simplification may be achieved by leveraging vector current conservation, which should hold at least for $\J^+_\mu(0)$, implying that in this frame $R_{\mu|\nu\alpha}$ vanishes for $\mu=0$.

Contrary to the treatment of lepton Bremsstrahlung, where we adopted several Mandelstam invariants, we keep the various momentum components explicit here. For the phase-space integral (see also Appendix~\ref{app:lips}) we use $d^3 \vec p$ and $d\omega_q$, from $d^3 \vec q = d\omega_q \omega_q^2 d\Omega_q$, to integrate out the $\delta^4(P-q-p-k)$, while the remaining $d\Omega_q$ produces an overall factor of $4\pi$. To calculate the contribution to the rate we must integrate over $\vec k$ and $s$. To guarantee four-momentum conservation one should restrict the integral over $s$ at fixed $\vec k$ up to
\begin{equation}
    s_\max(\vec k) = \sqrt{\mt^2 + |\vec k|^2} - \omega_k \,, 
\end{equation}
while $|\vec k| \in [0, \mt^\ast]$ and $\mt^\ast = \sqrt{\mt^2 - \mg^2}$. It follows that the contribution to the differential rate $d\Gamma^\mathrm{nonf}/ds$ is\footnote{With a slight abuse of notation the first argument of $R_{\mu|\nu\alpha}$, the four-vector $p$, is replaced by $\sqrt s$.}
\begin{equation}
    -i \frac{\alpha}{4\pi} \frac{\GF^2 |\Vud|^2}{\sqrt 2 \pi^2 \mt} \, d^3 \vec k \, \mathcal G^{\mu\nu\alpha}_{\real}(\mg, \sqrt s, \vec k) \, R_{\mu |\nu \alpha}(\sqrt s,\omega_k, \vec k) \, \theta(\mt - |\vec k|) \,,
    \label{eq:brems_real}
\end{equation}
with the dimensionless kernel\footnote{To keep a compact notation we use $(q+p)\cdot k$ in the denominator, with the understanding that the photon is on-shell, $k_0=\omega_k$, and therefore $(q+p)\cdot k$ is a function of $\sqrt s$ and $\vec k$. The vertical line is used to remark that this process obeys the four-particle kinematics constraints. In fact one can show that all the arguments of the integrand depend only on $\vec k$ and $\sqrt s$.}
\begin{equation}
    \mathcal G^{\mu\nu\alpha}_{\real}(\mg, \sqrt s, \vec k) = \frac{\omega_q}{2\omega_k \sqrt s} \frac{\mathcal L^{\mu\nu\alpha}(P,q,k)}{2 (q+p) \cdot k + \mg^2} \bigg|_{P=q+p+k} \theta(s_\max(\vec k) - s) \,, \quad k^2 = \mg^2 \,.
\end{equation}
Note that we took into account the time dilation caused by the unusual frame, such that the rate corresponds to the standard definition in the rest frame of the decaying particle (see also Appendix~\ref{app:lips}).
Following the discussion in Section~\ref{sec:anatomy}, we add the hermitian conjugate of Eq.~\eqref{eq:brems_real}, multiply with $1/\Gamma_e$ and apply the operator $\mathfrak L$ to find the corresponding Euclidean correlator
\begin{equation}
\begin{split}
    \frac{\alpha}{4\pi} \frac{8}{\sqrt 2 \pi \mt^4} 2 \mathrm{Im} & \int d^3 \vec k \, \theta(\mt^\ast - |\vec k|)  \\ & \times \int d s \, \frac{\sqrt s}{2} \frac{\theta(1 - \hat s)}{\kappa(\hat s)} e^{-\sqrt s t}  
    \mathcal G^{\mu\nu\alpha}_{\real}(\mg, \sqrt s, \vec k) \, R_{\mu |\nu \alpha}(\sqrt s,\omega_k, \vec k) \,. 
\end{split}
\end{equation}

\subsection{Virtual correction}

The infrared divergence driven by the denominator $2(q+p) \cdot k + \mg^2$ in $\mathcal G^{\mu\nu\alpha}_{\real}$ is canceled by the corresponding virtual amplitude
\begin{equation}
\begin{split}
    \M^{\virt}_f(P, q, p_1 \cdots  \, p_{n_f}) = & \, -(-ie)^2 \frac{4 \GF \Vud^\ast}{\sqrt 2} \, \bar u (q) \, \gamma^{\mu,L} \, \int \frac{d^4k}{(2\pi)^4} \frac{i(\slashed P + \slashed k + \mt)}{(P+k)^2-\mt^2} \, \gamma^\alpha \, u (P) \, i \Delta_{\alpha\beta}(k) \\ \times &  \int d^4x \, e^{i k x} \,
    \langle \out, \vec p_1 \cdots \vec p_{n_f} | T\big\{ \J^{\beta,\gamma}(x) \J^{L,-}_\mu (0) \big\} | 0 \rangle  \,. 
    \label{eq:M_virt}
\end{split}
\end{equation}
Note that we took the momentum in the loop as leaving the hadronic system, to have a phase $e^{ikx}$ consistent with the Bremsstrahlung part. Alternatively we could have chosen $e^{-ikx}$ together with $P-k$ in the lepton propagator. The interference with the Born amplitude in Eq.~\eqref{eq:Mf} is
\begin{equation}
    \frac{1}{4 \mt} \sum_{\mathrm{spin,pol}} \M_f^{\virt} \M_f^\dagger = -i \frac{e^2}{4 \mt} \frac{8 \GF^2 |\Vud|^2}{\sqrt 2} \int \frac{d^4k}{(2\pi)^4 i} \frac{\mathcal L^{'\mu\nu\alpha}(P,q,k)}{2P \cdot k + k^2} \frac{\mathcal H_{f,\mu|\nu\alpha}(p_1 \dots p_{n_f},k)}{k^2 - \mg^2} \,,
\end{equation}
with
\begin{equation}
    \mathcal L^{'\mu\nu\alpha}(P,q,k) = 2 \mathcal L^{\mu\nu}(P,q) P^\alpha + \Tr[\gamma^{\mu,L} \slashed q \gamma^{\nu,L} \slashed k \gamma^\alpha \slashed P] \,.
\end{equation}
After integrating over the hadronic phase space and summing over all (vector) states we arrive at
\begin{equation}
    -i \frac{e^2}{4 \mt} \frac{8 \GF^2 |\Vud|^2}{\sqrt 2} \int \frac{d^4k}{(2\pi)^4 i} \frac{\mathcal L^{'\mu\nu\alpha}(P,q,k)}{2P \cdot k + k^2} \frac{R_{\mu|\nu\alpha}(p,k_0,\vec k)}{k^2 - \mg^2} \,.
\end{equation}
At this stage, we do not regulate yet the UV divergence and focus instead on the IR behavior (note the appearance of a leptonic tensor similar to Eq.~\eqref{eq:L_mu_nu_alpha}) by examining the integral over the temporal component $k_0$ (see also Appendix~\ref{app:GEM} for additional details)
\begin{equation}
    \int dx_0 \frac{d k_0}{2\pi i} \frac{e^{ik_0 x_0}}{2 P \cdot k + k^2 + i \varepsilon} \, \frac{\mathcal L^{'\mu\nu\alpha}(P,q,k)}{k^2 - \mg^2 + i\varepsilon}  \langle 0 \vert \widetilde \J_\mu^+(0, \vec 0) \, \delta(\hat H - p_0) \,  T\big\{ \widetilde \J_\alpha^\gamma(x_0, \vec k) \J^{L,-}_\nu (0) \big\} | 0 \rangle \,,
    \label{eq:int_k0}
\end{equation}
where we have (re-)inserted the $i \varepsilon$ prescription in the propagators and we have adopted the hadronic rest frame to match the convention above. Depending on whether $x_0$ is positive or negative we close the contour integral in the complex $k_0$ plane in the upward or downward directions respectively, selecting different poles. For this reason we split the hadronic density in the two time orderings (see Appendix~\ref{app:tproducts} for more details and Ref.~\cite{Frezzotti:2023nun} for a similar discussion)
\begin{align}
    \label{eq:rhopos}
    \rhopos_{\mu\nu\alpha}(\sqrt s, k_0, \vec k) = &\, \langle 0 | \widetilde \J_\mu^+(0, \vec 0) \, \delta(\hat H - \sqrt s) \widetilde \J_\alpha^\gamma(0,\vec k) \, \frac{-i}{\hat H - \sqrt s - k_0 - i \varepsilon} \, \J_\nu^{L,-}(0) | 0 \rangle \,, \\
    \rhoneg_{\mu\nu\alpha}(\sqrt s, k_0, \vec k) = &\, \langle 0 | \widetilde \J_\mu^+(0, \vec 0) \, \delta(\hat H - \sqrt s) \J_\nu^{L,-}(0) \, \frac{-i}{\hat H + k_0 - i \varepsilon} \, \widetilde \J_\alpha^\gamma(0,\vec k) | 0 \rangle \,.
\end{align}
Let us start from the poles of the photon propagator. For $x_0>0$ by closing the contour integral in the upper complex plane, due to the prefactor $e^{i k_0 x_0}$, we encircle the poles located at $k_0 = -\omega_k + i\varepsilon$ and find
a contribution to Eq.~\eqref{eq:int_k0} resembling the real photon emission
\begin{equation}
\begin{split}
    - \frac{\mathcal L^{'\mu\nu\alpha}(P,q,k)}{2 P \cdot k + \mg^2}  \, \frac{1}{2\omega_k} \rhopos_{\mu\nu\alpha}(\sqrt s, -\omega_k, \vec k) \,.
\end{split}
\end{equation}
As we will see below, it amounts to putting the photon (in the loop) on its mass shell which in the $\mg \to 0$ limit generates the corresponding (virtual) IR divergence. By adding also the second pole from the photon propagator, relevant when $x_0<0$, the rate receives a term of the form
\begin{equation}
    i \frac{\alpha}{4\pi} \frac{\GF^2 |\Vud|^2}{\sqrt 2 \pi^2 \mt} ds \, d^3 \vec k \, \mathcal G^{\mu\nu\alpha}_{\virt,0}(\mg, \sqrt s, \vec k) \big( \rhopos_{\mu\nu\alpha}(\sqrt s, -\omega_k, \vec k) - \rhoneg_{\mu\nu\alpha}(\sqrt s, \omega_k, \vec k) \big)
\end{equation}
with
\begin{equation}
    \mathcal G^{\mu\nu\alpha}_{\virt,0}(\mg, \sqrt s, \vec k) = \frac{\omega_q}{2 \omega_k \, \sqrt s} \frac{\mathcal L^{'\mu\nu\alpha}(P,q,k)}{2 P \cdot k + \mg^2} \bigg|_{P=q+p} \,, \quad k^2 = \mg^2 \,.
    \label{eq:Gvirt0}
\end{equation}
Contrary to before the kernel $\mathcal G^{\mu\nu\alpha}_{\virt,0}$ satisfies the two-particle kinematic constraints. Moreover, the integral over $s$ is independent from the one over $\vec k$.\\

The last non-factorizable term is given by the residues generated by the $\tau$ propagator.
For $x_0>0$ we encircle the pole located at $k_0 = -\omega_P - \omega_{P+k} + i\varepsilon$, with $\omega_{P+k} = \sqrt{\mt^2 + (\vec P + \vec k)^2}$, which returns another part of Eq.~\eqref{eq:int_k0}
\begin{equation}
\begin{split}
    - \frac{\mathcal L^{'\mu\nu\alpha}(P,q,k)}{(\omega_P + \omega_{P+k})^2 - \omega_k^2}  \, \frac{1}{2\omega_{P+k}} \rhopos_{\mu\nu\alpha}(\sqrt s, -\omega_P-\omega_{P+k}, \vec k) \,,
\end{split}
\end{equation}
whereas for $x_0<0$ the relevant pole is $k_0 = - \omega_P + \omega_{P+k} - i \varepsilon$. 
With little algebra we find the last term
\begin{equation}
    i \frac{\alpha}{4\pi} \frac{\GF^2 |\Vud|^2}{\sqrt 2 \pi^2 \mt} ds \, d^3 \vec k \, \sum_{\mathsf{s}=\pm} \mathsf s \, G^{\mu\nu\alpha}_{\virt,\mathsf{s}}(\mg, \sqrt s, \vec k)  \, \rho^\mathsf{s}_{\mu\nu\alpha}(\sqrt s, -\omega_P - \mathsf{s} \, \omega_{P+k}, \vec k) 
\end{equation}
with the new kernel\footnote{We note that by explicitly calculating the residues from the integral over $dk_0$ we have effectively obtained the expression in time-ordered perturbation theory (TOPT). In fact by further expanding the denominators inside the various kernel functions, $\mathcal G_{\virt,0}$ and $\mathcal G_{\virt,\pm}$, we would recover all possible time orderings in what is often called old-fashioned perturbation theory. Recently in Ref.~\cite{Sterman:2023xdj} TOPT has been analyzed in the context of resummation of IR divergences, a problem very similar to the one treated here.}
\begin{equation}
    \mathcal G^{\mu\nu\alpha}_{\virt,\pm}(\mg, \sqrt s, \vec k) = \frac{\omega_q}{2 \omega_{P+k} \sqrt s} \frac{\mathcal L^{'\mu\nu\alpha}(P,q,k)}{ (\omega_P \pm \omega_{P+k})^2 - \omega_k^2} \bigg|_{P=q+p} \,.
\end{equation}

\subsection{Rate}

\textit{Cancellation of IR divergences} --- To verify the cancellation of the IR divergence we restrict ourselves to the soft-photon limit and $\rhopos_{\mu\nu\alpha}$: in this approximation the integrand in Eq.~\eqref{eq:brems_real} simplifies to
\begin{equation}
    \frac{\omega_q}{2 \omega_k \sqrt s} 2 \mathcal L^{\mu\nu}(P,q) (q+p)^\alpha \rhopos_{\mu\nu\alpha}(\sqrt s,\omega_k,\vec k) \frac{1}{2 (q+p) \cdot k + \mg^2} \bigg|_{P=q+p+k} 
\end{equation}
thanks to the conservation of the EM current. For the virtual loop one finds exactly the same expression from $\mathcal G_{\virt, 0}$ evaluated for $P=q+p$ and with $\rhopos_{\mu|\nu\alpha}$ as a function of $-\omega_k$, but with an overall minus sign in front. By adding the two terms together we find that the integrand (over $d^3 \vec k$) tends to zero in $|\vec k| \to 0$ limit, since the two hadronic tensors become identical and the four-particle kinematics tends to the three-particle case (see Appendix~\ref{app:lips}). Hence the integrand is well defined and we may remove the IR regulator $\mg$ (even though it may still be very beneficial to keep it in an actual calculation). The origin of the IR divergence may be traced by examining the kernels together with the hadronic tensors. In fact, when the hadronic states  propagating between $\widetilde \J_\alpha^\gamma$ and $\J^{L,-}_\nu$ inside $\rhopos_{\mu\nu\alpha}$ have an energy equal to $\sqrt s$ there is a pole of order $1/|\vec k|$.
In this situation by further considering the soft-photon limit of the kernel we find the well-known IR logarithm 
\begin{equation}
    d^3 \vec k \times \mathcal{G}_{\real} \times \rhopos_{\mu\nu\alpha} \overset{|\vec k|\to 0}{\simeq} d|\vec k| |\vec k|^2 \times \frac{1}{|\vec k|^2} \times \frac{1}{|\vec k|} \to \log(|\vec k|) \,.
\end{equation}
Instead since $\rhoneg_{\mu\nu\alpha}$ is evaluated at $+\omega_k$ for both $\mathcal G_{\real}$ and $\mathcal G_{\virt,0}$ it never generates a pole with negative powers of $|\vec k|$, also thanks to the mass gap in QCD. Its contribution is therefore IR safe and no cancellation is required in this case.

Finally, with little algebra, $\mathcal G_{\virt,\pm}$ can be shown to be remain finite when $|\vec k| \to 0$. Combining this with the corresponding hadronic tensors does not spoil its IR safety and as a consequence we may take $\mg \to 0$.\\

\textit{Regularization} ---
Since the virtual Bremsstrahlung originates from the exchange of a virtual photon, there is no upper limit on the integration over $|\vec k|$, which can be problematic. For large photon momenta both time orderings, $\rhopos_{\mu\nu\alpha}$ and $\rhoneg_{\mu\nu\alpha}$, scale as $1/|\vec k|$ generating the well-known UV logarithms once integrated with the inverse powers of $|\vec k|$ inside the kernels.

By adopting the extended $W$-scheme from Eq.~\eqref{eq:PV} we note that $\mathcal G_{\virt,0}$ is unchanged when the pole at $-\omega_k + i \varepsilon$ is encircled, but there is a new pole located at $k_0 = -\sqrt{\mu^2 + |\vec k|^2} + i \varepsilon$ whose residue regulates the integral, see Appendix~\ref{app:GEM}.
However the pole structure of the PV propagator slightly complicates the expression with this additional residue, making the proposed smooth cutoff scheme more appealing in this respect. In this case $\mathcal G_{\virt,0}$ is simply multiplied by the function $\Theta_\Delta(\mu - |\vec k|)$, which does not spoil the soft limit and automatically regulates the UV one.\\

\textit{Euclidean form} ---
Finally, we have all the ingredients to calculate the contribution of $d\Gamma^\mathrm{nonf}/ds(\mu)$ to $G^{W,\exp}(t,\mu)$. After some algebra we find
\begin{equation}
\begin{split}
    \mathfrak{L}(t) \cdot \frac{1}{\Gamma_e} \frac{d\Gamma^\mathrm{nonf}}{ds}(\mu) = &\, 2 \bigg[\frac{Z_\triangle(\mu^2)}{Z_\triangle^\scheme(\mu^2)} - 1 \bigg] \mathfrak{L}(t) \cdot \frac{1}{\Gamma_e} \frac{d\Gamma}{ds} \\ & + \frac{\alpha}{4\pi} \frac{8}{\sqrt 2 \pi \mt^2} 2 \mathrm{Im} \int d^3 \vec k
    \int d\sqrt s \,  \frac{\theta(1 - \hat s)}{\kappa(\hat s)} \, \hat s \,  e^{-\sqrt s t} \times  \\ & 
    \bigg\{ \mathcal G^{\mu\nu\alpha}_{\real}(\mg) \, \theta(\mt^\ast - |\vec k|) \left(\rhopos_{\mu\nu\alpha}(\omega_k) + \rhoneg_{\mu\nu\alpha}(\omega_k) \right)
    \\ &
    - \mathcal G^{\mu\nu\alpha}_{\virt,0}(\mg) \,\Theta_\Delta(\mu -|\vec k|) \left(\rhopos_{\mu\nu\alpha}(-\omega_k) - \rhoneg_{\mu\nu\alpha}(\omega_k) \right) 
    \\ & 
    - \sum_{\mathsf{s}=\pm} \mathsf s \, \mathcal G^{\mu\nu\alpha}_{\virt,\mathsf{s}}(\mg) \,\Theta_\Delta(\mu -|\vec k|) \, \rho^\mathsf{s}_{\mu\nu\alpha}(-\omega_P - \mathsf{s} \, \omega_{P+k})
     \bigg\} \,,
    \label{eq:dGamma_nonf}
\end{split}
\end{equation}
where we dropped the dependence on $\sqrt s$ and $\vec k$ inside the curly brackets, for better readability.

We remind the reader that the ratio of renormalization factors in the first line has been discussed in Section~\ref{sec:short-distance} and is scale independent. Above $\scheme$ denotes the smooth cutoff scheme and the dependence of the overall amplitude $d\Gamma^\mathrm{nonf}(\mu)$ on $\mu$ resides in the curly brackets. In particular by choosing $\Theta_\Delta \overset{\Delta \to 0}{\to} \theta$ and by setting $\mu=\mt^\ast$, we find that all terms can be integrated together over the same range making the calculation more transparent and the choice of the prescription based on a hard cutoff more natural.
The main reason for introducing instead a smooth step function are finite-volume effects. While we do not discuss them here, as they would require a separate and dedicated treatment, we simply point out that in general we expect a smooth step function to behave better, possibly exponentially~\cite{Bresciani:2026kjv}, when the volume of the corresponding Lattice calculation is increased.

\subsection{Analytic continuation}

At this point the real question to be answered is whether the hadronic densities can be calculated from the Euclidean theory. Since this is now our focus, we omit the multiplicative factors required for local currents for better readability.
Following the arguments presented in Subsection~\ref{sec:inverse}, we immediately notice a difference between $\rhoneg_{\mu\nu\alpha}$ and $\rhopos_{\mu\nu\alpha}$. Both may be seen as densities smeared with a Cauchy kernel with a simple pole. For $\rhoneg_{\mu\nu\alpha}$ and $k_0>0$ the pole is always located on the left of the hadronic branch cut generated by the weak and EM currents, implying that a direct analytic calculation in Euclidean space-time is possible (and $\varepsilon$ is superfluous). Instead this is not the case for $\rhopos_{\mu\nu\alpha}$, where an inverse problem is present.

So let us begin from the simpler case, $\rhoneg_{\mu\nu\alpha}$. Using a few algebraic steps reported in Appendix~\ref{app:tproducts}, and $n_{\mu\mu'} = \mathrm{diag}(1,-i,-i,-i)$ (see Appendix~\ref{app:conventions}) one can show that
\begin{equation}
\begin{split}
    i \int d\sqrt s \, e^{-\sqrt s t} \, \rhoneg_{\mu\nu\alpha}(\sqrt s,\omega_k,\vec k) = &\, \int_{-\infty}^0 d\tau \, e^{\omega_k \tau} \, \langle \widetilde \J_\mu^+(-it, \vec 0) \, \J_\nu^{L,-}(0) \widetilde \J_\alpha^\gamma(-i\tau, \vec k) \,  \rangle \\ = &\, 
    n_{\mu\mu'} n_{\nu\nu'} n_{\alpha\alpha'} \int_{-\infty}^0 d\tau \, e^{\omega_k \tau} \, \langle \widetilde j_{\mu'}^+(t, \vec 0) \, j_{\nu'}^{L,-}(0) \widetilde j_{\alpha'}^\gamma(\tau, \vec k) \,  \rangle
    \,.
    \label{eq:R<_eucl}
\end{split}
\end{equation}
If we replace $\omega_k$ with $\omega_{P+k}-\omega_P$, the integral on the r.h.s. it still converging (at $-\infty$), since $\omega_{P+k} \geq \omega_P$, and the equation above holds for $\rhoneg_{\mu\nu\alpha}(\sqrt s,\omega_{P+k}-\omega_P,\vec k)$ as well. 
Beyond the large $\tau$ limit one should also consider the case $\tau \to 0$, as this could lead to potential non-integrable singularities due to contact terms. To understand whether this is the case we examine the operator product expansion (OPE) of the weak and EM currents. The flavor-changing nature of the weak current forces the minimal structure to be of the form $\bar u \cdots d$, with a short-distance coefficient scaling at most with $1/|x|^3$, where $x$ now denotes the separation of the two operators $x_\mu=(\tau, \vec x)$. Close to $x\simeq0$ we can set the exponential factors to one and after integrating over $d^3 \vec x d\tau$ we obtain a finite result. At non-zero lattice spacing discretization errors are present and we defer their study to future work.

\begin{figure}[ht]
    \centering
    \includegraphics[width=0.49\linewidth]{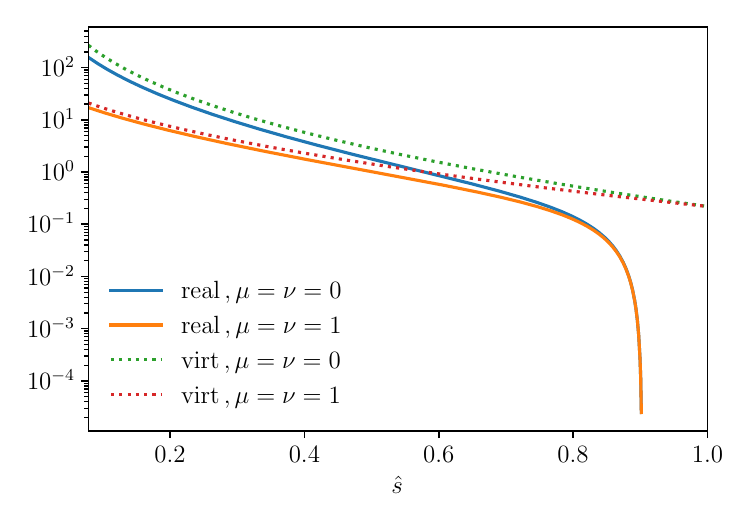}
    \includegraphics[width=0.49\linewidth]{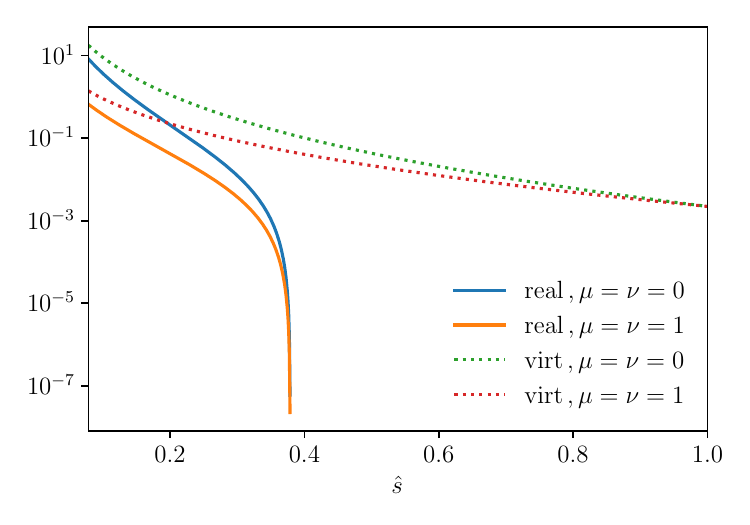}    
    \caption{Plot of the absolute value of the kernel in Eq.~\eqref{eq:g} for both $\mathcal G_{\real}$ and $\mathcal G_{\virt,0}$. The leptonic tensor is approximated to the soft-photon limit and the index $\alpha$ is fixed to the temporal component, $\alpha=0$. The photon mass is approximately $90~\mathrm{MeV}$ and $\mt t=5$. The horizontal axis spans the integration range in $s$, which goes up to $s_\max(|\vec k|)$ for $\mathcal G_{\real}$. \textit{Left}: $\vec k =(0,0,0) $. \textbf{Right}: $\vec k = (0,0,\mt/2)$.}
    \label{fig:Gbrems}
\end{figure}

Therefore it seems that only a single inverse problem is present, the latter being represented by the need to extract $\rhoneg_{\mu\nu\alpha}$ from the l.h.s. of Eq.~\eqref{eq:R<_eucl} for a fixed $\sqrt s$. This defines a fairly hard inverse problem since it attempts to select hadronic intermediate states (between $j_\mu^+$ and $j_\nu^{L,-}$) with fixed energy, or said differently it tries to find the inverse Laplace transform of Eq.~\eqref{eq:R<_eucl}. 
However this is not needed since our approach is based on 
the inclusive hadronic integral defined by Eq.~\eqref{eq:dGamma_nonf} a fact that could lead to a potential simplification. To visualize it let us consider the virtual-photon kernel $\mathcal G_{\virt,0}$ as a representative example whose associated inverse problem is formally defined by
\begin{equation}
    \int_{0}^\infty dt' \, g^{\mu\nu\alpha}_{\virt,0}(t,t'| \mg, \vec k) \, e^{-\sqrt s t'} = \frac{\hat s}{\mt^2} e^{-\sqrt s t} \frac{\theta(1 - \hat s)}{\kappa(\hat s)} \, \mathcal{G}^{\mu\nu\alpha}_{\virt,0}(\mg, \sqrt s, \vec k) \,.
    \label{eq:g}
\end{equation}
By leveraging our arguments from Subsection~\ref{sec:inverse} we see that the non-analytic structure introduced by the step function is sufficient to prevent a direct analytic continuation.
However since Eq.~\eqref{eq:g} attempts to turn one Euclidean correlator into another we expect the numerical complexity of the solution of Eq.~\eqref{eq:g} to be significantly reduced compared to the direct extraction of $\rhoneg_{\mu\nu\alpha}$ from Eq.~\eqref{eq:R<_eucl}.
By combining Eq.~\eqref{eq:R<_eucl} together with Eq.~\eqref{eq:g} we derive an expression for the calculation of one term of $d\Gamma^\mathrm{nonf}/ds$ in Eq.~\eqref{eq:dGamma_nonf}
\begin{equation}
\begin{split}
    -i \int_{0}^\infty dt' \, g^{\mu\nu\alpha}_{\virt,0}(t,t'|\mg,\vec k) \, n_{\mu\mu'} n_{\nu\nu'} n_{\alpha\alpha'} \int_{-\infty}^0 d\tau \, e^{\omega_k \tau} \, \langle \widetilde j_{\mu'}^+(t', \vec 0) \, j_{\nu'}^{L,-}(0) \widetilde j_{\alpha'}^\gamma(\tau, \vec k) \,  \rangle \,.
\end{split}
\end{equation}
The operation above can be repeated for the other kernels as well, which could be treated altogether as a single inverse problem, entirely in Euclidean space.
Also in this case the short-distance behavior of the integral in $t'$ requires dedicated considerations. Here we need to study the OPE of two currents which, contrary to above, generate short-distance singularities~\cite{Chetyrkin:1985kn}: this is readily seen by noticing that $j^+ \cdot j^-$ has vacuum quantum numbers and in fact one finds the typical $1/t'^3$ behavior of the traditional HVP correlator. As we discussed in Section~\ref{sec:anatomy} this is balanced by the $(g-2)_\mu$ weight function~\cite{Bernecker_2011}.
Above we kept the explicit dependence on $t$ to underline the generality of the approach which could be useful in other contexts, but since we are interested in $a_\mu$ the reconstruction of the correlator in the new coordinate $t$ is not necessary: one can directly integrate Eq.~\eqref{eq:g} with the muon kernel, and possibly a window, and define new coefficients that do not depend on $t$.  By doing this operation we expect to regulate the OPE at distances $\mt t' \lesssim 1$, but we defer a concrete study to future work when real data will be available. Additionally, as shown in Fig.~\ref{fig:Gbrems}, the kernel from Eq.~\eqref{eq:g} is a well-behaved function in the relevant range in $s$ and the presence of the step function further suppresses the high-energy regime.
We conclude that the terms proportional to $\rhoneg_{\mu\nu\alpha}$ can be calculated from Lattice QCD, and before continuing to the other density, we discuss possible strategies to further improve its determination.

After inserting a complete set of states between $j_\mu^+$ and $j_\nu^{L,-}$, assuming a discrete spectrum in a finite-volume for simplicity, 
\begin{equation}
    \sum_n \langle j_\mu^+(0) | n, \vec 0 \rangle e^{-t E_n(\vec 0)} \times \int_{-\infty}^0 d\tau \,e^{\omega_k \tau}\, \langle n, \vec 0 | j_\nu^{L,-}(0) \widetilde j_\alpha^{\gamma}(\tau, \vec k)   | 0 \rangle  \,,
\end{equation}
we find the same energies ($E_n(\vec 0)$) and matrix elements ($\langle j_\mu^+(0) | n, \vec 0 \rangle$) that also contribute to the isovector part of the (neutral) HVP and that have been determined by several groups in dedicated studies of the long-distance window contribution to $a_\mu$~\cite{RBC:2024fic,Djukanovic:2024cmq,FermilabLatticeHPQCD:2024ppc}. By re-using this knowledge in a fit of the above expression to the data, with only the matrix elements on the right as free parameters, one would directly calculate the integral over $d\sqrt s$ in Eq.~\eqref{eq:dGamma_nonf}, up to the most energetic state that one is able to extract. The presence of the $\theta$ function in Eq.~\eqref{eq:dGamma_nonf} is very beneficial since it restricts the energy range which becomes even narrower for the Bremsstrahlung part as $|\vec k|$ increases (see Fig.~\ref{fig:Gbrems}), making the exclusive reconstruction a promising direction to bypass the inverse problem\footnote{Solving the inverse problem is in any case needed for the remaining part of the spectrum not covered by the explicit reconstruction, using a subtracted correlator as proposed in Ref.~\cite{Bruno:2020kyl}.}.\\

In the time ordering where the weak current acts on the vacuum, $\rhopos_{\mu\nu\alpha}$, a less trivial kernel (of Cauchy type) appears inside the hadronic tensor. As already discussed and further shown in Appendix~\ref{app:tproducts}, it cannot be obtained in isolation from a simple integral over Euclidean time coordinates. Partly this can be understood from the need of the $i \varepsilon$ prescription in its definition, which regulates the case where the hadronic states propagating between $\J^\gamma_\alpha$ and $\J_\nu^{L,-}$ have an energy equal to $\sqrt s \pm \omega_k$. 
It follows that there is a second inverse problem, which cannot be straightforwardly factorized from the first one, due to the presence of $\sqrt s$ both in the $\delta$-function and in the Cauchy kernel in Eq.~\eqref{eq:rhopos}. Hence, the ``double'' inverse problem is formally defined by (at fixed $\mg, \vec k, \varepsilon$)
\begin{equation}
\begin{split}
    \int_{0}^\infty dt' \, \int_{-\infty}^0 d\tau \,  h_{\virt,0}^{\mu\nu\alpha} & \, (\tau,t',t|\mg,\vec k,\varepsilon) \, e^{-\sqrt s t' + \omega \tau} = \\
    & \frac{\hat s}{\mt^2} e^{-\sqrt s t} \frac{\theta(1-\hat s)}{\kappa(\hat s)} \, \frac{\mathcal{G}^{\mu\nu\alpha}_{\virt,0}(\mg, \sqrt s, \vec k)}{\omega - \sqrt s + \omega_k - i\varepsilon} \,.
    \label{eq:h}
\end{split}
\end{equation}
It admits a solution, at least numerically, and we defer an in-depth study to future work, once the corresponding numerical data will be available. 
From Eq.~\eqref{eq:h} (and by taking the limit $\varepsilon \to 0$) one can extract the virtual-photon contribution proportional to $\rhopos_{\mu\nu\alpha}$ in Eq.~\eqref{eq:dGamma_nonf} from the Euclidean correlator 
\begin{equation}
    (-i) \, n_{\mu\mu'} n_{\nu\nu'} n_{\alpha\alpha'} \int_{0}^\infty dt' \, \int_{-\infty}^0 d\tau \, h_{\virt,0}^{\mu\nu\alpha}(\tau,t',t|\mg,\vec k,\varepsilon) \, \langle \widetilde j^+_{\mu'}(t',\vec 0) \, j^\gamma_{\alpha'}(0) \, \widetilde j_{\nu'}^{L,-}(\tau,-\vec k) \rangle \,.
\end{equation}
Following similar considerations as above, we deduce that short-distance singularities are generated only when both $t'$ and $\tau$ are close to zero and therefore particular care is required near these integration limits. We conclude that the determination of the part proportional to $\rhopos_{\mu\nu\alpha}$ via Eq.~\eqref{eq:h} constitutes a challenging calculation, requiring dedicated efforts, since it involves several delicate limits (e.g. $\varepsilon \to 0$), but some mitigation is in principle possible\footnote{A form factor decomposition, provided that the basis is sufficiently contained, could be an interesting alternative.}.

From a quick inspection of the relevant quantum numbers we find that the non-zero contributions from the product $j^\gamma \cdot j^{L,-}$ are $j^{\gamma,1} \cdot j^-$ and $j^{\gamma,0} \cdot j^{A,-}$, with $j^{A,-}$ the axial part of the weak current. By isolating the iso-vector part of the central EM current (expected to dominate), we insert two complete sets of states (again taken in a finite-box for convenience) to find
\begin{equation}
    \sum_{n,m} \langle j_\mu^+(0) | n, \vec 0 \rangle e^{-t E_n(\vec 0)} \langle n, \vec 0 | \widetilde j_\alpha^{\gamma,1}(0) | m, \vec k \rangle e^{\tau E_m(\vec k)} \langle m, \vec k |  j_\nu^-(0) | 0 \rangle \,.
\end{equation}
As before energies and matrix elements on the left can be taken from the exclusive reconstruction of the vector-vector correlator, which would require an extension to moving frames to determine $E_m(\vec k)$ and the matrix elements on the right, for several values of $\vec k$. With this knowledge one could fit the expression above with the matrix elements in the middle as unknowns. We note however that the case $\vec k =0$ expected to be the more relevant contribution, since it has the larger hadronic phase-space and is related to the IR divergence cancellation, can be studied straightforwardly once again from the knowledge inherited from the exact reconstruction, which also here proves to be beneficial. \\

In summary, one time ordering can be entirely determined from a non-perturbative Lattice QCD calculation in an inclusive manner by evaluating the three-point correlator $\langle \widetilde j_\mu^+(t,\vec 0) \, j_\nu^{L,-}(0) \, \widetilde j_\alpha(\tau, \vec k) \rangle$ for several values of $\vec k$, by performing the necessary Euclidean time integrations and then concluding with the final integration over $\vec k$ (alternatively one could also perform the calculation in coordinate space~\cite{DiCarlo:2026kpv}).
The other hadronic tensor $\rhopos_{\mu\nu\alpha}$ is fairly challenging for an Euclidean setup, as it entails solving a new class of ``double'' inverse problems; here the interplay with the ongoing effort based on the dispersive approach~\cite{Colangelo:2025ivq} will be very important.

\section{Final-state corrections}
\label{sec:fin-state}

Final state radiative corrections, amount to the exchange of a photon within the hadronic sector in all possible ways, as schematically shown in Fig.~\ref{fig:fin-state}. As explained in Section~\ref{sec:anatomy}, rather than the absolute size of radiative corrections in $\tau$ decays, we are only interested in the difference with the neutral channel leading to $\delta G_{11}$ which is calculable in the Euclidean theory. In addition, our focus on the HVP contribution to $a_\mu$ makes it particularly suited for Lattice simulations given the absence of an inverse problem.

We begin by reviewing the leading order isospin-breaking corrections to $G^\gamma(t)$, defined in terms of several new three- and four-point correlation functions, which we report in Fig.~\ref{fig:QED}, for the QED sector, and in Fig.~\ref{fig:SIB}, for the SIB one. Several groups~\cite{\latib} are actively working on the calculation of these diagrams to estimate the QED and SIB corrections to $a_\mu^\mathrm{HVP,LO}$ and our plan is to leverage this ongoing effort to address final-state corrections, making it clearer why we designed our strategy around the decomposition introduced in Eq.~\eqref{eq:dG11}.
We remind the reader that the photon propagator has to be taken in a given gauge, consistent with the initial and non-factorizable effects, here chosen to be the Feynman gauge.
When performing the calculation in a finite periodic box, subtleties related to its definition arise. Discussing the various prescriptions and the associated benefits goes beyond the scope of this paper, whose primary focus lies in the identification of the relevant contributions to define a concrete strategy, and we refer the interested reader to Ref.~\cite{Patella:2017fgk}. One proposal that has been extensively used in the literature is the $\QEDL$ formalism~\cite{Hayakawa:2008an}
where the spatial zero-modes $|\vec k|^2 =0$ are excluded from the sum.
Other appealing approaches are the infinite-volume QED or $\QEDinf$~\cite{Blum:2018mom} and the additional alternatives in Refs.~\cite{Endres:2015gda,Lucini:2015hfa,DiCarlo:2025uyj}. 

\def\WT{0.19\textwidth}
\def\S{2.7em}
\def\SS{3.1em}
\def\SSS{4.4em}
\def\SSSS{5.9em}
\begin{figure}[ht]
    \centering
    \begin{subfigure}[t]{\WT}
        \centering
        \includegraphics[height=\S]{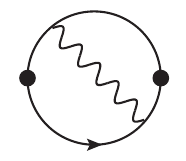}
        \caption{$(V)$}
        \label{fig:V}
    \end{subfigure}
    \begin{subfigure}[t]{\WT}
        \centering
        \includegraphics[height=\S]{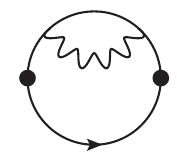}
        \caption{$(S)$}
        \label{fig:S}
    \end{subfigure}
    \begin{subfigure}[t]{\WT}
        \centering
        \includegraphics[width=\SS]{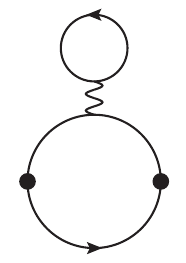}
        \caption{$(T)$}
        \label{fig:T}
    \end{subfigure}
    \begin{subfigure}[t]{\WT}
        \centering
        \includegraphics[width=\SS]{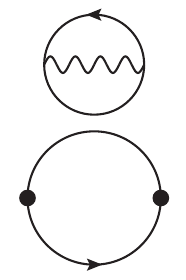}
        \caption{$(D1)$}
        \label{fig:D1}
    \end{subfigure}
    \begin{subfigure}[t]{\WT}
        \centering
        \includegraphics[height=\SSS]{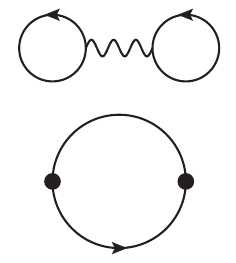}
        \caption{$(D2)$}
        \label{fig:D2}
    \end{subfigure}
    \vskip 2ex
    \begin{subfigure}[t]{\WT}
        \centering
        \includegraphics[height=\S]{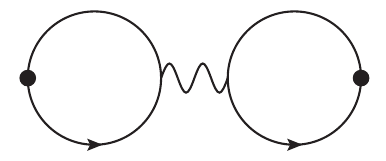}
        \caption{$(F)$}
        \label{fig:F}
    \end{subfigure}
    \begin{subfigure}[t]{\WT}
        \centering
        \includegraphics[height=\S]{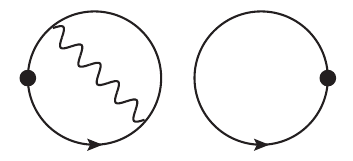}
        \caption{$(D3)$}
        \label{fig:D3}
    \end{subfigure}
    \begin{subfigure}[t]{\WT}
        \centering
        \includegraphics[height=\SSS]{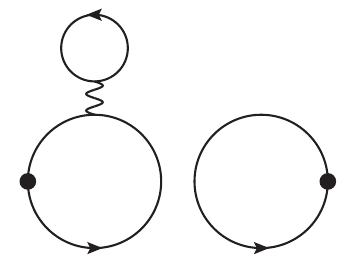}
        \caption{$(Td)$}
        \label{fig:Td}
    \end{subfigure}
    \begin{subfigure}[t]{\WT}
        \centering
        \includegraphics[width=\SSSS]{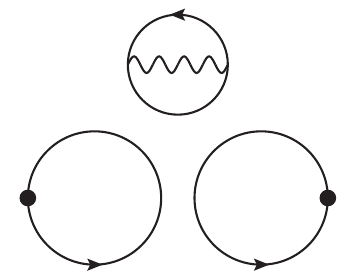}
        \caption{$(D1d)$}
        \label{fig:D1d}
    \end{subfigure}
    \begin{subfigure}[t]{\WT}
        \centering
        \includegraphics[height=\SSS]{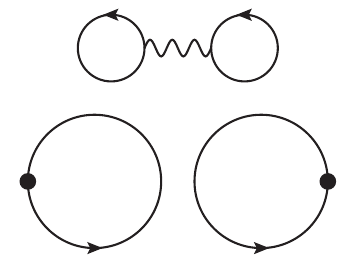}
        \caption{$(D2d)$}
        \label{fig:D2d}
    \end{subfigure}
    
    \caption{List of QED corrections to $G^\gamma$. The two black dots represent the external currents. We draw only fermion lines and for disconnected diagrams the appropriate subtractions are understood. In the first and second rows we plot the QED corrections to the connected and disconnected parts respectively. We do not draw the tadpole topologies associated with the point-split conserved vector currents.}
    \label{fig:QED}
\end{figure}

\begin{figure}[ht]
    \centering
    \begin{subfigure}[t]{\WT}
        \centering
        \includegraphics[height=\S]{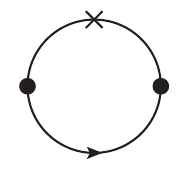}
        \caption{$(M)$}
        \label{fig:M}
    \end{subfigure}
    \begin{subfigure}[t]{\WT}
        \centering
        \includegraphics[height=\S]{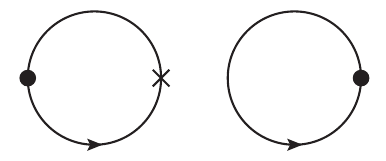}
        \caption{$(O)$}
        \label{fig:O}
    \end{subfigure}
    \begin{subfigure}[t]{\WT}
        \centering
        \includegraphics[width=\SS]{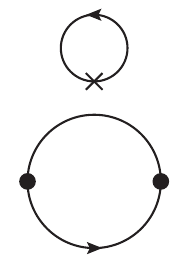}
        \caption{$(R)$}
        \label{fig:R}
    \end{subfigure}
    \begin{subfigure}[t]{\WT}
        \centering
        \includegraphics[width=\SSSS]{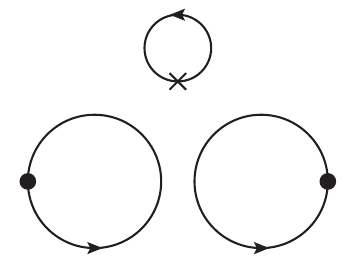}
        \caption{$(Rd)$}
        \label{fig:Rd}
    \end{subfigure}
    
    \caption{List of SIB corrections to $G^\gamma$. The cross represents the insertion
    of the scalar operator.}
    \label{fig:SIB}
\end{figure}

Diagrams with disconnected loops or tadpoles are in general expected to be suppressed, with the most significant contributions coming from diagrams $(V)$ and $(S)$, and from $(F)$ and $(D3)$, for the QED corrections to the connected and disconnected parts respectively, together with diagrams $(M)$ and $(O)$ for SIB effects.
As before, charge factors and minus signs are separate from the definition of the diagram; as an example we provide the expression of the Wick contraction labeled $(V)$ in Fig.~\ref{fig:V}
\begin{equation}
\begin{split}
    (V)_{rs}(\Eucl x_4) = \frac{1}{3} \sum_k \int d^3 {\vec x} \,d^4 \Eucl y \, d^4 \Eucl z \
    \langle \tr &\,  \big[
    \Eucl \gamma_k D^{-1}_r (\Eucl x,\Eucl y) \Eucl \gamma_\alpha D^{-1}_r (\Eucl y,0)  \Eucl \gamma_k \\ & \, \times
    D^{-1}_s (0,\Eucl z) \Eucl \gamma_\beta D^{-1}_s (\Eucl z,\Eucl x) \big] \rangle 
    \Eucl \Delta_{\alpha\beta}(\Eucl y,\Eucl z) \,.
    \label{eq:V}
\end{split}
\end{equation}
At the order considered here, the flavor subscripts in $(V)_{rs}(\Eucl x_4)$ are not needed since we focus on the light quarks only, which are degenerate, so we drop them from the notation.
Using the decomposition introduced in Figs.~\ref{fig:QED} and \ref{fig:SIB}, we expand the three isospin components of $G^\gamma$ in Eq.~\eqref{eq:Ggamma-iso} to first-order in the isospin-breaking parameters, obtaining (c.f. Eq.~\eqref{eq:overlineQ})
\begin{equation}
\begin{split}
    G_{00}^\gamma =  \frac{(\Qu+\Qd)^2}{4} (Z_V^\QCD)^2 \bigg\lbrace  & \, \frac{Z_{\gamma,0}^2}{(Z_V^\QCD)^2}  \Big[2(c) - 4 (d)\Big]
    - 2 \overline {\Delta m} \Big[(M)- 2(O) -(R) + 2 (Rd)\Big]
    \\ & - e^2 (Z_V^\QCD)^2 \Big[\overline Q_2 ((V) + 2(S) - (D1) + 2 (D1d) - 4 (D3)) \\&  - \overline Q^2 ((F) + 2 (T) - 4 (Td) - (D2) + 2 (D2d)) \Big] 
    \bigg\rbrace  \,,
    \label{eq:Ggamma00}
\end{split}
\end{equation}
\begin{equation}
\begin{split}
    \label{eq:Ggamma01}
    G_{01}^\gamma = \frac{(\Qu^2 - \Qd^2)}{4} & \, (Z_V^\QCD)^2 \Big[ - 2 (\Delta \m{u} - \Delta \m{d}) ((M)-(O)) \\ &
    - e^2 (\Qu^2 - \Qd^2) (Z_V^\QCD)^2 ((V)+ 2(S) - 2(T) - (F) - 2 (D3) + 2 (Td)) \Big] \,,
\end{split}
\end{equation}
and
\begin{equation}
\begin{split}
    G_{11}^\gamma =  \frac{ (\Qu - \Qd)^2}{4} &\, (Z_V^\QCD)^2 \bigg\lbrace  2  \frac{Z_{\gamma,1}^2}{(Z_V^\QCD)^2} (c) + e^2 (Z_V^\QCD)^2 \Big[ (\Qu-\Qd)^2  (F) \\ &
    + \overline Q^2 (2 (T) - (D2))  
    - \overline Q_2 ((V)+2(S) - (D1)) 
    \Big] - 2 \overline{\Delta m} ((M) -(R)) 
    \bigg\rbrace \,.
    \label{eq:Ggamma11}
\end{split}
\end{equation}
In the equations above we omit the trivial dependence of all correlations functions on the Euclidean-time separation $t$ between the two external currents.
The renormalization of the three neutral correlators above is in general protected by the Ward Identity, but at non-zero lattice spacing and for local currents the appropriate renormalization factors $Z_{\gamma,1}$ and $Z_{\gamma,0}$ including isospin-breaking effects are required for the isovector and isoscalar currents (see also Refs.~\cite{Blum:2018mom,Erb:2025nxk} for similar discussions). For the latter, we do not give an explicit expression which can be derived by following the arguments in Section~\ref{sec:short-distance} and by adding the relevant topologies missing in the $j^{\gamma,1}$ case.

Using the same decomposition also for $G^W(t,\mu)$ introduced earlier in Eq.~\eqref{eq:GWt}, we find
\begin{equation}
\begin{split}
    G^W(t,\mu) = &\, \frac{1}{2} (Z_V^\QCD)^2 \Bigg\lbrace  \frac{Z_\mathrm{V,\up\down}^2(\mu^2)}{(Z_V^\QCD)^2} (c)(t) +  \frac{e^2}{2} (Z_V^\QCD)^2 \Big[
     \overline Q^2 (2 (T)  - (D2))(t) \\ &  - 2 \Qu \Qd (V)(t)  - \overline Q_2 (2 (S) - (D1))(t) \Big]
     - \overline \Delta m ((M)-(R))(t) \Bigg\rbrace \,.
     \label{eq:G11W}
\end{split}
\end{equation}
Note that even if in a practical Lattice calculation regulating the photon propagator with a PV scheme can be beneficial, e.g. in terms of cutoff effects as observed in Refs.~\cite{Biloshytskyi:2022ets,Erb:2025nxk}, in the end the appropriate conversion factor to the scheme used for the Wilson coefficient is necessary, as discussed earlier in Section~\ref{sec:short-distance}.

By combining Eq.~\eqref{eq:Ggamma11} ($\Qu-\Qd=1$) with Eq.~\eqref{eq:G11W} we calculate $\delta G_{11}(t,\mu)$, defined in Eq.~\eqref{eq:dG11}, and observe large cancellations\footnote{This was initially observed in our earlier work~\cite{Bruno:2018ono} and confirmed in Ref.~\cite{Biloshytskyi:2022ets}.} except for the topologies given by $(V)$ and $(F)$, leading to
\begin{equation}
    \delta G_{11}(t,\mu) = Z_V^\QCD (Z_{\gamma,1} - Z_{V,\up\down}(\mu^2)) (c)(t) + e^2 \frac{(\Qu - \Qd)^2}{4} (Z_V^\QCD)^4 ((F)-(V))(t) \,.
\end{equation}
Similarly to the renormalization factor of the charged current in Eqs.~\eqref{eq:ZVud} and \eqref{eq:Zg1_minus_ZVud}, we observe here that only the topologies $V$ and $F$ survive from the difference, highlighting the fact that $\Gam{V}$ and $\Gam{F}$ precisely renormalize, at this order in the EM coupling, the combination above.
As another consistency check we note that one could start directly from $\delta G_{11}$ (as in Ref.~\cite{Bruno:2018ono}) finding immediately the combination $(F)-(V)$ and for conserved currents, since $G_{11}^\gamma$ is protected by gauge invariance, it follows that $G^W$ renormalizes only with the topologies described by diagrams $(V)$ and $(F)$.
A quick inspection of $(F)$ (we remind the reader that gluons are exchanged between the two bubbles) reveals that it is UV safe~\cite{Biloshytskyi:2022ets}, while $(V)$ is not, implying that $\Gam{F}$ amounts to a finite renormalization.

In summary $\delta G_{11}$ can be calculated from Lattice QCD+QED simulations and depends on the renormalization scale and scheme, together with the QED gauge. Only the combination $\delta G_{11}(t,\mu) + G^W(t,\mu)$ is physical. Once $\delta G_{11}$ is known, one can calculate the shift
\begin{equation}
    \Delta a_\mu^\mathrm{win,\tau}(\mu) = \int dt \, w(t,m_\mu) \, \delta G_{11}(t,\mu) \, \Theta^\mathrm{win}(t) \,,
\end{equation}
to be added to the intermediate window obtained from $G^{W,\exp}(t,\mu)$. We remind the reader that the intermediate window is just a study case, which has however certain benefits. From the point of view of the Lattice calculation it suppresses long distances, where the statistical noise grows, and the short distances, where discretization effects are more prominent. From the phenomenological perspective it is also quite appealing because it reduces the contributions of the higher multiplicity channels, making the inclusive calculation more relevant for a comparison with previous results restricted to the two final-state pions. A dedicated study considering the accuracy of all the individual parts examined here could lead to an optimal window which reduces the overall error.\\

We conclude this Section with a few considerations. By providing a phenomenological description of the spectral densities that define the various correlators above, one could imagine to fix its corresponding free parameters by fitting directly the Euclidean correlators calculated from a Lattice QCD first-principles simulation. While we do not advocate for this exercise, especially in the context of the muon anomaly where high precision and model independence is our priority, it is quite instructive to think in such terms to better understand, from a phenomenological point of view, the content of the individual correlators.

$G_{01}^\gamma$ can be studied in isolation and is well defined in the continuum limit. It contains the physics of the $\rho-\omega$ mixing but since we are using an Euclidean approach, which is fully inclusive, it is affected by both $2\pi$ and $3\pi$ states, the lowest most important channels. The phenomenological analysis reported in  Ref.~\cite{Hoferichter:2023sli} shows that they contribute to $a_\mu$ with similar magnitude but opposite sign, hindering the possibility to disentangle the two effects in $G_{01}^\gamma$. Nevertheless, we do not exclude that studying several window quantities, or attempting to determine its underlying (smeared) density by solving the corresponding Laplace problem, could be helpful. The interested reader may check Ref.~\cite{Erb:2025nxk} for a recent calculation in the 3-flavor theory.

$G_{00}$, which is already non-zero in the isospin limit, is dominated by the $3\pi$ channel (at long distances). At first order in isospin breaking it contains the electro-magnetic and SIB shifts to the various resonances in that channel.

Finally, we examine $\delta G_{11}$, which is purely electromagnetic (at the order considered here). It is reminiscent of the pion mass difference, where exactly the same two topologies contribute. In fact, in an unphysical theory where the $\rho$ meson is stable, one can extract the $\rho$ mass difference using the same approach used for the pions~\cite{deDivitiis:2013xla} (i.e. by isolating the linear term of the time derivative). Since this is not the case, one could consider matching a given model of the underlying neutral and charged spectral densities to the results of a first-principles Lattice QCD calculation. In fact, under the assumption that only two intermediate pions contribute, $\delta G_{11}$ describes the difference between the two-pion form factors in the neutral and charged channels (see the more recent Refs.~\cite{Davier:2023cyp,Castro:2024prg} for a phenomenological model-dependent estimate), one of the major sources of uncertainties in the current prediction of $a_\mu$ from $\tau$ data~\cite{Aliberti:2025beg}. Let us remark again that $\delta G_{11}(t,\mu)$ is prescription dependent and therefore any comparison should be made among  quantities renormalized in the same manner.
While this can be informative in the context of ispospin-breaking corrections in $\tau$ decays, one should also consider a different angle, in particular in light of the recent considerations on this topic reported in Ref.~\cite{Aliberti:2025beg}.
Window quantities, such as the intermediate window contribution to $a_\mu$, have proven to be an excellent meeting point for different approaches. The same should be done here as well taking into account the inclusive nature of our study.

\section{Conclusions}

In this manuscript we have examined isospin-breaking corrections in hadronic $\tau$ decays. Our goal is to design a strategy to attack this problem in a model-independent systematically-improvable approach based on Lattice QCD calculations. The ultimate physics motivation is the prediction of the HVP contribution to the muon anomalous magnetic moment by correcting the experimental data with the corresponding theoretical input. Given the current tensions in the data-driven determinations of the HVP from electron-positron experiments, it is very important to pursue an approach based on a different dataset, also in light of the ongoing experimental efforts for $\tau$ decays at Belle II~\cite{Zani:2023ngd}.

After introducing the basic framework for the theoretical description of hadronic $\tau$ decays and for the calculation of the HVP contribution to $a_\mu$, we have defined a separation of radiative contributions, both real and virtual, in three classes which are infrared safe but gauge, scheme and scale dependent. To understand this dependence, we have examined short-distance corrections using recent developments in the perturbative calculation of the Wilson coefficient in RI schemes~\cite{Gorbahn:2022rgl,Boyle:2026xls}. We have studied the various topologies at $O(\alpha,\Delta m)$ and demonstrated that only two topologies (denoted by letters $V$ and $F$ in our convention) are necessary for the renormalization of the charged quark bilinear. By examining the various components, including the triangle diagram, we have defined the prescription necessary to calculate the individual IR-safe classes such that their sum is physical.

Initial-state radiation is obtained in a factorized form and we provide explicit analytic expressions both with and without the soft-photon approximation, thereby complementing earlier results in the literature~\cite{Cirigliano:2001er,Cirigliano:2002pv}. 

We have examined non-factorizable contributions and showed the expected cancellation of IR divergences at order $\alpha$ in a non-perturbative way. By adopting an inclusive approach and by studying the two time orderings involved (on the hadronic side) we have expressed their effect in terms of two hadronic densities, $\rhopos_{\mu\nu\alpha}$ and $\rhoneg_{\mu\nu\alpha}$, involving three different currents. 
The corresponding Euclidean correlators can develop a double inverse problem related to the two (Euclidean) time separations among the three currents.
For $\rhoneg_{\mu\nu\alpha}$ we have demonstrated that one time coordinate can be analytically continued without issues, and if one further relies on previous results for the explicit reconstruction of the vector-vector correlator, the remaining inverse problem is practically avoided.
The same data can be used for $\rhopos_{\mu\nu\alpha}$ as well but only for the soft-photon limit. For a general treatment we have observed more difficulties and have highlighted the r\^ole of recent developments in the solution of the inverse Laplace transform. 

Finally we have studied final-state corrections directly in Euclidean space from the corresponding correlation functions. We have collected the relevant topologies contributing to the isospin-decomposed HVP correlator and to the (vector) charged correlator. At first order in isospin-breaking, large cancellations occur in the difference of the isovector correlators, $\delta G_{11}$, making it suitable for a direct Lattice calculation. We identified self-consistency checks between the short-distance behavior of the correlation function and its renormalization, which we derived in Section~\ref{sec:short-distance}. Despite being fully inclusive, we advocate here for a detailed study of several windows of $\delta G_{11}$ where higher-multiplicity channels beyond the two-pion one are sufficiently suppressed: in this regime one could compare windows with other phenomenological determinations.

The strategy outlined in this manuscript provides a path towards a first-principles calculation of isospin-breaking effects in hadronic $\tau$ decays, with the specific goal of reducing the current model-dependence and associated uncertainty in the $\tau$-based determination of $a_\mu$. Several considerations and analytic derivations presented in our work are also relevant in other contexts, such as the inclusive determination of CKM parameters from $\tau$ data~\cite{ETMCtau23,ETMCtau24,DiCarlo:2026kpv}.

\begin{acknowledgments}

We thank our colleagues of the RBC and UKQCD collaborations for many valuable discussions and joint efforts over the years.
We thank K.~Maltman, M.~Golterman, G.~Colangelo, P.~Stoffer, M.~Hoferichter, M.~Davier, B.~Malescu and Z.~Zhang for several useful discussions over the years on this topic.
M.~B. is particularly indebted to V.~Cirigliano for many stimulating and illuminating conversations. 
At the beginning of the project, M.~B. was supported by the national program for young researchers
“Rita Levi Montalcini”. M.~B. was (partially) supported by
ICSC - Centro Nazionale di Ricerca in High Performance
Computing, Big Data and Quantum Computing, funded
by European Union – NextGenerationEU.
This work was performed under the auspices of the U.S. Department of Energy by Lawrence Livermore National Laboratory under Contract DE-AC52-07NA27344
and the U.S.~Department of Energy, Office of Science, under the Neutrino Theory Network Program Grant No.\ DE-AC02-07CHI11359 and No.\ DE-SC0020250 (A.S.M.). X.Y.T has been supported by US DOE Contract
DESC0012704(BNL).

\end{acknowledgments}

\appendix

\section{Euclidean and Minkowski conventions}
\label{app:conventions}

In this work we adopt the mostly-minus metric
\begin{equation}
    g_{\mu\nu} = \mathrm{diag}(1,-1,-1,-1) \,.
\end{equation}
To perform the rotation to the Euclidean metric we extend the coordinate system by introducing $x_4 = e^{i \theta} x_0$ from Wick's rotation performed from $\theta=0$ up to $\theta=\pi/2$. Norms of four vectors change accordingly
\begin{equation}
    x^\mu x_\mu = x_0^2 - |\vec x|^2 = - x_4^2 - | \vec x|^2 = - \Eucl x^\mu \Eucl x_\mu \,,
\end{equation}
and we introduce the Euclidean vector $\Eucl x_\mu = (x_1, x_2, x_3, x_4)$ and the Euclidean metric $\delta_{\mu\nu}$.
To preserve slashed terms, Dirac's matrices have to be changed accordingly. Here we follow the common convention
\begin{equation}
    \gamma_0 = \Eucl \gamma_4 \,, \quad -\gamma_k = \gamma^k = i \Eucl \gamma^k = i \Eucl \gamma_k \,,
\end{equation}
where $\Eucl \gamma_\mu$ denotes Dirac matrices with Euclidean signature. From the equation above, and since $\overline \psi$ does not change, we define the following mapping for the vector currents
\begin{equation}
    \J_0(-i x_4, \vec x)  = j_4(x_4, \vec x) \,, \quad
    \J_i(-i x_4, \vec x) = -i j_i(x_4, \vec x) \,.
    \label{eq:J_to_j}
\end{equation}
To simplify the notation we introduce the tensor $n_{\mu\mu'} = \mathrm{diag}(1,-i,-i,-i)$ such that
\begin{equation}
    \J_\mu(-ix_4, \vec x) = n_{\mu\mu'} j_{\mu'}(x_4, \vec x) \,.
\end{equation}
The Euclidean matrices obey the anti-commutator relation $\{\Eucl \gamma_\mu, \Eucl \gamma_\nu \} = 2 \delta_{\mu\nu}$ and, by using $I_n$ to denote the $n \times n$ identity matrix, from $\gamma_0^2 = I_4$, $\gamma_i^2 = - I_4$ and $\gamma_i^\dagger = - \gamma_i$ it follows that
\begin{equation}
     \Eucl \gamma_\mu^2 = I_4 \,, \quad \text{and}\quad  \Eucl \gamma_\mu^\dagger = \Eucl \gamma_\mu \,.
\end{equation}
For a generic slashed term one finds
\begin{equation}
    \slashed x = \gamma^0 x^0 - \gamma^i x^i = - i \Eucl \gamma^4 \Eucl x^4 - i \Eucl \gamma^i x^i = - i \Eucl{\slashed x} \,.
\end{equation}
The fifth Dirac matrix is defined as
\begin{equation}
    \gamma^5 = i \gamma^0 \gamma^1 \gamma^2 \gamma^3 \,,
\end{equation}
and two choices are typically performed for the Euclidean signature
\begin{align}
    \text{Weyl/chiral} & \quad \Eucl \gamma^5 \equiv  \Eucl \gamma^4 \Eucl \gamma^1 \Eucl \gamma^2 \Eucl \gamma^3 = \gamma^5 \\
    \text{DeGrand-Rossi} & \quad \Eucl \gamma^1 \Eucl \gamma^2 \Eucl \gamma^3 \Eucl \gamma^4 = -\gamma^5 \,. 
\end{align}
We specified the different DeGrand-Rossi basis~\cite{DeGrand:1990dk} because it is quite common to find it in several Lattice QCD codes (see also Refs.~\cite{Bhattacharya:2021lol,ljin_g5}). When this is the case left and right chiral projectors should be adjusted accordingly. In Weyl's basis, the one adopted here, since $\gamma_5$ is unchanged, the four-fermion operator $\O(x)$ can be mapped seamlessly to Euclidean signature
\begin{equation}
    4 \, \gamma^{\mu,L} \otimes \gamma_\mu^L = \Eucl \gamma^\mu (1-\gamma_5) \otimes \Eucl \gamma_\mu (1-\gamma_5) \,.
\end{equation}
The same holds for the RI projectors. Since $\gamma_5$ is unchanged Eq.~\eqref{eq:J_to_j} applies to all currents studied in this work. Using our definitions, we find that
\begin{equation}
    \delta^{\mu\nu} j_\mu(x) j_\nu(0) = g^{\mu\nu} \J_\mu(x) \J_\nu(0) \,.
\end{equation}
Since the spectral density may be calculated as
\begin{equation}
    \rho(s) = -\frac{1}{3s} g^{\mu\nu} \rho_{\mu\nu}(p) \,,
\end{equation}
one finds the alternative definition of the correlator $G^\gamma(t)$
\begin{equation}
    G^\gamma(t) = -\frac13 g^{\mu\nu} \int d^3 \vec x  \langle \J_\mu^\gamma(-it, \vec x) \J_\nu^\gamma(0) \rangle = - \frac{1}{3} \delta^{\mu\nu} \int d^3 \vec x \langle j_\mu(t,\vec x) j_\nu(0) \rangle \,.
\end{equation}

\section{Lorentz invariant phase spaces}
\label{app:lips}

\textit{Two-pion case} --- 
Relativistic single pion states are normalized as
\begin{equation}
    \langle \pi^a_{\vec p} \vert \pi^b_{\vec p'} \rangle = \delta_{ab} (2\pi)^3 2 \omega_{\vec p} \delta^3(\vec p - \vec p')\,,
\end{equation}
and for the isovector vector form factor we use the definition
\begin{equation}
    \langle \pi^-(p_1) \, \pi^0(p_2) \vert \J_\mu^-(0) \vert 0 \rangle = -i \, (p_1 - p_2)_\mu \, F_V(s) \,,
\end{equation}
with $|F_V(0)|^2=1$.
For simplicity we consider the rest frame of the hadronic system, $p=(\sqrt s, \vec 0)$. To calculate the two-pion (scalar) spectral density we consider
\begin{equation}
    g^{\mu\nu} \rho_{\mu\nu}(p) = -3 s \rho(s) = \frac{1}{2\pi} \int \frac{d^3 \vec p_1}{(2\pi)^3 2 \omega_1} \frac{d^3 \vec p_2}{(2\pi)^3 2 \omega_2} (2\pi)^4 \delta^4(p-p_1-p_2) g^{\mu\nu} \mathcal H_{2\pi,\mu\nu}(p_1,p_2) \,,
\end{equation}
and restrict hadronic contributions to the two-pion channel in Eq.~\eqref{eq:H_munu}. 
Using a few simplifications (e.g. $s = 2m_\pi^2 + 2 p_1 \cdot p_2$), we arrive at
\begin{equation}
\begin{split}
    \rho(s) = & \, -\frac{1}{3s} \frac{1}{(2\pi)^3} \int \frac{d \omega_1 d\Omega_{\vec p_1}}{2} \vert \vec p_1 \vert \, \frac{d^3 \vec p_2}{2 \omega_2}  \, \delta^4(p-p_1-p_2) \, (4m_\pi^2 - s) \vert F_V(s) \vert^2 \\ = & \, \frac{1}{24 \pi^2} \int \frac{d \omega_1 }{\omega_2} \vert \vec p_1 \vert \, \delta(\sqrt s - 2 \omega_1) \, \frac{(s- 4m_\pi^2)}{s} \vert F_V(s) \vert^2 \,, \quad \left[|\vec p_1|^2 = \frac{s}{4}\left(1 - \frac{4m_\pi^2}{s}\right) \right]\\ = & \, \frac{1}{48 \pi^2} \, \left(1 - \frac{4m_\pi^2}{s} \right)^{3/2} \vert F_V(s) \vert^2 \,.
\end{split}
\end{equation}
By inserting the result above in Eq.~\eqref{eq:dGammads} we obtain the well-known expression for the two-pion decay of the $\tau$ lepton~\cite{Davier:2005xq}
\begin{equation}
    \frac{1}{\Gamma_e} \frac{d\Gamma}{d s} = \frac{|\Vud|^2}{2 \mt^2} \kappa(\hat s) \left(1-\frac{4m_\pi^2}{s} \right)^{3/2} |F_V(s)|^2 \,.
\end{equation}

\textit{Two final-state particles} --- In the rest frame of the hadronic system we have
\begin{equation}
    \omega_P = \omega_q + \sqrt s \,, \quad \vec P = \vec q \,.
\end{equation}
For the scalar product of $P \cdot q$ we find
\begin{equation}
    (P-q)^2 = p^2 = s \quad \to \quad P \cdot q = \frac{\mt^2 - s}{2} \,.
\end{equation}
By choosing $q_\mu=(\omega_q,0,0,\omega_q)$ we obtain
\begin{equation}
    (P-p)^2 = q^2 = 0 \quad \to \quad \omega_P = \frac{\mt^2 + s}{2 \sqrt s} \,, \quad \omega_q = \omega_P - \sqrt s = \frac{\mt^2 - s}{2 \sqrt s} \,.
\end{equation}
The decay rate is typically defined in the rest frame of the decaying particle, as signaled by the prefactor $1/(2 \mt)$. By choosing a different frame such prefactor becomes $1/(2 \omega_P)$. Following the rule for time dilation the two are related by multiplying $\omega_P / \mt$~\cite{Weinberg_1995}, which we automatically include in our derivation above.

\textit{Three final-state particles} --- We take the rest frame of the hadronic system and start from
\begin{equation}
    \omega_P = \omega_q + \omega_k + \sqrt s \,, \quad \vec P = \vec q + \vec k \,,
\end{equation}
and choose a frame where $q_\mu=(\omega_q,0,0,\omega_q)$. It follows that $\vec q \cdot \vec k = \omega_q |\vec k| \cos\theta$
with $\theta$ the angle between the neutrino and photon momenta. Setting $k^2 = \mg^2$ we find
\begin{equation}
    |\vec P|^2 = \omega_q^2 + \omega_k^2 - \mg^2 + 2 \omega_q |\vec k| \cos\theta \,.
\end{equation}
Using it inside the relation for energy conservation we can calculate $\omega_q$ first, given by
\begin{equation}
    \omega_q = \frac{1}{2} \frac{\mt^2 - \mg^2 - s - 2 \sqrt{ s} \omega_k}{ \sqrt s + \omega_k - |\vec k| \cos \theta} \,,
\end{equation}
from which we find $\omega_P$.
One can easily verify that in the soft limit ($|\vec k| \to 0$ and $\mg \to 0$), $\omega_q$ and $\omega_P$ tend to their values in the one-to-two particle kinematics, a relevant observation to prove the cancellation of IR divergences. 
Once in polar coordinates, the integration over the norm $|\vec k|$ goes from 0 up to its maximal value, dictated by the maximal energy carried by the photon.

\section{Time-ordered products and spectral densities}
\label{app:tproducts}

In this Appendix, we collect a few general facts on the relation between time-ordered products and spectral densities. We begin by considering the two-point correlator (Minkowski) using Heinsenberg's picture for $x_0>0$,
\begin{equation}
    \langle A | \O_1(x_0) \O_2(0) | B \rangle = \langle A | \O_1(0) \,   e^{-i x_0(\hat H - E_A - i \varepsilon)} \, \O_2(0) | B \rangle \,,
\end{equation}
which we express as the Fourier transform of a certain spectral density. By taking the Fourier transform of the correlator, we find
\begin{equation}
    \int_0^\infty d x_0 \, e^{i k_0 x_0} \, \langle A | \O_1(x_0) \O_2(0) | B \rangle = \langle A | \O_1(0) \, \frac{-i}{\hat H - E_A - k_0 - i \varepsilon} \, \O_2(0) | B \rangle \,.
\end{equation}
If we now consider the case $x_0<0$,
\begin{equation}
    \langle A | \O_2(0) \O_1(x_0) | B \rangle = \langle A | \O_2(0) \,   e^{i x_0(\hat H - E_B - i \varepsilon)} \, \O_1(0) | B \rangle \,,
\end{equation}
we find 
\begin{equation}
    \int_{-\infty}^0 d x_0 \, e^{i k_0 x_0} \, \langle A | \O_2(0) \O_1(x_0) | B \rangle = \langle A | \O_2(0) \, \frac{-i}{\hat H - E_B + k_0 - i \varepsilon} \, \O_1(0) | B \rangle \,.
\end{equation}
It follows that for fixed external states $|A\rangle$ and $|B\rangle$, the time-ordered product is re-written as
\begin{equation}
\begin{split}
    i \int  dx_0 \, e^{ik_0 x_0} \, \langle A| T \{\O_1(x_0) O_2(0) \} | B \rangle = & \, 
    \langle A | \O_1(0) \, \frac{1}{\hat H - E_A - k_0 - i \varepsilon} \, \O_2(0) | B \rangle \\ & 
    + \langle A | \O_2(0) \, \frac{1}{\hat H - E_B + k_0 - i \varepsilon} \, \O_1(0) | B \rangle \,.
\end{split}
\end{equation}
Above we have used invariance under translations, in fact the most general product would have been $\O_1(x_0) \O_2(y_0)$. In that case an overall $\delta$-function for energy conservation would have appeared, which means that in our approach this is implicitly assumed.
Using this result, we rewrite the most relevant expectation value in the context of the non-factorizable corrections, as follows 
\begin{equation}
\begin{split}
    W = i & \, \int dx_0 \, e^{i k_0 x_0} \langle 0 | \O_3(0) \, \delta(\hat H - \omega) \, T \{\O_1(x_0) \, \O_2(0) \} | 0 \rangle = \\
    & \langle 0 | \O_3(0) \, \delta(\hat H - \omega) \bigg[ \, \O_1(0) \, \frac{1}{\hat H - \omega - k_0 - i \varepsilon} \, \O_2(0) + \O_2(0) \, \frac{1}{\hat H + k_0 - i \varepsilon} \, \O_1(0) \bigg] | 0 \rangle \,.
\end{split}
\end{equation}
Using the relation
\begin{equation}
    \left( \frac{i}{a+i b} \right)^\dagger = \frac{i}{-a + ib} \,,
\end{equation}
one can derive the explicit expression for $W^\dagger$ as well.

Let us turn to the Euclidean metric. With little algebra one can show that the following relations hold ($t$ and $\tau$ are Euclidean times)
\begin{equation}
\begin{split}
    \int_0^t d\tau \, e^{k_0 (\tau-t)} \, \O_1(t) \O_2(\tau) =  \int d\omega \, e^{t \hat H} \O_1(0) \Big[ &\,  \delta(\hat H - \omega) e^{-t(\omega + k_0)} \O_2(0) \frac{1}{\hat H - \omega - k_0} \\ & 
    +  \frac{1}{\hat H + k_0 - \omega} \O_2(0) \delta(\hat H - \omega) e^{-t \omega} \Big] \,,
    \label{eq:integral_eucl_2pt}
    \end{split}
\end{equation}
and for $k_0>-E_0$ (with $E_0$ the relevant lowest state of $\hat H$)
\begin{equation}
    \int_{-\infty}^t d\tau e^{-k_0 (t-\tau)} \O_1(t) \O_2(\tau) | 0 \rangle = 
    e^{t \hat H} \O_1(0) \frac{1}{\hat H + k_0} \O_2(0) | 0 \rangle \,,
    \label{eq:integral_eucl_minf}
\end{equation}
\begin{equation}
    \int_t^{\infty} d\tau e^{-k_0 (\tau-t)} \langle 0 | \O_2(\tau) \O_1(t) = 
    \langle 0 |\O_2(0) \frac{1}{\hat H + k_0} \O_1(0) e^{-t \hat H} \,.
    \label{eq:integral_eucl_pinf}
\end{equation}
These simple two-operator relations can be extended and used inside three-point functions to prove the following additional identities (under the same restrictions on $k_0$):
\begin{equation}\label{eq:integral_eucl_minf2}
    \int_{-\infty}^0 d\tau \, e^{k_0 \tau} \langle 0 | \O_3(t) \O_1(0) \O_2(\tau) | 0\rangle = \int d\omega \, e^{-\omega t} \langle 0 | \O_3(0) \delta(\hat H - \omega)  \O_1(0) \frac{1}{\hat H + k_0} \O_2(0) | 0 \rangle \,,
\end{equation}
\begin{equation}
    \int_{t}^\infty d\tau \, e^{-k_0 (\tau-t)} \langle 0 | \O_2(\tau) \O_3(t) \O_1(0) | 0\rangle = \int d\omega \, \langle 0 | \O_2(0) \frac{1}{\hat H + k_0} \O_3(0) \delta(\hat H - \omega) \O_1(0) | 0 \rangle e^{-\omega t} \,,
\end{equation}
and
\begin{equation}\label{eq:integral_eucl_2pt2}
\begin{split}
    \int_0^t d\tau \, e^{k_0 \tau} \langle 0 | \O_3(t) \O_2(\tau) \O_1(0) | 0\rangle = &\, \int d\omega \, e^{-\omega t} \langle 0 | \O_3(0) \bigg[ \delta(\hat H - \omega)  \O_2(0) \frac{1}{\hat H - \omega - k_0} \\ & + \, e^{k_0 t} \frac{1}{\hat H -\omega + k_0} \O_2(0) \delta(\hat H - \omega)   \bigg] \O_1(0) | 0 \rangle \,.
\end{split}
\end{equation}
Neglecting for a moment the exponential factor $e^{-\omega t}$, i.e. assuming that we solved its corresponding inverse problem, we observe that Eq.~\eqref{eq:integral_eucl_minf2} reproduces the second part of $W$. Similarly Eq.~\eqref{eq:integral_eucl_2pt2} produces the first term of $W$ with an additional unwanted contribution. The latter is important to guarantee the convergence of the Euclidean integral, thereby signaling the presence of a second inverse problem, related to the time coordinate $\tau$.

\section{Useful relations for non-factorizable effects}
\label{app:GEM}

\textit{Pole structure ---} In the following we examine in more details the pole structure of the integrand over $dk_0$ used to calculate the virtual non-factorizable effects.
Starting from the textbook expansion
\begin{equation}
    \frac{1}{k^2 - \mg^2 + i \varepsilon} = \frac{1}{2 \omega_k} \left[\frac{1}{k_0 - (\omega_k - i \varepsilon)} - \frac{1}{k_0 - (-\omega_k + i \varepsilon)}\right],
\end{equation}
we see that by encircling the pole located at $k_0 = -\omega_k + i\varepsilon$ we are setting the photon on its mass shell, $k^2=\mg^2$, while picking the residue $-1/(2\omega_k)$. The residue of the other pole instead is $1/(2\omega_k)$.
By applying the same decomposition also to the denominator of the lepton propagator appearing in the loop correction, we find
\begin{equation}
    \frac{1}{(P+k)^2 - \mt^2 + i \varepsilon} = \frac{1}{2 \omega_{P+k}} \left[\frac{1}{k_0 - (-P_0 + \omega_{P+k} - i \varepsilon)} - \frac{1}{k_0 - (-P_0 -\omega_{P+k} + i \varepsilon)}\right] \,.
\end{equation}
When we encircle the pole with a positive imaginary component we select a residue with a negative sign
\begin{equation}
    k_0 = -P_0 - \omega_{P+k} + i\varepsilon \to - \frac{1}{2 \omega_{P+k}} \,.
\end{equation}

\textit{Pauli-Villars case ---} Here, we revisit the case with Pauli-Villars regulators in the triangle diagram.
After re-inserting the $i\varepsilon$ prescription in the propagators, by closing the contour integral in the upper complex plane we encircle the pole located at $-\omega_k+i \varepsilon$ together with the one situated at  $-\sqrt{\mu^2 + |\vec k|^2} + i\varepsilon$. If we discard the third pole from the denominator $-2P\cdot k+k^2$, for $x_0>0$ we find
\begin{equation}
\begin{split}
\label{eq:appendix_PV_k0_integral}
    \int \frac{dk_0}{2\pi i} \frac{ e^{i k_0 x_0}}{-2P \cdot k + k^2 + i\varepsilon} &\, \frac{\mg^2 - \mu^2}{(k^2-\mg^2 + i \varepsilon)(k^2-\mu^2 + i\varepsilon)} \\ 
    = &  \frac{-1}{- 2 P \cdot k + \mg^2}  \, \frac{e^{-i \omega_k x_0}}{2\omega_k} - \frac{-1}{- 2 P \cdot k + \mu^2} \frac{e^{-i \sqrt{\mu^2 + |\vec k|^2} x_0}}{2 \sqrt{\mu^2 + |\vec k|^2}} + \dots
\end{split}
\end{equation}
As expected the second term regulates the ultraviolet behavior making the integral convergent in the limit $|\vec k| \to \infty$.

\bibliographystyle{JHEP}
\bibliography{biblio,biblio3,SM}

\end{document}